\documentclass[12pt,english]{article}
\usepackage{times}
\usepackage[T1]{fontenc}
\usepackage{geometry}
\usepackage{rotating}
\usepackage{graphicx}
\usepackage{setspace}
\usepackage{amssymb}

\makeatletter

\providecommand{\tabularnewline}{\\}

\usepackage{bm}

\usepackage{babel}
\makeatother
\begin{document}

\section*{\noindent Generalized Analysis and Unified Design of \emph{EM} Skins}

\noindent ~

\noindent \vfill

\noindent G. Oliveri,$^{(1)(2)}$, M. Salucci,$^{(1)(2)}$, and A.
Massa,$^{(1)(2)(3)(4)}$

\noindent \vfill

\noindent ~

\noindent {\footnotesize $^{(1)}$} \emph{\footnotesize ELEDIA Research
Center} {\footnotesize (}\emph{\footnotesize ELEDIA}{\footnotesize @}\emph{\footnotesize UniTN}
{\footnotesize - University of Trento)}{\footnotesize \par}

\noindent {\footnotesize DICAM - Department of Civil, Environmental,
and Mechanical Engineering}{\footnotesize \par}

\noindent {\footnotesize Via Mesiano 77, 38123 Trento - Italy}{\footnotesize \par}

\noindent \textit{\emph{\footnotesize E-mail:}} {\footnotesize \{}\emph{\footnotesize giacomo.oliveri,
marco.salucci, andrea.massa}{\footnotesize \}@}\emph{\footnotesize unitn.it}{\footnotesize \par}

\noindent {\footnotesize Website:} \emph{\footnotesize www.eledia.org/eledia-unitn}{\footnotesize \par}

\noindent {\footnotesize ~}{\footnotesize \par}

\noindent {\footnotesize $^{(2)}$} \emph{\footnotesize CNIT - \char`\"{}University
of Trento\char`\"{} ELEDIA Research Unit }{\footnotesize \par}

\noindent {\footnotesize Via Sommarive 9, 38123 Trento - Italy}{\footnotesize \par}

\noindent {\footnotesize Website:} \emph{\footnotesize www.eledia.org/eledia-unitn}{\footnotesize \par}

\noindent {\footnotesize ~}{\footnotesize \par}

\noindent {\footnotesize $^{(3)}$} \emph{\footnotesize ELEDIA Research
Center} {\footnotesize (}\emph{\footnotesize ELEDIA}{\footnotesize @}\emph{\footnotesize UESTC}
{\footnotesize - UESTC)}{\footnotesize \par}

\noindent {\footnotesize School of Electronic Engineering, Chengdu
611731 - China}{\footnotesize \par}

\noindent \textit{\emph{\footnotesize E-mail:}} \emph{\footnotesize andrea.massa@uestc.edu.cn}{\footnotesize \par}

\noindent {\footnotesize Website:} \emph{\footnotesize www.eledia.org/eledia}{\footnotesize -}\emph{\footnotesize uestc}{\footnotesize \par}

\noindent {\footnotesize ~}{\footnotesize \par}

\noindent {\footnotesize $^{(4)}$} \emph{\footnotesize ELEDIA Research
Center} {\footnotesize (}\emph{\footnotesize ELEDIA@TSINGHUA} {\footnotesize -
Tsinghua University)}{\footnotesize \par}

\noindent {\footnotesize 30 Shuangqing Rd, 100084 Haidian, Beijing
- China}{\footnotesize \par}

\noindent {\footnotesize E-mail:} \emph{\footnotesize andrea.massa@tsinghua.edu.cn}{\footnotesize \par}

\noindent {\footnotesize Website:} \emph{\footnotesize www.eledia.org/eledia-tsinghua}{\footnotesize \par}

\noindent \vfill

\noindent \textbf{\emph{This work has been submitted to the IEEE for
possible publication. Copyright may be transferred without notice,
after which this version may no longer be accessible.}}

\noindent \vfill

\newpage
\section*{Generalized Analysis and Unified Design of \emph{EM} Skins}

~

~

~

\begin{flushleft}G. Oliveri, M. Salucci, and A. Massa\end{flushleft}

\vfill

\begin{abstract}
\noindent A generalized formulation is derived for the analysis of
the field manipulation properties of electromagnetic skins (\emph{EMS}s)
in the working regimes of interest for wireless communications. Based
on such a theoretical framework, a unified method for the design of
anomalous-reflecting and focusing \emph{EMS}s is presented. Representative
results, from a wide set of numerical experiments, are reported and
validated with full-wave \emph{HFSS} simulations to give the interested
readers some insights on the accuracy, the effectiveness, and the
computational efficiency of the proposed analysis/synthesis tools.

\vfill
\end{abstract}
\noindent \textbf{Key words}: Static Passive \emph{EM} Skins; Inverse
Scattering; Inverse Problems; Smart Electromagnetic Environment; \emph{EM}
Holography; Next-Generation Communications; Metamaterials, Metasurfaces.

\newpage
\section{Introduction and Motivation\label{sec:Introduction}}

\noindent The methodological approaches commonly adopted for designing
wireless systems are currently subject to a deep revisiting due to
the recent introduction of the Smart ElectroMagnetic Environment (\emph{SEME})
paradigm \cite{Di Renzo 2019}-\cite{Barbuto 2022} where the \emph{propagation
environment} is no longer an uncontrollable {}``element'' of the
wireless system that affects/limits its performance, but it can be
actually profitably exploited by the designer \cite{Di Renzo 2019}\cite{Massa 2021}\cite{Oliveri 2021c}.
Towards this end, suitable wave manipulating devices such as the static-passive
\emph{EMS}s (\emph{SP-EMS}s) or the reconfigurable-passive \emph{EMS}s
(\emph{RP-EMS}s), called reconfigurable intelligent surfaces (\emph{RIS}s),
as well, have been introduced \cite{Di Renzo 2019}\cite{Massa 2021}\cite{Oliveri 2021c}
and non-negligible improvements in the performance of the wireless
system have been already demonstrated in several applicative areas
including communications \cite{Di Renzo 2019}-\cite{Massa 2021}\cite{Benoni 2021},
sensing \cite{Dardari 2022}, and wireless power transmission \cite{Yang 2021}.

\noindent It is worth pointing out that the use of \emph{EMS}s (i.e.,
both static and reconfigurable non-amplifying structures) has been
proposed to improve the wireless coverage and the quality-of-service
(\emph{QoS}), while minimizing the energy requirements and the needs,
in terms of network architecture and protocols, for their integration
within the wireless infrastructure \cite{Massa 2021}-\cite{Rocca 2022}.
On the other hand, passive reflecting structures can only redirect
the wireless energy they collect \cite{Massa 2021}\cite{Oliveri 2021c},
thus \emph{EMS}s with electrically-large apertures are necessary in
many applicative scenarios \cite{Dardari 2022}\cite{Mei 2022}-\cite{Danufane 2021}
to gather/deliver a sufficient amount of power.

\noindent Since the far-field (\emph{FF}) region of any electromagnetic
(\emph{EM}) device starts at a distance proportional to the square
of its size \cite{Balanis 1997}, \emph{EMS}s with apertures of hundreds
of wavelengths will often serve receiver terminals located in the
Fresnel region {[}i.e., the so-called radiative near-field (\emph{R-NF}){]}
\cite{Dardari 2022}\cite{Mei 2022}-\cite{Danufane 2021}. Therefore,
the extension/generalization of the formulation derived for \emph{FF}
conditions \cite{Diaz 2021} has been recently considered for a reliable
\emph{EM} prediction of the \emph{EMS}s behaviour. For instance, asymptotic
path loss formulas have been derived for \emph{R-NF/FF} cases \cite{Danufane 2021},
while closed-form approximations to determine the ergodic capacity
of \emph{R-NF}/\emph{FF} \emph{RIS}-aided wireless communication systems
have been defined \cite{Zhang 2022}. Furthermore, the exploitation
of the \emph{R-NF} properties of reflecting panels has been investigated
in \cite{Dardari 2022} to yield a robust localization. The impact
of the \emph{NF} focusing \emph{}on the cumulative distribution function
of the \emph{NF} gain of \emph{RIS}s has been also evaluated by means
of a phase-control strategy based on ideal spherical wavefronts radiated
by each meta-atom of the \emph{RIS} \cite{Mei 2022}. However, to
the best of the authors' knowledge, no rigorous and complete \emph{EM}-based
derivation is still available for (\emph{i}) a generalized method
for the analysis of the field manipulation properties of large \emph{EMS}s
in the working regimes of interest for wireless communications (i.e.,
the \emph{R-NF} and the \emph{FF} regions) starting from their micro-scale
properties (i.e., the electric and magnetic surface currents as controlled
by the local susceptibility of the \emph{EMS} \cite{Yang 2019}) and
for (\emph{ii}) the physically-driven synthesis of large \emph{EMS}s.

\noindent This paper is then aimed at complementing previous state-of-the-art
works on the topic \cite{Dardari 2022}\cite{Mei 2022}-\cite{Danufane 2021}\cite{Diaz 2021}
(\emph{i}) to derive rigorous yet simple expressions for the \emph{EM}
field reflected by large \emph{EMS}s in both the \emph{R-NF} and the
\emph{FF} regions, hence extending/generalizing the existing approaches
in the \emph{SEME} literature only based on \emph{FF} approximations
\cite{Oliveri 2021c}\cite{Diaz 2021}\cite{Yang 2019}, (\emph{ii})
to introduce a unified design approach for the synthesis of \emph{EMS}s%
\footnote{\noindent As a matter of fact, the derived theory applies regardless
of the meta-atom technology (i.e., static or reconfigurable passive
unit cells). For the sake of simplicity, the numerical validation
will refer to \emph{SP-EMS}s.%
} in wireless communications, and (\emph{iii}) to prove the advantages,
in terms of power focusing, of using such a unified strategy when
considering different setups, \emph{EMS} apertures, transmitting/receiving
antennas, distances, and \emph{Snell} or \emph{anomalous} (i.e., \emph{Non-Snell})
directions of reflection of the impinging wave.

\noindent The main methodological innovations of the proposed work
with respect to the state-of-the-art include:

\begin{itemize}
\item the derivation and the validation of a generalized analytic expressions
for the \emph{EM} field reflected by arbitrary \emph{EMS}s used in
wireless communications, which is based on the \emph{Generalized Sheet
Transition Condition} (\emph{GSTC}) theoretical framework \cite{Yang 2019};
\item the introduction of a unified method for the design of anomalous-reflecting
wave manipulating devices used in wireless communications;
\item the numerical assessment, also against commercial full-wave simulators
(i.e., \emph{Ansys} \emph{HFSS} \cite{HFSS 2021}), of the reliability
of the generalized analysis method and of the effectiveness of arising
synthesis approach.
\end{itemize}
\noindent The outline of the paper is as follows. After the generalization
of the method \emph{}for the analysis of the \emph{EM} radiation from
\emph{EMS}s to include all the working conditions of interest for
wireless communications (Sect. \ref{sec:Analysis}), a unified technique
for the synthesis of arbitrary \emph{EMS}s is derived and detailed
in Sect. \ref{sec:Synthesis}. A set of representative results from
an exhaustive numerical analysis is presented and discussed in Sect.
\ref{sec:Results}, also in comparison with full-wave simulations,
to illustrate the features and to assess the reliability of the proposed
theoretical formulation as well as to provide some insights on the
effectiveness and the computational efficiency of the arising \emph{EMS}
design method. Some concluding remarks follow (Sect. \ref{sec:Conclusions-and-Remarks}).

\section{\noindent Generalized \emph{EMS}-Analysis \label{sec:Analysis}}

\noindent Let us consider the reference scenario in Fig. 1(\emph{a})
where a primary source, which models a base station (\emph{BTS}) or
a generic wireless transmitter located at $\mathbf{r}_{TX}=\left(r_{TX},\theta_{TX},\varphi_{TX}\right)$,
illuminates a passive \emph{EMS} in the origin of the \emph{EMS} local
coordinate system $\left(x,y,z\right)$ with an incident time-harmonic
field having electric/magnetic component $\mathbf{F}_{inc}^{o}\left(\mathbf{r}\right)$
($o\in\left\{ e,h\right\} $)%
\footnote{\noindent For notation simplicity, the time-dependence term $\exp\left(j\omega t\right)$
is omitted hereinafter.%
}.

\noindent According to the \emph{GSTC} formulation \cite{Oliveri 2021c}\cite{Yang 2019}\cite{Oliveri 2022},
the electric and the magnetic surface currents induced on the aperture
$\Omega$ of the \emph{EMS} are given by\begin{equation}
\mathbf{J}^{e}\left(\mathbf{r}\right)=j\omega\mathbf{P}_{t}^{e}\left(\mathbf{r}\right)-\widehat{\bm{\nu}}\times\nabla_{t}P_{\nu}^{h}\left(\mathbf{r}\right)\label{eq:GSTC Je}\end{equation}
\begin{equation}
\mathbf{J}^{h}\left(\mathbf{r}\right)=j\omega\mu_{0}\mathbf{P}_{t}^{h}\left(\mathbf{r}\right)+\frac{1}{\varepsilon_{0}}\widehat{\bm{\nu}}\times\nabla_{t}P_{\nu}^{e}\left(\mathbf{r}\right)\label{eq:GSTC Jm}\end{equation}
where $\widehat{\bm{\nu}}$ is the normal to the \emph{EMS} surface,
$\nabla_{t}$ is the transverse gradient operator, $\varepsilon_{0}$
and $\mu_{0}$ are the free-space electric permittivity and magnetic
permeability, respectively. Moreover, $\mathbf{P}^{o}\left(\mathbf{r}\right)$
($o\in\left\{ e,h\right\} $) is the electric/magnetic polarization
surface density, $P_{\nu}^{o}\left(\mathbf{r}\right)$ and $P_{t}^{o}\left(\mathbf{r}\right)$
being the normal component {[}$P_{\nu}^{o}\left(\mathbf{r}\right)\triangleq\mathbf{P}^{o}\left(\mathbf{r}\right)\cdot\widehat{\bm{\nu}}${]}
and the transversal one {[}$\mathbf{P}_{t}^{o}\left(\mathbf{r}\right)\triangleq\mathbf{P}^{o}\left(\mathbf{r}\right)-P_{\nu}^{o}\left(\mathbf{r}\right)\widehat{\bm{\nu}}${]},
respectively, that assumes the following expression \cite{Oliveri 2021c}\cite{Yang 2019}\cite{Oliveri 2022}\begin{equation}
\begin{array}{c}
\mathbf{P}^{e}\left(\mathbf{r}\right)\approx\sum_{m=1}^{M}\sum_{n=1}^{N}\left[\varepsilon_{0}\overline{\overline{\psi}}^{e}\left(\mathbf{g}_{mn}\right)\cdot\left.\mathbf{F}_{mn}^{e}\right\rfloor _{avg}\right]\Upsilon_{mn}\left(\mathbf{r}\right)\\
\mathbf{P}^{h}\left(\mathbf{r}\right)\approx\sum_{m=1}^{M}\sum_{n=1}^{N}\left[\overline{\overline{\psi}}^{h}\left(\mathbf{g}_{mn}\right)\cdot\left.\mathbf{F}_{mn}^{h}\right\rfloor _{avg}\right]\Upsilon_{mn}\left(\mathbf{r}\right)\end{array}\label{eq:polarization density}\end{equation}
for sufficiently symmetric unit cells of the \emph{EMS} that consists
of $M\times N$ meta-atoms distributed on a uniform lattice with domains
of area $\Delta_{x}\times\Delta_{y}$, centered at the positions \{$\mathbf{r}_{mn}=\left(x_{m},y_{n}\right)\in\Omega$;
$m=1,...,M$; $n=1,...,N$\}, and characterized by $L$ geometrical/electrical
descriptors, $\mathbf{g}_{mn}\triangleq\left\{ g_{mn}^{l};l=1,...,L\right\} $
($n=1,...,N$, $m=1,...,M$), that control the diagonal tensors of
the electric/magnetic local surface susceptibilities \cite{Yang 2019}
$\overline{\overline{\psi}}^{o}\left(\mathbf{g}_{mn}\right)\triangleq\sum_{a=x,y,z}\psi_{aa}\left(\mathbf{g}_{mn}\right)\widehat{\mathbf{a}}\widehat{\mathbf{a}}$
($o\in\left\{ e,h\right\} $). Moreover, $\Upsilon_{mn}$ {[}$\Upsilon_{mn}\left(\mathbf{r}\right)\triangleq\left\{ 1\, if\,\mathbf{r}\in\Omega_{mn},\,0\, if\,\mathbf{r}\notin\Omega_{mn}\right\} ${]}
is the ($m$, $n$)-th ($m=1,...,M$; $n=1,...,N$) basis function
defined on the support $\Omega_{mn}$ {[}$\Omega_{mn}$ $\triangleq$
\{$x_{m}-\frac{\Delta_{x}}{2}\leq x\leq x_{m}+\frac{\Delta_{x}}{2}$;
$y_{n}-\frac{\Delta_{y}}{2}\leq y\leq y_{n}+\frac{\Delta_{y}}{2}$\};
$\Omega=\bigcup_{n=1}^{N}\bigcup_{m=1}^{M}\Omega_{mn}${]}, while
$\left.\mathbf{F}_{mn}^{o}\right\rfloor _{avg}$ is the corresponding
surface averaged field given by \cite{Oliveri 2021c}\cite{Yang 2019}\cite{Oliveri 2022}\begin{equation}
\left.\mathbf{F}_{mn}^{o}\right\rfloor _{avg}=\frac{\int_{x_{m}-\frac{\Delta_{x}}{2}}^{x_{m}+\frac{\Delta_{x}}{2}}\int_{y_{n}-\frac{\Delta_{y}}{2}}^{y_{n}+\frac{\Delta_{y}}{2}}\left[\mathbf{F}_{inc}^{o}\left(\mathbf{r}\right)+\overline{\overline{\Gamma}}_{mn}\cdot\mathbf{F}_{inc}^{o}\left(\mathbf{r}\right)\right]dxdy}{2\times\Delta_{x}\times\Delta_{y}},\label{eq:avg field}\end{equation}
$\overline{\overline{\Gamma}}_{mn}\triangleq\left[\begin{array}{cc}
\Gamma_{\bot\bot}\left(\mathbf{g}_{mn}\right) & \Gamma_{\parallel\bot}\left(\mathbf{g}_{mn}\right)\\
\Gamma_{\bot\parallel}\left(\mathbf{g}_{mn}\right) & \Gamma_{\parallel\parallel}\left(\mathbf{g}_{mn}\right)\end{array}\right]$ being the local reflection tensor.

\noindent According to the Love's equivalence principle \cite{Balanis 1997},
the field reflected by an \emph{EMS} in an arbitrary location $\mathbf{r}$,
\{$\mathbf{F}_{ref}^{o}\left(\mathbf{r}\right)$, $o\in\left\{ e,h\right\} $\},
turns out to be equal to that radiated in free-space by the equivalent
current $\mathbf{J}^{o}\left(\mathbf{r}\right)$ ($o\in\left\{ e,h\right\} $)
in (\ref{eq:GSTC Je})-(\ref{eq:GSTC Jm}). Moreover, $\mathbf{F}_{ref}^{o}\left(\mathbf{r}\right)$
($o\in\left\{ e,h\right\} $) can be expressed in terms of the auxiliary
electric\begin{equation}
\mathbf{A}^{e}\left(\mathbf{r}\right)=\frac{\mu_{0}}{4\pi}\int_{\Omega}\mathbf{J}^{e}\left(\mathbf{r}'\right)\frac{\exp\left(-jk_{0}\left|\mathbf{r}-\mathbf{r}'\right|\right)}{\left|\mathbf{r}-\mathbf{r}'\right|}d\mathbf{r}'\label{eq:electric potential}\end{equation}
and magnetic\begin{equation}
\mathbf{A}^{h}\left(\mathbf{r}\right)=\frac{\varepsilon_{0}}{4\pi}\int_{\Omega}\mathbf{J}^{h}\left(\mathbf{r}'\right)\frac{\exp\left(-jk_{0}\left|\mathbf{r}-\mathbf{r}'\right|\right)}{\left|\mathbf{r}-\mathbf{r}'\right|}d\mathbf{r}'\label{eq:magnetic potential}\end{equation}
vector potentials as follows \cite{Balanis 1997} \begin{equation}
\mathbf{F}_{ref}^{e}\left(\mathbf{r}\right)=-j\omega\mathbf{A}^{e}\left(\mathbf{r}\right)-\frac{j}{\omega\mu_{0}\varepsilon_{0}}\nabla\left(\nabla\cdot\mathbf{A}^{e}\left(\mathbf{r}\right)\right)-\frac{1}{\varepsilon_{0}}\nabla\times\mathbf{A}^{h}\left(\mathbf{r}\right)\label{eq:reflected electric field}\end{equation}
\begin{equation}
\mathbf{F}_{ref}^{h}\left(\mathbf{r}\right)=-\frac{1}{j\omega\mu_{0}}\nabla\times\mathbf{F}_{ref}^{e}\left(\mathbf{r}\right),\label{eq:reflected magnetic field}\end{equation}
$k_{0}\triangleq\omega\sqrt{\varepsilon_{0}\mu_{0}}$ being the free-space
wavenumber. Therefore, field distribution $\mathbf{F}_{ref}^{o}\left(\mathbf{r}\right)$
($o\in\left\{ e,h\right\} $) may be computed by substituting (\ref{eq:GSTC Je})-(\ref{eq:GSTC Jm})
in (\ref{eq:electric potential}) and (\ref{eq:magnetic potential}),
and numerically integrating the arising \{$\mathbf{A}^{o}\left(\mathbf{r}\right)$;
$o\in\left\{ e,h\right\} $\} expressions to be then used in (\ref{eq:reflected electric field}).
However, such an approach is here avoided since it is numerically
cumbersome and it is difficult to derive simple guidelines for the
\emph{EMS} synthesis. Otherwise, the prediction of $\mathbf{F}_{ref}^{o}$
is carried out as detailed in the following.

\noindent Let us start from the observation that the exponential term
in (\ref{eq:electric potential}) and (\ref{eq:magnetic potential})
can be approximated as \cite{Balanis 1997}\begin{equation}
\frac{\exp\left(-jk_{0}\left|\mathbf{r}-\mathbf{r}'\right|\right)}{\left|\mathbf{r}-\mathbf{r}'\right|}\approx\frac{\exp\left(-jk_{0}r\right)\exp\left(jk_{0}r'\cos\chi\right)\exp\left(-jk_{0}\frac{\left(r'\sin\chi\right)^{2}}{2r}\right)}{r},\label{eq:fresnel approximation}\end{equation}
$\chi$ being the angle between $\mathbf{r}$ and $\mathbf{r}'$,
while $r\triangleq\left|\mathbf{r}\right|$ and $r'\triangleq\left|\mathbf{r}'\right|$,
when the {}``\emph{Fresnel condition}'' holds true\begin{equation}
\left\{ \begin{array}{l}
r\geq10D\\
r\geq0.62\sqrt{\frac{D^{3}}{\lambda_{0}}}\end{array}\right..\label{eq:fresnel condition 1}\end{equation}
In (\ref{eq:fresnel condition 1}), $\lambda_{0}=\frac{2\pi}{k_{0}}$
is the free-space wavelength and $D$ is the diameter of the smallest
sphere containing the \emph{EMS} surface current $\mathbf{J}^{o}\left(\mathbf{r}\right)$
($o\in\left\{ e,h\right\} $), that is\begin{equation}
D\triangleq\sqrt{\left(M\times\Delta_{x}\right)^{2}+\left(N\times\Delta_{y}\right)^{2}}.\label{eq:source diameter}\end{equation}
It is worth pointing out that (\ref{eq:fresnel condition 1}) defines
the \emph{R-NF region}, which contains, as a sub-case, the \emph{FF}
one \cite{Balanis 1997}. Therefore, the proposed formulation is expected
to hold true for both \emph{R-NF} and \emph{FF} regions. Further well-known
approximations of (\ref{eq:fresnel approximation}) can be derived
for wider $r$ values (i.e., if $r\geq\frac{2D^{2}}{\lambda}$ \cite{Balanis 1997}),
but they will be neglected owing to the objective of deriving a generalized
mathematical model for the \emph{EM} behavior of \emph{EMS}s used
in wireless communications.

\noindent By substituting (\ref{eq:fresnel approximation}) in (\ref{eq:electric potential}),
one obtains that\begin{equation}
\mathbf{A}^{e}\left(\mathbf{r}\right)\approx\frac{\mu_{0}}{4\pi}\frac{\exp\left(-jk_{0}r\right)}{r}\int_{\Omega}\mathbf{J}^{e}\left(\mathbf{r}'\right)\exp\left(jk_{0}r'\cos\chi\right)\exp\left(-jk_{0}\frac{\left(r'\sin\chi\right)^{2}}{2r}\right)d\mathbf{r}'\label{eq:fresnel potential}\end{equation}
and a similar expression for $\mathbf{A}^{h}\left(\mathbf{r}\right)$
by simply replacing $\mathbf{J}^{e}\left(\mathbf{r}'\right)\leftarrow\mathbf{J}^{h}\left(\mathbf{r}'\right)$
and $\mu_{0}\leftarrow\varepsilon_{0}$. 

\noindent Successively, subject to the condition\begin{equation}
r\geq10\,\lambda\label{eq:fresnel condition 2}\end{equation}
and using (\ref{eq:fresnel potential}) in (\ref{eq:reflected electric field}),
it can be proved (see {}``Appendix A.1'') that

\noindent \begin{equation}
\mathbf{F}_{ref}^{e}\left(\mathbf{r}\right)\approx-\frac{j\exp\left(-jk_{0}r\right)}{2\lambda_{0}r}\left\{ \left[\eta_{0}S_{\theta}^{e}\left(\mathbf{r}\right)+S_{\varphi}^{h}\left(\mathbf{r}\right)\right]\widehat{\bm{\theta}}+\left[\eta_{0}S_{\varphi}^{e}\left(\mathbf{r}\right)-S_{\theta}^{h}\left(\mathbf{r}\right)\right]\widehat{\bm{\varphi}}\right\} \label{eq:near field E expression}\end{equation}
where \begin{equation}
\mathbf{S}^{o}\left(\mathbf{r}\right)\triangleq\int_{\Omega}\mathbf{J}^{o}\left(\mathbf{r}'\right)\exp\left(jk_{0}r'\cos\chi\right)\exp\left(-jk_{0}\frac{\left(r'\sin\chi\right)^{2}}{2r}\right)d\mathbf{r}'\label{eq:near field radiation vector}\end{equation}
is the generalization of the radiation vector that here, unlike the
traditional \emph{FF} radiation theory \cite{Balanis 1997}, depends
on $r$, as well.

\noindent The expression (\ref{eq:near field E expression}) allows
one to derive in algebraic-form the field reflected by an arbitrary
\emph{EMS} in any scenario starting from the induced electric and
magnetic surface currents (\ref{eq:GSTC Je})-(\ref{eq:GSTC Jm})
once (\ref{eq:near field radiation vector}) has been numerically
evaluated. Towards this end, the surface current distribution $\mathbf{J}^{o}$
($o\in\left\{ e,h\right\} $) is firstly discretized by using the
same set of basis functions in (\ref{eq:polarization density})\begin{equation}
\mathbf{J}^{o}\left(\mathbf{r}\right)=\sum_{a=x,y}\sum_{m=1}^{M}\sum_{n=1}^{N}\left(J_{a}^{o}\right)_{mn}\Upsilon_{mn}\left(\mathbf{r}\right)\widehat{\mathbf{a}}\label{eq:discretized current}\end{equation}
where $\left(J_{a}^{o}\right)_{mn}$ is the coefficient of the $a$-th
($a\in\left\{ x,y\right\} $) component of the $o$-th surface ($o\in\left\{ e,h\right\} $)
current in the ($m$, $n$)-th ($m=1,...,M$; $n=1,...,N$) \emph{EMS}
sub-domain.

\noindent By replacing (\ref{eq:discretized current}) in (\ref{eq:near field radiation vector})
and reordering the linear operators, it turns out that\begin{equation}
\mathbf{S}^{o}\left(\mathbf{r}\right)\triangleq\sum_{a=x,y}\widehat{\mathbf{a}}\left[\sum_{m=1}^{M}\sum_{n=1}^{N}\left(J_{a}^{o}\right)_{mn}\gamma_{mn}\left(\mathbf{r}\right)\right]\label{eq:discretized radiation vector}\end{equation}
where\begin{equation}
\gamma_{mn}\left(\mathbf{r}\right)\triangleq\int_{x_{m}-\frac{\Delta_{x}}{2}}^{x_{m}+\frac{\Delta_{x}}{2}}\int_{y_{n}-\frac{\Delta_{y}}{2}}^{y_{n}+\frac{\Delta_{y}}{2}}\exp\left(jk_{0}r'\cos\chi\right)\exp\left(-jk_{0}\frac{\left(r'\sin\chi\right)^{2}}{2r}\right)d\mathrm{x}'d\mathrm{y}'.\label{eq:coefficient radiation vector}\end{equation}
The discretized expression of the radiation vector in (\ref{eq:discretized radiation vector})
is then decomposed in its spherical components\begin{equation}
\begin{array}{c}
S_{\theta}^{o}\left(\mathbf{r}\right)\triangleq\sum_{m=1}^{M}\sum_{n=1}^{N}\left\{ \left[\left(J_{x}^{o}\right)_{mn}\cos\theta\cos\varphi+\left(J_{y}^{o}\right)_{mn}\cos\theta\sin\varphi\right]\gamma_{mn}\left(\mathbf{r}\right)\right\} \\
S_{\varphi}^{o}\left(\mathbf{r}\right)\triangleq\sum_{m=1}^{M}\sum_{n=1}^{N}\left\{ \left[-\left(J_{x}^{o}\right)_{mn}\sin\varphi+\left(J_{y}^{o}\right)_{mn}\cos\varphi\right]\gamma_{mn}\left(\mathbf{r}\right)\right\} \end{array}\label{eq: Rossiglione theta}\end{equation}
to be substituted in (\ref{eq:near field E expression}) for yielding
through simple manipulations (see {}``Appendix A.2'') the final
closed-form of the electric field reflected by an \emph{EMS}

\noindent \begin{equation}
\begin{array}{l}
\mathbf{F}_{ref}^{e}\left(\mathbf{r}\right)\approx-\frac{j\exp\left(-jk_{0}r\right)}{2\lambda_{0}r}\sum_{m=1}^{M}\sum_{n=1}^{N}\mathrm{sinc}\left[\frac{\pi\Delta_{x}}{\lambda_{0}}u\right]\mathrm{sinc}\left[\frac{\pi\Delta_{y}}{\lambda_{0}}v\right]\Delta_{x}\Delta_{y}\\
\exp\left\{ j\frac{\pi}{\lambda_{0}}\left(2\left(x_{m}u+y_{n}v\right)-\frac{1}{r}\left[\left(x_{m}w\right)^{2}+\left(y_{n}w\right)^{2}+\left(x_{m}v-y_{n}u\right)^{2}\right]\right)\right\} \\
\left\{ \left[\eta_{0}\cos\theta\cos\varphi\left(J_{x}^{e}\right)_{mn}+\eta_{0}\cos\theta\sin\varphi\left(J_{y}^{e}\right)_{mn}-\sin\varphi\left(J_{x}^{h}\right)_{mn}+\cos\varphi\left(J_{y}^{h}\right)_{mn}\right]\widehat{\bm{\theta}}+\right.\\
\left.+\left[-\eta_{0}\sin\varphi\left(J_{x}^{e}\right)_{mn}+\eta_{0}\cos\varphi\left(J_{y}^{e}\right)_{mn}+\cos\theta\cos\varphi\left(J_{x}^{h}\right)_{mn}+\cos\theta\sin\varphi\left(J_{y}^{h}\right)_{mn}\right]\widehat{\bm{\varphi}}\right\} .\end{array}\label{eq:final expression field}\end{equation}
where $u\triangleq\sin\theta\cos\varphi$, $v\triangleq\sin\theta\sin\varphi$,
and $w\triangleq\cos\theta$ \cite{Balanis 1997}.

\noindent Such an expression (\ref{eq:final expression field}) allows
one to faithfully predict the \emph{EMS} reflection as a simple double
summation of the surface currents when the conditions (\ref{eq:fresnel condition 1})
and (\ref{eq:fresnel condition 2}) are fulfilled. Moreover, it can
be easily proved that (\ref{eq:final expression field}) reduces to
the relation for the \emph{FF} approximation \cite{Oliveri 2021c}
when $r\to\infty$, since \begin{equation}
\gamma_{mn}\left(r,u,v,w\right)\to\exp\left[j\frac{2\pi}{\lambda_{0}}\left(x_{m}u+y_{n}v\right)\right]\Delta_{x}\Delta_{y}\mathrm{sinc}\left[\frac{\pi\Delta_{x}}{\lambda_{0}}u\right]\mathrm{sinc}\left[\frac{\pi\Delta_{y}}{\lambda_{0}}v\right].\label{eq:}\end{equation}
On the other hand, it must be noted that (\ref{eq:final expression field})
does not hold true in the \emph{reactive} near field region \cite{Balanis 1997}.
While this latter scenario does not generally arise in wireless communications,
the prediction of the corresponding field distribution, $\mathbf{F}_{ref}^{e}\left(\mathbf{r}\right)$,
can be numerically performed by using (\ref{eq:reflected electric field})
along with the discretized versions of (\ref{eq:electric potential})
and (\ref{eq:magnetic potential}), but without benefiting of simple
algebraic closed-form expressions as in the other regimes (\ref{eq:final expression field}).

\section{\noindent Unified \emph{EMS}-Synthesis Method\label{sec:Synthesis}}

\noindent According to the guidelines in \cite{Oliveri 2021c}\cite{Yang 2019}\cite{Oliveri 2022}
for wireless communications, the design of an \emph{EMS} able to maximize
the power received by a terminal located at $\mathbf{r}=\mathbf{r}_{RX}$
{[}$\mathbf{r}_{RX}\triangleq\left(r_{RX},\theta_{RX},\varphi_{RX}\right)${]}
can be formulated as that of finding the \emph{EMS} descriptors such
that the received power \cite{Balanis 1997} \begin{equation}
\Psi_{RX}=\frac{\lambda_{0}^{2}G_{RX}^{\max}}{8\pi\eta_{0}}\left|\mathbf{F}_{ref}^{e}\left(\mathbf{r}_{RX}\right)\right|^{2}\label{eq:received power}\end{equation}
is maximized, $G_{RX}^{\max}$ being the maximum gain of the receiver,
which is assumed to be aligned and perfectly matched in polarization.
This problem is here addressed by first (\emph{a}) synthesizing the
ideal surface currents, $\widetilde{\mathbf{J}}^{o}\left(\mathbf{r}\right)$
($o\in\left\{ e,h\right\} $), that maximize the incident power at
the receiver

\noindent \begin{equation}
\left\{ \widetilde{\mathbf{J}}^{o}\left(\mathbf{r}\right);o\in\left\{ e,h\right\} \right\} =\arg\left\{ \max\left[\left|\mathbf{F}_{ref}^{e}\left(\mathbf{r}_{RX}\right)\right|^{2}\right]\right\} ,\label{eq:current design objective}\end{equation}
and subsequently by (\emph{b}) identifying the optimal setting of
the \emph{EMS} descriptors\begin{equation}
\mathcal{G}^{opt}\triangleq\left\{ \mathbf{g}_{mn}^{opt};m=1,...,M,\, n=1,...,N\right\} \label{eq:}\end{equation}
that support the surface currents, \{$\mathbf{J}^{o}\left(\mathbf{r}\right)$;
$o\in\left\{ e,h\right\} $\}, whose \emph{phase} distributions within
the \emph{EMS} aperture $\Omega$ closely match the ideal ones by
minimizing the phase mismatch cost function\begin{equation}
\Phi\left(\mathcal{G}\right)=\sum_{o=e,h}\sum_{a=x,y}\left[\int_{\Omega}\left|\Delta J_{a}^{o}\left(\mathbf{r};\mathcal{G}\right)\right|^{2}d\mathbf{r}\right]\label{eq:cost function matching}\end{equation}
where $\Delta J_{a}^{o}\left(\mathbf{r};\mathcal{G}\right)\triangleq\arg\left[\widetilde{J}_{a}^{o}\left(\mathbf{r}\right)\right]-\arg\left[J_{a}^{o}\left(\mathbf{r};\mathcal{G}\right)\right]$
(i.e., $\mathcal{G}^{opt}=\arg\left\{ \min_{\mathcal{G}}\Phi\left(\mathcal{G}\right)\right\} $).

\noindent As a matter of fact, owing to the passive nature of an \emph{EMS},
the optimization of its micro-scale structure does not actually allow
a fine control of the current magnitude, $\left|\mathbf{J}^{o}\left(\mathbf{r}\right)\right|$
($o\in\left\{ e,h\right\} $). More specifically, it turns out that
a suitable arrangement of the \emph{EMS} meta-atoms carefully shapes
only the phase distribution of the surface currents, while the magnitudes
are inherited by the field radiated from the primary source on the
\emph{EMS} aperture \cite{Oliveri 2021c}\cite{Yang 2019}\cite{Oliveri 2022}.

\noindent Let us now detail the Step (\emph{a}) concerned with the
synthesis of $\widetilde{\mathbf{J}}^{o}\left(\mathbf{r}\right)$
($o\in\left\{ e,h\right\} $) (\ref{eq:current design objective}).
An approach different from the usual \emph{FF}-focusing \emph{EMS}
strategy \cite{Oliveri 2021c}\cite{Oliveri 2022} is required since
the expression of $\left|\mathbf{F}_{ref}^{e}\left(\mathbf{r}_{RX}\right)\right|^{2}$
in (\ref{eq:received power}) to be maximized is now (\ref{eq:final expression field}),
which extends and generalizes the most commonly adopted \emph{FF}-approximated
one \cite{Oliveri 2021c}\cite{Oliveri 2022}. Towards this end, the
well-known phase-conjugation technique \cite{Mailloux 2005} is used
so that, by cancelling the exponential terms in (\ref{eq:final expression field}),
all current terms are added \emph{in-phase} at the focusing point.
Accordingly, it turns out that if\begin{equation}
\arg\left(\widetilde{J}_{a}^{o}\right)_{mn}=-\arg\left[\gamma_{mn}\left(r_{RX},u_{RX},v_{RX},w_{RX}\right)\right]\label{eq:NF focusing}\end{equation}
($o\in\left\{ e,h\right\} $; $a\in\left\{ x,y\right\} $) where $u_{RX}=\sin\theta_{RX}\cos\varphi_{RX}$,
$v_{RX}=\sin\theta_{RX}\sin\varphi_{RX}$ and $w_{RX}=\cos\theta_{RX}$,
then the addends of the summation in (\ref{eq:final expression field})
can be elided. Thus, by substituting (\ref{eq:coefficient radiation solved})
in (\ref{eq:NF focusing}), the following closed-form rule for the
synthesis of the optimal phase distribution of the \emph{EMS} currents
components ($o\in\left\{ e,h\right\} $; $a\in\left\{ x,y\right\} $)
is derived

\noindent \begin{equation}
\begin{array}{r}
\arg\left(\widetilde{J}_{a}^{o}\right)_{mn}=\frac{\pi}{\lambda_{0}r_{RX}}\left(\left(x_{m}w_{RX}\right)^{2}+\left(y_{n}w_{RX}\right)^{2}+\left(x_{m}v_{RX}-y_{n}u_{RX}\right)^{2}\right)+\\
-\frac{2\pi}{\lambda_{0}}\left(x_{m}u_{RX}+y_{n}v_{RX}\right).\end{array}\label{eq:final NF focusing}\end{equation}
It is worth remarking that (\ref{eq:final NF focusing}) allows one
to analytically synthesize the surface currents without local/global
optimization processes. Moreover, it reduces to the well-known \emph{FF}
steering when $r_{RX}\to\infty$ (i.e., $\arg\left(\widetilde{J}_{a}^{o}\right)_{mn}\to-\frac{2\pi}{\lambda_{0}}\left(x_{m}u_{RX}+y_{n}v_{RX}\right)$,
$o\in\left\{ e,h\right\} $, $a\in\left\{ x,y\right\} $), as theoretically
expected.

\noindent Finally, the power pattern in the focusing point $\mathbf{r}=\mathbf{r}_{RX}$
is determined by substituting (\ref{eq:final NF focusing}) in (\ref{eq:final expression field})\begin{equation}
\begin{array}{l}
\left|\mathbf{F}_{ref}^{e}\left(\mathbf{r}_{RX}\right)\right|^{2}\approx\left(\frac{1}{2\lambda_{0}r}\right)^{2}\sum_{m=1}^{M}\sum_{n=1}^{N}\\
\left\{ \left[\eta_{0}\cos\theta\cos\varphi\left|\left(J_{x}^{e}\right)_{mn}\right|+\eta_{0}\cos\theta\sin\varphi\left|\left(J_{y}^{e}\right)_{mn}\right|-\sin\varphi\left|\left(J_{x}^{h}\right)_{mn}\right|+\cos\varphi\left|\left(J_{y}^{h}\right)_{mn}\right|\right]^{2}+\right.\\
\left.+\left[-\eta_{0}\sin\varphi\left|\left(J_{x}^{e}\right)_{mn}\right|+\eta_{0}\cos\varphi\left|\left(J_{y}^{e}\right)_{mn}\right|+\cos\theta\cos\varphi\left|\left(J_{x}^{h}\right)_{mn}\right|+\cos\theta\sin\varphi\left|\left(J_{y}^{h}\right)_{mn}\right|\right]^{2}\right\} .\end{array}\label{eq:focused beampattern}\end{equation}
As for the {}``Step (\emph{b})'', the \emph{System-by-Design} (\emph{SbD})
\cite{Massa 2021b} implementation, proposed in \cite{Oliveri 2021c}\cite{Oliveri 2022},
is adopted and summarized in the following. A succession of $I$ \emph{SbD}
iterations ($i=1,...,I$) is performed \cite{Massa 2021b} to find
$\mathcal{G}^{opt}$. At each $i$-th ($i=1,...,I$) iteration, $B$
guess \emph{EMS} layouts, \{$\mathcal{G}_{b}^{\left(i\right)}\triangleq\left\{ \left.\mathbf{g}_{mn}\right\rfloor _{b}^{\left(i\right)};m=1,...,M,\, n=1,...,N\right\} $;
$b=1,...,B$\}, are generated by (\emph{i}) the \emph{solution-space-exploration}
block, which is implemented according to the \emph{particle swarm}
evolutionary algorithm \cite{Massa 2021b}. The corresponding $B$
diagonal tensors of the local surface susceptibilities, \{$\overline{\overline{\psi}}^{o}\left(\left.\mathbf{g}_{mn}\right\rfloor _{b}^{\left(i\right)}\right)$
($o\in\left\{ e,h\right\} $); $b=1,...,B$\}, are then selected in
(\emph{ii}) the \emph{local susceptibility look-up table}, whose entries
have been offline-computed with the full-wave Ansys HFSS \cite{HFSS 2021}
commercial SW, and inputted into (\emph{iii}) the \emph{electric/magnetic
surface current computation} block that implements (\ref{eq:polarization density})
to compute the polarization surface densities, \{$\mathbf{P}^{o}\left(\mathbf{r};\mathcal{G}_{b}^{\left(i\right)}\right)$
($o\in\left\{ e,h\right\} $); $b=1,...,B$\}, to yield \{$\mathbf{J}^{o}\left(\mathbf{r};\mathcal{G}_{b}^{\left(i\right)}\right)$
($o\in\left\{ e,h\right\} $); $b=1,...,B$\} through (\ref{eq:GSTC Je})-(\ref{eq:GSTC Jm}).
The surface currents are then provided to (\emph{iv}) the \emph{physical
linkage} block that computes the corresponding cost function values
(\ref{eq:cost function matching}), \{$\Phi\left(\mathcal{G}_{b}^{\left(i\right)}\right)$;
$b=1,...,B$\}. The iterative loop of (\emph{i})-(\emph{iv}) is repeated
until the maximum number of iterations is reached, $i=I$, or alternative
convergence conditions (e.g., the stationarity of the cost function
value \cite{Massa 2021b}) are met. The final layout of the \emph{EMS}
is then defined as $\mathcal{G}^{opt}=\arg\left\{ \min_{i=1,...,I;b=1,...,B}\left[\Phi\left(\mathcal{G}_{b}^{\left(i\right)}\right)\right]\right\} $.

\section{\noindent Numerical Validation and Performance Assessment \label{sec:Results}}

\noindent The objective of this Section is twofold. On the one hand
(Sect. \ref{sub:EMS-Analysis-Validation}), it is aimed at validating
the {}``Generalized \emph{EMS}-Analysis'' presented in Sect. \ref{sec:Analysis}
for predicting the \emph{EM} field reflected by an \emph{EMS} in wireless
communication scenarios. On the other hand (Sect. \ref{sub:EMS-Synthesis-Assessment}),
it is devoted to assess the effectiveness and the efficiency of the
arising {}``\emph{Unified EMS}-\emph{Synthesis Method} {}`` (\emph{USM})
(Sect. \ref{sec:Synthesis}) with respect to the state-of-the-art
\emph{FF}-oriented method \cite{Oliveri 2021c}\cite{Oliveri 2022}
(\emph{FFM}) by considering full-wave simulations as reference ground
truth.

\noindent Towards this end, the benchmark \emph{SEE} scenario consists
of a source operating at $f=17.5$ {[}GHz{]} and equipped with a linearly-polarized
pyramidal horn antenna {[}Fig. 1(\emph{d}){]} that illuminates the
\emph{EMS} from $\left(\theta_{TX},\varphi_{TX}\right)=\left(30,180\right)$
{[}deg{]}. To avoid any bias related to the exploitation of advanced/complex
\emph{EMS} unit cells, a canonical square-shaped patch of side $\Delta l$
(i.e., $\Delta l_{x}=\Delta l_{y}=\Delta l$) with no via holes {[}Fig.
1(\emph{c}){]} has been chosen as elementary \emph{EMS} meta-atom
{[}$\to$ $L=1$ $\Rightarrow$ $g_{mn}\triangleq\left(\Delta l\right)_{mn}$
($n=1,...,N$, $m=1,...,M$){]}. This latter has been modeled in \emph{HFSS}
by considering a single-layer Rogers \emph{RO4350} substrate with
thickness $\tau=7.62\times10^{-4}$ {[}m{]} and a half-wavelength
(i.e., $\Delta_{x}=\Delta_{y}=\Delta=8.565\times10^{-3}$ {[}m{]})
uniform lattice. Moreover, the calibration parameters of the \emph{SbD}-based
synthesis process have been set according to the guidelines in \cite{Oliveri 2021c}\cite{Oliveri 2022}:
$B=10$ and $I=10^{4}$.

\subsection{\emph{EMS}-Analysis Validation\label{sub:EMS-Analysis-Validation}}

\noindent In order to assess the reliability and the accuracy of the
prediction method in Sect. \ref{sec:Analysis}, comparisons with full-wave
simulations, carried out with \emph{Ansys} \emph{HFSS} \cite{HFSS 2021},
have been considered and the results from the analysis of three representative
\emph{EMS}s%
\footnote{\noindent The \emph{EMS} apertures have been chosen to keep reasonable/feasible
the time/memory requirements for their \emph{HFSS} simulation on a
standard workstation (e.g., a workstation equipped with $512$ GB
of RAM and a $32$-core $3.5$ GHz CPU).%
} will be illustrated hereinafter.

\noindent The first test case deals with an \emph{EMS} {[}Fig. 3(\emph{a}){]}
defined on a lattice of $M\times N=48\times48$ domains \emph{}(i.e.,
a square panel of side $\mathcal{L}\approx0.41$ {[}m{]}), composed
by uniform square meta-atoms of side $\Delta l=5.0\times10^{-3}$
{[}m{]}, and illuminated by a $G_{TX}^{\max}=20.4$ {[}dBi{]} horn
antenna (see Tab. I - second column), which is located a distance
of $r_{TX}=50$ {[}m{]} from the \emph{EMS} and fed by a $\Xi_{TX}=20$
{[}dBm{]}-power transmitter, that generates on the \emph{EMS} aperture
$\Omega$ an incident $y$-polarized electric field whose magnitude
and phase distributions are shown in Fig. 2(\emph{a}) and Fig. 2(\emph{b}),
respectively. The \emph{EM} field reflected by the \emph{EMS} has
been predicted with the generalized theory (\ref{eq:final expression field})
and the traditional \emph{FF} radiation theory \cite{Balanis 1997},
while it has been \emph{HFSS}-computed by modeling the finite structure
of the \emph{EMS} and the transmitting horn antennas with the \emph{Finite
Element Boundary Integral} (\emph{FEBI}) approach.

\noindent Figure 4 shows the co-polar field magnitude, $\left|F_{ref,\varphi}^{e}\left(\mathbf{r}\right)\right|$,
in the two planes $\Theta_{NF}$ and $\Theta_{FF}$ of Fig. 1(\emph{b})
along the angular direction $\theta=30$ {[}deg{]}. More specifically,
the former is located in the \emph{R-NF} region of the \emph{EMS,}
$r=6$ {[}m{]} ($\mathbf{r}\in\Theta_{NF}$, $r$ $>$ $r_{NF}$,
$r_{NF}=5.82$ {[}m{]} since $r_{NF}\triangleq\min\left\{ 10\, D;\,10\,\lambda;\,0.62\sqrt{\frac{D^{3}}{\lambda}}\right\} $)
far from the \emph{EMS} aperture $\Omega$. The other is $r=40$ {[}m{]}
from $\Omega$ in the \emph{EMS} \emph{FF} region ($\mathbf{r}\in\Theta_{FF}$,
$r$ $>$ $r_{FF}$, $r_{FF}=39.50$ {[}m{]}, $r_{FF}$ being $r_{FF}\triangleq\min\left\{ 10\, D;\,10\,\lambda;\,\frac{2\times D^{2}}{\lambda}\right\} $).
As expected, while both predictions are very close to the actual/\emph{HFSS}-computed
\emph{EM} distribution in the \emph{FF} plane $\Theta_{FF}$ {[}Figs.
4(\emph{d})-4(\emph{e}) vs. Fig. 4(\emph{f}){]} since (\ref{eq:final expression field})
reduces to the relation for the \emph{FF} approximation \cite{Oliveri 2021c}
when $r\to\infty$, the generalized approach gives a better estimate
of the field reflected by the \emph{EMS} in the \emph{R-NF} plane
$\Theta_{NF}$ {[}Fig. 4(\emph{b}) vs. Figs. 4(\emph{c}){]} as highlighted
by the corresponding map of the prediction error, which is defined
as\begin{equation}
\Delta F_{ref,\varphi}^{e}\left(\mathbf{r}\right)\triangleq\frac{\left|\left.F_{ref,\varphi}^{e}\left(\mathbf{r}\right)\right|_{Predicted}-\left.F_{ref,\varphi}^{e}\left(\mathbf{r}\right)\right|_{HFSS-Computed}\right|^{2}}{\max_{\mathbf{r}\in\Theta}\left|\left.F_{ref,\varphi}^{e}\left(\mathbf{r}\right)\right|_{HFSS-Computed}\right|^{2}},\label{eq:}\end{equation}
in Fig. 5(\emph{b}), as compared to that in Fig. 5(\emph{a}) for the
traditional \emph{FF} theory \cite{Balanis 1997}. It is worth pointing
out that the marginal deviations from the \emph{HFSS} values, as shown
in Fig. 5(\emph{b}), are caused by the unavoidable approximations
due to the discretization of the surface currents with a $\Delta_{x}\times\Delta_{y}$
step (\ref{eq:discretized current}) as well as by the truncation/edge
scattering effects of the finite \emph{EMS} at hand.

\noindent The other two test cases refer to the non-uniform \emph{EMS}s
shown in Figs. 3(\emph{b})-3(\emph{c}) arranged on a uniform lattice
of $M\times N=84\times84$ (i.e., $\mathcal{L}=0.72$ {[}m{]}) $\Delta$-sized
domains. By keeping the same setup of the previous experiment and
performing the same analysis, it turns out that similar conclusions
on the reliability and the accuracy of the generalized approach can
still be drawn. Indeed, the maps of $\Delta F_{ref,\varphi}^{e}\left(\mathbf{r}\right)$
in $\Theta_{NF}$ ($r=10.5$ {[}m{]} and $\theta=10$ {[}deg{]} being
$r_{NF}=10.18$ {[}m{]}) significantly differ {[}Fig. 6(\emph{a})
vs. Fig. 6(\emph{b}) and Fig. 6(\emph{c}) vs. Fig. 6(\emph{d}){]},
while those in $\Theta_{FF}$ ($r=121$ {[}m{]} and $\theta=10$ {[}deg{]}
being $r_{FF}=120.96$ {[}m{]}) look almost equal {[}Fig. 7(\emph{a})
vs. Fig. 7(\emph{b}) and Fig. 7(\emph{c}) vs. Fig. 7(\emph{d}){]}.

\noindent As for the computational efficiency of the proposed approach,
a proof is that an \emph{EM} prediction of the field reflected by
an $M\times N=84\times84$ (i.e., $\mathcal{L}=0.72$ {[}m{]}) \emph{EMS}
with the generalized formulation is yielded in $\Delta t^{Generalized\,\, Theory}\approx2.1$
{[}s{]} with a non-optimized Matlab implementation, while the corresponding
full-wave computation needs $\Delta t^{HFSS}\approx2.6\times10^{6}$
{[}s{]}.

\subsection{\emph{EMS}-Synthesis Assessment\label{sub:EMS-Synthesis-Assessment}}

\noindent The first test case of the numerical assessment of the synthesis
method in Sect. \ref{sec:Synthesis} is concerned with the design
of a $M\times N=120\times120$ \emph{EMS} ($\to$ $\mathcal{L}\approx1.029$
{[}m{]}) illuminated by a source, located at a distance of $r_{TX}=50$
{[}m{]} from $\Omega$ and modeled with a $G_{TX}^{\max}=13.7$ {[}dBi{]}
horn antenna (see Tab. I - first column) that generates on the \emph{EMS}
an incident $y$-polarized electric field whose magnitude and phase
distributions are shown in Fig. 2(\emph{c}) and Fig. 2(\emph{d}),
respectively. The \emph{EMS} is required to establish a wireless link
with a terminal located $r_{RX}=15$ {[}m{]} far from the \emph{EMS}
along the anomalous/non-Snell angular direction $\left(\theta_{RX},\varphi_{RX}\right)=\left(10,0\right)$
{[}deg{]} where a horn antenna of the same type of the transmitting
one has been used as receiver%
\footnote{\noindent For the sake of simplicity, both transmitter and receiver
have been modeled with the same elementary radiator. However, the
\emph{EMS} design process is independent on the receiving antenna.%
}. Since both (\ref{eq:fresnel condition 1}) and (\ref{eq:fresnel condition 2})
conditions are fulfilled and the \emph{R-NF} region of the \emph{EMS}
at hand starts at $r>r_{NF}=14.54$ {[}m{]}, while the \emph{FF} one
initiates at $r>r_{FF}=\frac{2D^{2}}{\lambda_{0}}=246.86$ {[}m{]},
the theoretical approach formulated in Sect. \ref{sec:Analysis} holds
true and the arising \emph{EMS} synthesis method (Sect. \ref{sec:Synthesis})
can be reliably applied. Accordingly, the reference phase distribution
of the surface currents on the \emph{EMS}, $\arg\left(\widetilde{J}_{a}^{o}\right)_{mn}$,
$o\in\left\{ e,h\right\} $, $a\in\left\{ x,y\right\} $, is firstly
{[}Step (\emph{a}){]} computed with (\ref{eq:final NF focusing}).
As expected ($r\approx r_{NF}$), the arising phase profile {[}Fig.
8(\emph{b}) - $o=e$, $a=x${]} significantly differs from that obtained
with a standard \emph{FF} focusing process \cite{Yang 2019} {[}Fig.
8(\emph{a}){]}. The subsequent Step (\emph{b}) is then aimed at determining
the \emph{EMS} descriptors, $\mathcal{G}^{opt}\triangleq\left\{ g_{mn}\triangleq\left(\Delta l\right)_{mn};\, m=1,...,M,\, n=1,...,N\right\} $,
that allow the wave incident on the \emph{EMS} (Fig. 2) to induce
a surface current, $\mathbf{J}^{o}$, matching the reference one,
$\widetilde{\mathbf{J}}^{o}$, by minimizing the phase mismatch cost
function $\Phi\left(\mathcal{G}\right)$ (\ref{eq:cost function matching}).
Owing to the simplicity of the meta-atom at hand, the synthesized
phase profile {[}Fig. 8(\emph{d}){]} does not perfectly reproduce
the reference one {[}Fig. 8(\emph{b}){]} in each unit cell location,
but the overall behavior is complied with {[}Fig. 8(\emph{d}) vs.
Fig. 8(\emph{b}){]}. For the same reason, a similar mismatch can be
also observed in the case of the \emph{FF} design {[}Fig. 8(\emph{c})
vs. Fig. 8(\emph{a}){]} as quantitatively confirmed by the values
of the corresponding cumulative distribution function of the phase
mismatch, $\left|\Delta J_{y}^{e}\left(\mathbf{r};\mathcal{G}^{opt}\right)\right|$,
in Fig. 8(\emph{e}). On the contrary, the \emph{EMS} layouts synthesized
in the Step (\emph{b}) are not alike {[}Fig. 9(\emph{a}) vs. Fig.
9(\emph{b}){]}. This implies that there are significant differences
{[}Fig. 9(\emph{d}) vs. Fig. 9(\emph{c}){]} in the magnitude of the
dominant component of $\mathbf{F}_{ref}^{e}\left(\mathbf{r}\right)$
reflected in the receiver plane $\Theta_{RX}$ {[}Fig. 1(\emph{b}){]}.
As a matter of fact, the focusing of the field reflected by the \emph{EMS}
synthesized with the \emph{USM} (Sect. \ref{sec:Synthesis}) significantly
improves and the received power $\Psi_{RX}$ increases up to $\Psi_{RX}^{USM}\approx-33.05$
{[}dBm{]} from the value of $\Psi_{RX}^{FFM}\approx-41.34$ {[}dBm{]}
yielded with the \emph{FFM} despite the same $\mathcal{L}$ and transmitting
source.

\noindent The second experiment of this section (Sect. \ref{sub:EMS-Synthesis-Assessment})
is then aimed at assessing the dependence of the \emph{USM} performance
on the receiver distance $r_{RX}$ by keeping the same design setup
of the first test case, but varying $r_{RX}$. Figure 10 shows that
$\Psi_{RX}^{USM}\geq\Psi_{RX}^{FFM}$ regardless of the distance of
the receiver from the \emph{EMS}. The power improvement guaranteed
by the \emph{USM}, $\Delta\Psi_{RX}$ ($\Delta\Psi_{RX}\triangleq\frac{\Psi_{RX}^{USM}-\Psi_{RX}^{FFM}}{\Psi_{RX}^{FFM}}$),
reduces as $r_{RX}$ increases (e.g., $\left.\Delta\Psi_{RX}\right\rfloor _{r_{RX}=15\, m}\approx7.59$
{[}dB{]} vs. $\left.\Delta\Psi_{RX}\right\rfloor _{r_{RX}=50\, m}\approx-7.63$
{[}dB{]} - Fig. 10). This is theoretically expected since both (\ref{eq:final expression field})
and (\ref{eq:NF focusing}) reduce to their \emph{FF} counterparts
\cite{Oliveri 2021c} when $r\to\infty$ (i.e., \emph{FF} conditions
occur).

\noindent The effectiveness of the \emph{USM} to deal with different
\emph{EMS} apertures is addressed next. Towards this end, the design
process has been carried out by ranging the \emph{EMS} size and the
receiver distance in the range $12\times12\leq M\times N\leq240\times240$
(i.e., $1.03\times10^{-1}$ {[}m{]} $\le$ $\mathcal{L}$ $\le$ $2.057$
{[}m{]}) and $30$ {[}m{]} $\le$ $r_{RX}$ $\le$ $150$ {[}m{]},
respectively. The results yielded by applying the \emph{USM} and the
\emph{FFM} are summarized in Fig. 11 in terms of the power received
by the terminal, $\Psi_{RX}$. By comparing the maps of $\Psi_{RX}^{FFM}$
{[}Fig. 11(\emph{a}){]} and $\Psi_{RX}^{USM}$ {[}Fig. 11(\emph{b}){]}
in the domain ($\mathcal{L}$, $r_{RX}$), it turns out that the two
approaches \emph{}perform similarly \emph{}when the receiver is in
the \emph{FF} of the \emph{EMS} {[}i.e., $\Psi_{RX}^{USM}\approx\Psi_{RX}^{FFM}$
$\to$ $\Delta\Psi_{RX}\approx-20$ {[}dB{]} - Fig. 11(\emph{c}){]}.
Otherwise, it is more and more convenient (i.e., $\Delta\Psi_{RX}\uparrow$)
to use the \emph{USM} as $\mathcal{L}\uparrow$ and $r_{RX}\downarrow$
{[}e.g., $r_{RX}<40$ {[}m{]} and $\mathcal{L}>1.3$ {[}m{]} - Fig.
11(\emph{c}){]}, $\Delta\Psi_{RX}\approx30$ {[}dB{]} being the maximum
{}``gain'' within the range of values of this parametric analysis
{[}Fig. 11(\emph{c}){]}.

\noindent For the sake of completeness, the synthesized \emph{EMS}
layouts {[}Figs. 12(\emph{a})-12(\emph{b}){]} and the corresponding
distributions of the co-polar field magnitude in the receiver cut
$\Theta_{RX}$ at $r_{RX}=30$ {[}m{]} {[}Figs. 12(\emph{c})-12(\emph{d}){]}
are reported for the case $M\times N=240\times240$ (i.e., $\mathcal{L}=2.057$
{[}m{]}). The \emph{USM} \emph{EMS} significantly outperforms the
\emph{FFM} one in terms of focusing ability {[}Fig. 12(\emph{d}) vs.
Fig. 12(\emph{c}){]} with a relevant enhancement of the power received
by the terminal ($\Psi_{RX}^{USM}\approx-27.04$ {[}dBm{]} vs. $\Psi_{RX}^{FFM}\approx-47.81$
{[}dBm{]}).

\noindent Next, the assessment of the \emph{USM} potentialities is
carried out with respect to the receiver location (Figs. 13-14). The
behaviour of $\Psi_{RX}$ as a function of $r_{RX}$ for both anomalous
($\theta_{RX}\neq30$ {[}deg{]}) and Snell ($\theta_{RX}=30$ {[}deg{]})
directions of reflection ($M\times N=120\times120$, $\mathcal{L}=1.029$
{[}m{]}) indicates that once more the \emph{USM} outperforms the traditional
\emph{FFM} independently on the $\theta_{RX}$ value since always
$\Psi_{RX}^{USM}\geq\Psi_{RX}^{FFM}$ {[}Fig. 13(\emph{a}){]}. Moreover,
as expected, the focusing improvement enabled by the \emph{USM}, $\Delta\Psi_{RX}$,
gets better and better as $r_{RX}\downarrow$ and $\theta_{RX}\downarrow$.
Furthermore, the values of $\Psi_{RX}^{USM}$ increase as $\theta_{RX}\to0$
{[}deg{]} consistently with the fact that the gain of any radiating
aperture is maximum along broadside. Differently, the behavior of
$\Psi_{RX}^{FFM}$ does not always comply with such a theoretical
expectation because of the sub-optimal focusing performance of the
\emph{FFM} solution in the \emph{R-NF} region {[}e.g., $\left.\Psi_{RX}^{FFM}\right\rfloor _{\theta_{RX}=0\,[deg]}<\left.\Psi_{RX}^{FFM}\right\rfloor _{\theta_{RX}=30\,[deg]}$
when $r_{RX}<18$ {[}m{]} - Fig. 13(\emph{a}){]}.

\noindent Let us now analyze more in detail two representative scenarios
of the test case at hand: ($r_{RX}=15$ {[}m{]}; $\theta_{RX}=\theta_{TX}=30$
{[}deg{]}) and ($r_{RX}=15$ {[}m{]}; $\theta_{RX}=50$ {[}deg{]}
$\to$ $\theta_{RX}\ne\theta_{TX}$). The layouts synthesized with
the \emph{FFM} {[}Figs. 13(\emph{b})-13(\emph{c}){]} and the \emph{USM}
{[}Figs. 13(\emph{d})-13(\emph{e}){]} strongly differ when both Snell
{[}Fig. 13(\emph{b}) vs. Fig. 13(\emph{d}){]} or anomalous {[}Fig.
13(\emph{c}) vs. Fig. 13(\emph{e}){]} reflections are forced. Analogously
and consequently, the field profiles in $\Theta_{RX}$ do not have
evident similarities {[}Fig. 14(\emph{a}) vs. Fig. 14(\emph{c}) and
Fig. 14(\emph{b}) vs. Fig. 14(\emph{d}){]} by pointing out the superior
performance of the \emph{USM} in steering the reflected power towards
the receiving terminal as quantitatively confirmed by the values of
$\Psi_{RX}$ plotted in Fig. 13(\emph{a}) {[}i.e., $\left.\Delta\Psi_{RX}\right\rfloor _{\theta_{RX}=30\,[deg]}^{r_{RX}=15\,[m]}\approx5.03$
{[}dB{]} and $\left.\Delta\Psi_{RX}\right\rfloor _{\theta_{RX}=50\,[deg]}^{r_{RX}=15\,[m]}\approx3.29$
{[}dB{]}).

\noindent Since the position of the illuminating source is not expected
to affect the previous conclusions on the effectiveness of the \emph{USM},
the \emph{EMS} design process has been repeated by only changing the
distance between the transmitter and the \emph{EMS} within the range
$15$ {[}m{]} $\le$ $r_{TX}$ $\le$ $100$ {[}m{]}. The plots of
the power received by the terminal, $\Psi_{RX}$, versus $r_{RX}$
in Fig. 15(\emph{a}) confirm that always $\Delta\Psi_{RX}\geq0$ as
well as that the improvement of the \emph{USM} with respect to the
\emph{FFM} is almost independent on the distance of the \emph{EMS}
from the \emph{BTS} (e.g., $\left.\Delta\Psi_{RX}\right\rfloor _{r_{TX}=15\,[m]}^{r_{RX}=15\,[m]}\approx7.56$
{[}dB{]} vs. $\left.\Delta\Psi_{RX}\right\rfloor _{r_{TX}=100\,[m]}^{r_{RX}=15\,[m]}\approx7.59$
{[}dB{]}). 

\noindent On the contrary, the position of the \emph{BTS} impacts
on the \emph{EMS} layout as highlighted by the comparison of the unit-cell
arrangements synthesized, when $r_{TX}=15$ {[}m{]} {[}Fig. 15(\emph{b})
and Fig. 15(\emph{c}){]} or $r_{TX}=50$ {[}m{]} {[}Fig. 9(\emph{a})
and Fig. 9(\emph{b}){]}, with either the \emph{FFM} {[}Fig. 15(\emph{b})
vs. Fig. 9(\emph{a}){]} or the \emph{USM} {[}Fig. 15(\emph{c}) and
Fig. 9(\emph{b}){]}\@. This is an obvious consequence of the two-step
\emph{EMS} synthesis strategy, since the incident field, $\mathbf{F}_{inc}^{o}\left(\mathbf{r}\right)$
($o\in\left\{ e,h\right\} $, together with the micro-scale descriptors
of the \emph{EMS}, $\mathcal{G}$, control the surface currents induced
on the \emph{EMS} aperture (see Sect. \ref{sec:Analysis}) to focus
the reflected wave towards the receiving terminal. On the other hand,
the reader can notice that, analogously to the case in Fig. 9 {[}Fig.
9(\emph{d}) vs. Fig. 9(\emph{c}){]}, also here the \emph{USM}-\emph{EMS}
better focus on the spot area than the \emph{FFM} one {[}Fig. 15(\emph{e})
vs. Fig. 15(\emph{d}){]}.

\noindent To further analyze the dependence of the \emph{USM} performance
on the features of the primary source, the relatively low-gain horn
(i.e., $G_{TX}^{\max}=13.7$ {[}dBi{]} - Tab. I) located at $r_{TX}=50$
{[}m{]} from an $M\times N=120\times120$ (i.e., $\mathcal{L}=1.029$
{[}m{]}) \emph{EMS} has been then substituted with the higher gain
horn (i.e., $G_{TX}^{\max}=20.4$ {[}dBi{]} - Tab. I) to mimic a more
directive transmitter. Moreover, the receiver has been still modeled
with the same device of the transmitter, while its position has been
varied as follows: $15$ {[}m{]} $\le$ $r_{RX}$ $\le$ $50$ {[}m{]}
and $\theta_{RX}\in\left\{ 10,30,50\right\} $ {[}deg{]}. The outcomes
of this numerical study are summarized in Fig. 16 where the plots
of $\Psi_{RX}$ versus $r_{RX}$ are shown. It turns out that there
is an advantage of using the \emph{USM} (i.e., $\Delta\Psi_{RX}\geq0$)
for any distance and whatever the angular position of the receiver
both along the Snell direction (i.e., $\theta_{RX}=30$ {[}deg{]})
or not (i.e., $\theta_{RX}\in\left\{ 10,50\right\} $ {[}deg{]}).
Moreover, as previously inferred, the improvement enabled by the \emph{USM}
is wider for reflection angles closer to the \emph{EMS} broadside
(e.g., $\left.\Delta\Psi_{RX}\right\rfloor _{\theta_{RX}=10\,[deg]}^{r_{RX}=15\,[m]}\approx7.59$
{[}dB{]} vs. $\left.\Delta\Psi_{RX}\right\rfloor _{\theta_{RX}=50\,[deg]}^{r_{RX}=15\,[m]}\approx3.31$
{[}dB{]}). These results, which are fully consistent with those arising
when using a less directive source {[}Fig. 16 vs. Fig. 13(\emph{a}){]},
assess that the performance of the \emph{USM}-based \emph{EMS} synthesis
are not affected by the illumination features.

\noindent The last experiment is devoted to prove that the analytic
prediction of the received-power improvement enabled by the \emph{USM}
through (\ref{eq:focused beampattern}) is accurate. Towards this
end, the same benchmark setup of Fig. 16 has been used, while the
synthesis of \emph{EMS}s with $M\times N=84\times84$ unit cells (i.e.,
$\mathcal{L}=0.72$ {[}m{]}) has been performed to steer the reflected
waves along the angular direction $\theta_{RX}=10$ {[}deg{]} in different
receiving spots within the interval $15$ {[}m{]} $\le$ $r_{RX}$
$\le$ $50$ {[}m{]}. Once the design process has been carried out
with the \emph{USM} and the \emph{FFM}, the reflection performance
of the arising layouts have been analytically predicted with (\ref{eq:focused beampattern})
in (\ref{eq:received power}), $\Psi_{RX}^{Predicted}$, or full-wave
simulated in \emph{Ansys} \emph{HFSS} \cite{HFSS 2021}, $\Psi_{RX}^{HFSS-Computed}$.
Both analytically-computed and full-wave simulated values of $\Delta\Psi_{RX}$
are reported in Fig. 17. It turns out that the maximum deviation of
the prediction from the actual value is below $0.5$ {[}dB{]} (i.e.,
$0.4$ {[}dB{]} $\le$ $\left|\Delta\Psi_{RX}^{HFSS-Computed}-\Delta\Psi_{RX}^{Predicted}\right|$
$\le$ $0.5$ {[}dB{]}) regardless of $r_{RX}$.

\section{\noindent Conclusions\label{sec:Conclusions-and-Remarks}}

\noindent Within the \emph{GSTC} theoretical framework and applying
the classical theory of radiated fields, a mathematically rigorous
derivation of a generalized expression of the \emph{EM} field reflected
by an \emph{EMS} in the working regimes of interest for wireless communications
has been presented to derive a unified method for the design of anomalous-reflecting
and focusing \emph{EMS}s.

\noindent Numerical results, along with comparisons with full-wave
simulations, have been illustrated to assess the reliability and the
accuracy of the proposed generalized method for the prediction of
the \emph{EM} behavior of \emph{EMS}s as well as to prove the advantages
of the arising unified strategy for the synthesis of \emph{EMS}s over
standard \emph{FF}-based state-of-the-art design techniques.

\noindent From the numerical validation and performance assessment,
the following outcomes can be drawn:

\begin{itemize}
\item \noindent the prediction of the \emph{EM} field reflected by an \emph{EMS}
coming from the proposed generalized method (\ref{eq:final expression field})
faithfully matches the full-wave computed one and it is obtained with
a non-negligible time saving (i.e., $\frac{\Delta t^{Prediction}}{\Delta t^{HFSS}}>10^{6}$);
\item \noindent the unified \emph{EMS}-synthesis strategy always outperforms
the state-of-the-art \emph{FF}-based method regardless of the benchmark
setup at hand (i.e., $\Psi_{RX}^{NF}\geq\Psi_{RX}^{FF}$), but the
advantage of using the \emph{USM} in terms of increment of the power
received by the terminal (i.e., $\Delta\Psi_{RX}$) is greater when
(\emph{a}) the distance of the receiver from the \emph{EMS} reduces
($r_{RX}\downarrow$ being $r_{RX}\ge r_{NF}$), (\emph{b}) the \emph{EMS}
aperture, $\Omega$, widens ($\mathcal{L}\uparrow$), and (\emph{c})
the reflection angle, $\theta_{RX}$, becomes closer and closer to
broadside ($\theta_{RX}\downarrow$).
\end{itemize}
\noindent Future works, beyond the scope of the current manuscript,
will be aimed at evaluating the unified \emph{EMS} synthesis strategy
when applied to the design of reconfigurable \emph{EMS}s. Currently,
the performance of \emph{USM}-based \emph{EMS}s with more complex
unit cells are under investigation and they will be the topic of future
publications.

\section*{Appendix}

\subsection*{A.1 - Proof of (\ref{eq:near field E expression})}

\noindent By substituting (\ref{eq:fresnel potential}) in (\ref{eq:reflected electric field})
and using (\ref{eq:near field radiation vector}), it turns out that\begin{equation}
\begin{array}{r}
\mathbf{F}_{ref}^{e}\left(\mathbf{r}\right)\approx-\frac{j\omega\mu_{0}}{4\pi}\frac{\exp\left(-jk_{0}r\right)}{r}\mathbf{S}^{e}\left(\mathbf{r}\right)-\frac{j}{4\pi\omega\varepsilon_{0}}\nabla\left(\nabla\cdot\left[\frac{\exp\left(-jk_{0}r\right)}{r}\mathbf{S}^{e}\left(\mathbf{r}\right)\right]\right)+\\
-\frac{1}{4\pi}\nabla\times\left[\frac{\exp\left(-jk_{0}r\right)}{r}\mathbf{S}^{h}\left(\mathbf{r}\right)\right].\end{array}\label{eq:field grad rot}\end{equation}
Such an expression (\ref{eq:field grad rot}) can be simplified into
(\ref{eq:near field E expression}) by rewriting the differential
operators in spherical coordinates and neglecting the non-dominant
radial terms (i.e., $\left(\frac{1}{r}\right)^{s}$, $s\ge2$) as
detailed in the following.

\noindent Let us first consider the explicit expression in spherical
coordinates of the divergence term in the right-hand side of (\ref{eq:field grad rot})\begin{equation}
\begin{array}{r}
\nabla\cdot\left[\frac{\exp\left(-jk_{0}r\right)}{r}\mathbf{S}^{e}\left(\mathbf{r}\right)\right]=\frac{1}{r^{2}}\frac{\partial}{\partial r}\left[r\exp\left(-jk_{0}r\right)S_{r}^{e}\left(\mathbf{r}\right)\right]+\frac{\exp\left(-jk_{0}r\right)}{r^{2}\sin\theta}\frac{\partial}{\partial\theta}\left[\sin\theta S_{\theta}^{e}\left(\mathbf{r}\right)\right]+\\
+\frac{\exp\left(-jk_{0}r\right)}{r^{2}\sin\theta}\frac{\partial}{\partial\varphi}\left[S_{\varphi}^{e}\left(\mathbf{r}\right)\right]\end{array}\label{eq:}\end{equation}
that, subject to the condition (\ref{eq:fresnel condition 2}) and
discarding the $s=2$-th order radial dependence, simplifies as follows

\noindent \begin{equation}
\nabla\cdot\left[\frac{\exp\left(-jk_{0}r\right)}{r}\mathbf{S}^{e}\left(\mathbf{r}\right)\right]\approx-jk_{0}\frac{\exp\left(-jk_{0}r\right)}{r}S_{r}^{e}\left(\mathbf{r}\right).\label{eq:divergence approximated}\end{equation}
The gradient of (\ref{eq:divergence approximated}) then turns out
to be

\noindent \begin{equation}
\begin{array}{r}
\nabla\left(\nabla\cdot\left[\frac{\exp\left(-jk_{0}r\right)}{r}\mathbf{S}^{e}\left(\mathbf{r}\right)\right]\right)\approx-jk_{0}\frac{\partial}{\partial r}\left[\frac{\exp\left(-jk_{0}r\right)}{r}S_{r}^{e}\left(\mathbf{r}\right)\right]\widehat{\mathbf{r}}-jk_{0}\frac{\exp\left(-jk_{0}r\right)}{r^{2}}\frac{\partial}{\partial\theta}\left[S_{r}^{e}\left(\mathbf{r}\right)\right]\widehat{\bm{\theta}}+\\
-jk_{0}\frac{\exp\left(-jk_{0}r\right)}{r^{2}\sin\theta}\frac{\partial}{\partial\varphi}\left[S_{r}^{e}\left(\mathbf{r}\right)\right]\widehat{\bm{\varphi}}\end{array}\label{eq:}\end{equation}
that, once again neglecting the non-dominant $\left(\frac{1}{r}\right)^{2}$
terms, reduces to\begin{equation}
\nabla\left(\nabla\cdot\left[\frac{\exp\left(-jk_{0}r\right)}{r}\mathbf{S}^{e}\left(\mathbf{r}\right)\right]\right)\approx-k_{0}^{2}\frac{\exp\left(-jk_{0}r\right)}{r}S_{r}^{e}\left(\mathbf{r}\right)\widehat{\mathbf{r}}.\label{eq:grad div}\end{equation}
As for the curl component in (\ref{eq:field grad rot}), the first-order
truncated expression is\begin{equation}
\nabla\times\left[\frac{\exp\left(-jk_{0}r\right)}{r}\mathbf{S}^{h}\left(\mathbf{r}\right)\right]\approx\frac{1}{r}\frac{\partial}{\partial r}\left[-\exp\left(-jk_{0}r\right)S_{\varphi}^{h}\left(\mathbf{r}\right)\widehat{\bm{\theta}}+\exp\left(-jk_{0}r\right)S_{\theta}^{h}\left(\mathbf{r}\right)\widehat{\bm{\varphi}}\right]\label{eq:}\end{equation}
that, after simple manipulation, becomes\begin{equation}
\nabla\times\left[\frac{\exp\left(-jk_{0}r\right)}{r}\mathbf{S}^{h}\left(\mathbf{r}\right)\right]\approx-jk_{0}\frac{\exp\left(-jk_{0}r\right)}{r}\left[S_{\varphi}^{h}\left(\mathbf{r}\right)\widehat{\bm{\theta}}+S_{\theta}^{h}\left(\mathbf{r}\right)\widehat{\bm{\varphi}}\right].\label{eq:rotor}\end{equation}
Equation (\ref{eq:near field E expression}) is finally yielded by
substituting (\ref{eq:grad div}) and (\ref{eq:rotor}) in (\ref{eq:field grad rot}).

\subsection*{\noindent A.2 - Proof of (\ref{eq:final expression field})}

\noindent By substituting (\ref{eq: Rossiglione theta}) in (\ref{eq:near field E expression}),
it turns out that\begin{equation}
\begin{array}{l}
\mathbf{F}_{ref}^{e}\left(\mathbf{r}\right)\approx-\frac{j\exp\left(-jk_{0}r\right)}{2\lambda_{0}r}\sum_{m=1}^{M}\sum_{n=1}^{N}\gamma_{mn}\left(\mathbf{r}\right)\times\\
\left\{ \left[\eta_{0}\cos\theta\cos\varphi\left(J_{x}^{e}\right)_{mn}+\eta_{0}\cos\theta\sin\varphi\left(J_{y}^{e}\right)_{mn}-\sin\varphi\left(J_{x}^{h}\right)_{mn}+\cos\varphi\left(J_{y}^{h}\right)_{mn}\right]\widehat{\bm{\theta}}+\right.\\
\left.+\left[-\eta_{0}\sin\varphi\left(J_{x}^{e}\right)_{mn}+\eta_{0}\cos\varphi\left(J_{y}^{e}\right)_{mn}+\cos\theta\cos\varphi\left(J_{x}^{h}\right)_{mn}+\cos\theta\sin\varphi\left(J_{y}^{h}\right)_{mn}\right]\widehat{\bm{\varphi}}\right\} .\end{array}\label{eq:Field MASCIULLA}\end{equation}
Let us then apply simple trigonometric operations to rewrite the arguments
of the exponential of $\gamma_{mn}$ (\ref{eq:coefficient radiation vector})
as follows

\noindent \begin{equation}
\left\{ \begin{array}{l}
r'\cos\chi=x'u+y'v\\
\left(r'\sin\chi\right)^{2}=\left(x'w\right)^{2}+\left(y'w\right)^{2}+\left(x'v-y'u\right)^{2}\end{array}\right.,\label{eq:coordinate change}\end{equation}
$u$, $v$, and $w$ being the direction cosine coordinates \cite{Balanis 1997}.

\noindent Accordingly, the new expression of $\gamma_{mn}\left(\mathbf{r}\right)$
turns out to be\begin{equation}
\begin{array}{l}
\gamma_{mn}\left(r,u,v,w\right)=\\
\int_{x_{m}-\frac{\Delta_{x}}{2}}^{x_{m}+\frac{\Delta_{x}}{2}}\int_{y_{n}-\frac{\Delta_{y}}{2}}^{y_{n}+\frac{\Delta_{y}}{2}}\exp\left[j\frac{2\pi}{\lambda_{0}}\left(x'u+y'v\right)\right]\exp\left[-j\frac{\pi}{\lambda_{0}r}\left(\left(x'w\right)^{2}+\left(y'w\right)^{2}+\left(x'v-y'u\right)^{2}\right)\right]d\mathrm{x}'d\mathrm{y}'\end{array}\label{eq:}\end{equation}
that reduces to\begin{equation}
\begin{array}{r}
\gamma_{mn}\left(r,u,v,w\right)\approx\exp\left[-j\frac{\pi}{\lambda_{0}r}\left(\left(x_{m}w\right)^{2}+\left(y_{n}w\right)^{2}+\left(x_{m}v-y_{n}u\right)^{2}\right)\right]\times\\
\exp\left[j\frac{2\pi}{\lambda_{0}}\left(x_{m}u+y_{n}v\right)\right]\Delta_{x}\Delta_{y}\mathrm{sinc}\left[\frac{\pi\Delta_{x}}{\lambda_{0}}u\right]\mathrm{sinc}\left[\frac{\pi\Delta_{y}}{\lambda_{0}}v\right]\end{array}\label{eq:coefficient radiation solved}\end{equation}
when using the approximation $\exp$ {[} $-j\frac{\pi}{\lambda_{0}r}$
$\left(\left(x'w\right)^{2}+\left(y'w\right)^{2}+\left(x'v-y'u\right)^{2}\right)$
{]} $\approx$ $\exp$ {[} $-j\frac{\pi}{\lambda_{0}r}$ $\left(\left(x_{m}w\right)^{2}+\left(y_{n}w\right)^{2}+\left(x_{m}v-y_{n}u\right)^{2}\right)$
{]} since $\left(x,y\right)\in\Omega_{mn}$. 

\noindent Finally, equation (\ref{eq:final expression field}) is
derived by substituting (\ref{eq:coefficient radiation solved}) in
(\ref{eq:Field MASCIULLA}).

\section*{\noindent Acknowledgements}

\noindent This work benefited from the networking activities carried
out within the Project {}``Cloaking Metasurfaces for a New Generation
of Intelligent Antenna Systems (MANTLES)'' (Grant No. 2017BHFZKH)
funded by the Italian Ministry of Education, University, and Research
under the PRIN2017 Program (CUP: E64I19000560001). Moreover, it benefited
from the networking activities carried out within the Project {}``SPEED''
(Grant No. 61721001) funded by National Science Foundation of China
under the Chang-Jiang Visiting Professorship Program.

\newpage
\section*{FIGURE CAPTIONS}

\begin{itemize}
\item \textbf{Figure 1.} \emph{Problem geometry}. Sketch of (\emph{a}) the
problem scenario, (\emph{b}) the coordinate system, (\emph{c}) the
\emph{EMS} meta-atom, and (\emph{d}) the TX/RX horn antenna.
\item \textbf{Figure 2.} \emph{Numerical Study} ($r_{TX}=50$ {[}m{]}) -
Plots of (\emph{a})(\emph{c}) the magnitude and (\emph{b})(\emph{d})
the phase distributions of the incident $y$-polarized electric field,
$F_{inc,y}^{e}\left(\mathbf{r}\right)$, generated on the \emph{EMS}
aperture, $\Omega$, by a transmitting horn antenna with gain (\emph{a})(\emph{b})
$G_{TX}^{\max}=20.4$ {[}dBi{]} or (\emph{c})(\emph{d}) $G_{TX}^{\max}=13.7$
{[}dBi{]}.
\item \textbf{Figure 3.} \emph{Numerical Study} - Pictures of the \emph{EMS}
layout when (\emph{a}) $M\times N=48\times48$ ($\to$ $\mathcal{L}\approx0.41$
{[}m{]}) and $\Delta l=5.0\times10^{-3}$ {[}m{]} and (\emph{b})(\emph{c})
$M\times N=84\times84$ ($\to$ $\mathcal{L}=0.72$ {[}m{]}).
\item \textbf{Figure 4.} \emph{EMS}-\emph{Analysis Validation} ($G_{TX}^{\max}=20.4$
{[}dBi{]}, $r_{TX}=50$ {[}m{]}, $\left(\theta_{TX},\varphi_{TX}\right)=\left(30,180\right)$
{[}deg{]}; $M\times N=48\times48$ $\to$ $\mathcal{L}\approx0.41$
{[}m{]}, $\Delta l=5.0\times10^{-3}$ {[}m{]}) - Plots of $\left|F_{ref,\varphi}^{e}\left(\mathbf{r}\right)\right|$
in (\emph{a})(\emph{b})(\emph{c}) $\mathbf{r}\in\Theta_{NF}$ ($r=6$
{[}m{]}, $\theta=30$ {[}deg{]}) or (\emph{d})(\emph{e})(\emph{f})
$\mathbf{r}\in\Theta_{FF}$ ($r=40$ {[}m{]}, $\theta=30$ {[}deg{]})
predicted with (\emph{a})(\emph{d}) the \emph{FF} radiation theory
\cite{Balanis 1997}, (\emph{b})(\emph{e}) the generalized theory
(\ref{eq:final expression field}), and (\emph{c})(\emph{f}) \emph{HFSS}-computed.
\item \textbf{Figure 5.} \emph{EMS}-\emph{Analysis Validation} ($G_{TX}^{\max}=20.4$
{[}dBi{]}, $r_{TX}=50$ {[}m{]}, $\left(\theta_{TX},\varphi_{TX}\right)=\left(30,180\right)$
{[}deg{]}; $M\times N=48\times48$ $\to$ $\mathcal{L}\approx0.41$
{[}m{]}, $\Delta l=5.0\times10^{-3}$ {[}m{]}) - Plots of $\Delta F_{ref,\varphi}^{e}\left(\mathbf{r}\right)$
in (\emph{a})(\emph{b}) $\mathbf{r}\in\Theta_{NF}$ ($r=6$ {[}m{]},
$\theta=30$ {[}deg{]}) or (\emph{c})(\emph{d}) $\mathbf{r}\in\Theta_{FF}$
($r=40$ {[}m{]}, $\theta=30$ {[}deg{]}) when using (\emph{a})(\emph{c})
the \emph{FF} radiation theory \cite{Balanis 1997} or (\emph{b})(\emph{d})
the generalized theory (\ref{eq:final expression field}).
\item \textbf{Figure 6.} \emph{EMS}-\emph{Analysis Validation} ($G_{TX}^{\max}=20.4$
{[}dBi{]}, $r_{TX}=50$ {[}m{]}, $\left(\theta_{TX},\varphi_{TX}\right)=\left(30,180\right)$
{[}deg{]}; $M\times N=84\times84$ $\to$ $\mathcal{L}=0.72$ {[}m{]};
$\mathbf{r}\in\Theta_{NF}$ ($r=10.5$ {[}m{]}, $\theta=10$ {[}deg{]}))
- Plots of $\Delta F_{ref,\varphi}^{e}\left(\mathbf{r}\right)$ for
(\emph{a})(\emph{b}) the layout in Fig. 3(\emph{b}) or (\emph{c})(\emph{d})
the layout in Fig. 3(\emph{c}) when using (\emph{a})(\emph{c}) the
\emph{FF} radiation theory \cite{Balanis 1997} or (\emph{b})(\emph{d})
the generalized theory (\ref{eq:final expression field}).
\item \textbf{Figure 7.} \emph{EMS}-\emph{Analysis Validation} ($G_{TX}^{\max}=20.4$
{[}dBi{]}, $r_{TX}=50$ {[}m{]}, $\left(\theta_{TX},\varphi_{TX}\right)=\left(30,180\right)$
{[}deg{]}; $M\times N=84\times84$ $\to$ $\mathcal{L}=0.72$ {[}m{]};
$\mathbf{r}\in\Theta_{FF}$ ($r=121$ {[}m{]}, $\theta=10$ {[}deg{]}))
- Plots of $\Delta F_{ref,\varphi}^{e}\left(\mathbf{r}\right)$ for
(\emph{a})(\emph{b}) the layout in Fig. 3(\emph{b}) or (\emph{c})(\emph{d})
the layout in Fig. 3(\emph{c}) when using (\emph{a})(\emph{c}) the
\emph{FF} radiation theory \cite{Balanis 1997} or (\emph{b})(\emph{d})
the generalized theory (\ref{eq:final expression field}).
\item \textbf{Figure 8.} \emph{EMS}-\emph{Synthesis Assessment} ($G_{TX}^{\max}=G_{RX}^{\max}=13.7$
{[}dBi{]}, $r_{TX}=50$ {[}m{]}, $\left(\theta_{TX},\varphi_{TX}\right)=\left(30,180\right)$
{[}deg{]}; $r_{RX}=15$ {[}m{]}, $\theta_{RX}=10$ {[}deg{]}; $M\times N=120\times120$
$\to$ $\mathcal{L}\approx1.029$ {[}m{]}) - Plots of the distributions
of (\emph{a})(\emph{b}) $arg\left[\widetilde{J}_{y}^{e}\left(\mathbf{r}\right)\right]$
and (\emph{c})(\emph{d}) $arg\left[J_{y}^{e}\left(\mathbf{r}\right)\right]$
on the \emph{EMS} aperture, $\Omega$, when applying (\emph{a})(\emph{c})
the \emph{FFM} or (\emph{b})(\emph{d}) the \emph{USM} along with (\emph{e})
the cumulative distribution function of the phase mismatch, $\left|\Delta J_{y}^{e}\left(\mathbf{r}\right)\right|$.
\item \textbf{Figure 9.} \emph{EMS}-\emph{Synthesis Assessment} ($G_{TX}^{\max}=G_{RX}^{\max}=13.7$
{[}dBi{]}, $r_{TX}=50$ {[}m{]}, $\left(\theta_{TX},\varphi_{TX}\right)=\left(30,180\right)$
{[}deg{]}; $r_{RX}=15$ {[}m{]}, $\theta_{RX}=10$ {[}deg{]}; $M\times N=120\times120$
$\to$ $\mathcal{L}\approx1.029$ {[}m{]}) - Plots of (\emph{a})(\emph{b})
the \emph{EMS} layout and of (\emph{c})(\emph{d}) $\left|F_{ref,\varphi}^{e}\left(\mathbf{r}\right)\right|$
in the receiver region $\Theta_{RX}$ when applying (\emph{a})(\emph{c})
the \emph{FFM} or (\emph{b})(\emph{d}) the \emph{USM}.
\item \textbf{Figure 10.} \emph{EMS}-\emph{Synthesis Assessment} ($G_{TX}^{\max}=G_{RX}^{\max}=13.7$
{[}dBi{]}, $r_{TX}=50$ {[}m{]}, $\left(\theta_{TX},\varphi_{TX}\right)=\left(30,180\right)$
{[}deg{]}; $\theta_{RX}=10$ {[}deg{]}; $M\times N=120\times120$
$\to$ $\mathcal{L}\approx1.029$ {[}m{]}) - Behavior of $\Psi_{RX}$
versus the receiver distance $r_{RX}$.
\item \textbf{Figure 11.} \emph{EMS}-\emph{Synthesis Assessment} ($G_{TX}^{\max}=G_{RX}^{\max}=13.7$
{[}dBi{]}, $r_{TX}=50$ {[}m{]}, $\left(\theta_{TX},\varphi_{TX}\right)=\left(30,180\right)$
{[}deg{]}; $\theta_{RX}=10$ {[}deg{]}) - Color-maps in the parametric
domain ($\mathcal{L}$, $r_{RX}$) $=$ ($1.03\times10^{-1}$ {[}m{]}
$\le$ $\mathcal{L}$ $\le$ $2.057$ {[}m{]} $\to$ $12\times12\leq M\times N\leq240\times240$;
$30$ {[}m{]} $\le$ $r_{RX}$ $\le$ $150$ {[}m{]}) of (\emph{a})(\emph{b})
the value of the power $\Psi_{RX}$ when applying (\emph{a}) the \emph{FFM}
or (\emph{b}) the \emph{USM} and of (\emph{c}) corresponding $\Delta\Psi_{RX}$
values.
\item \textbf{Figure 12.} \emph{EMS}-\emph{Synthesis Assessment} ($G_{TX}^{\max}=G_{RX}^{\max}=13.7$
{[}dBi{]}, $r_{TX}=50$ {[}m{]}, $\left(\theta_{TX},\varphi_{TX}\right)=\left(30,180\right)$
{[}deg{]}; $r_{RX}=30$ {[}m{]}, $\theta_{RX}=10$ {[}deg{]}; $M\times N=240\times240$
$\to$ $\mathcal{L}=2.057$ {[}m{]}) - Plots of (\emph{a})(\emph{b})
the \emph{EMS} layout and of (\emph{c})(\emph{d}) $\left|F_{ref,\varphi}^{e}\left(\mathbf{r}\right)\right|$
in the receiver region $\Theta_{RX}$ when applying (\emph{a})(\emph{c})
the \emph{FFM} or (\emph{b})(\emph{d}) the \emph{USM}.
\item \textbf{Figure 13.} \emph{EMS}-\emph{Synthesis Assessment} ($G_{TX}^{\max}=G_{RX}^{\max}=13.7$
{[}dBi{]}; $r_{TX}=50$ {[}m{]}, $\left(\theta_{TX},\varphi_{TX}\right)=\left(30,180\right)$
{[}deg{]}; $r_{RX}=15$ {[}m{]}; $M\times N=120\times120$ $\to$
$\mathcal{L}\approx1.029$ {[}m{]}) - Plots of (\emph{a}) the behavior
of $\Psi_{RX}$ versus $r_{RX}$ when $\theta_{RX}\in\left\{ 0,30,50\right\} $
{[}deg{]} and pictures of (\emph{a})(\emph{b}) the \emph{EMS} layouts
synthesized when (\emph{b})(\emph{d}) $\theta_{RX}=30$ {[}deg{]}
or (\emph{c})(\emph{e}) $\theta_{RX}=50$ {[}deg{]} with (\emph{b})(\emph{c})
the \emph{FFM} or (\emph{d})(\emph{e}) the \emph{USM}.
\item \textbf{Figure 14.} \emph{EMS}-\emph{Synthesis Assessment} ($G_{TX}^{\max}=G_{RX}^{\max}=13.7$
{[}dBi{]}; $r_{TX}=50$ {[}m{]}, $\left(\theta_{TX},\varphi_{TX}\right)=\left(30,180\right)$
{[}deg{]}; $r_{RX}=15$ {[}m{]}; $M\times N=120\times120$ $\to$
$\mathcal{L}\approx1.029$ {[}m{]}) - Plots of $\left|F_{ref,\varphi}^{e}\left(\mathbf{r}\right)\right|$
reflected in the receiver region $\Theta_{RX}$ when (\emph{a})(\emph{c})
$\theta_{RX}=30$ {[}deg{]} or (\emph{b})(\emph{d}) $\theta_{RX}=50$
{[}deg{]} by the \emph{EMS} synthesized with (\emph{a})(\emph{b})
the \emph{FFM} {[}Figs. 13(\emph{b})-13(\emph{c}){]} or (\emph{b})(\emph{d})
the \emph{USM} {[}Figs. 13(\emph{d})-13(\emph{e}){]}.
\item \textbf{Figure 15.} \emph{EMS}-\emph{Synthesis Assessment} ($G_{TX}^{\max}=G_{RX}^{\max}=13.7$
{[}dBi{]}; $\left(\theta_{TX},\varphi_{TX}\right)=\left(30,180\right)$
{[}deg{]}; $\theta_{RX}=10$ {[}deg{]}; $M\times N=120\times120$
$\to$ $\mathcal{L}\approx1.029$ {[}m{]}) - Plots of (\emph{a}) the
behavior of $\Psi_{RX}$ versus $r_{RX}$ when $r_{TX}\in\left\{ 15,100\right\} $
{[}m{]} and of (\emph{b})(\emph{c}) the \emph{EMS} layout along with
(\emph{d})(\emph{e}) the corresponding $\left|F_{ref,\varphi}^{e}\left(\mathbf{r}\right)\right|$
distribution in the receiver region $\Theta_{RX}$ when setting $r_{RX}=r_{TX}=15$
{[}m{]} and using (\emph{b})(\emph{d}) the \emph{FFM} or (\emph{c})(\emph{e})
the \emph{USM}.
\item \textbf{Figure 16.} \emph{EMS}-\emph{Synthesis Assessment} ($G_{TX}^{\max}=G_{RX}^{\max}=20.4$
{[}dBi{]}, $r_{TX}=50$ {[}m{]}, $\left(\theta_{TX},\varphi_{TX}\right)=\left(30,180\right)$
{[}deg{]}; $M\times N=120\times120$ $\to$ $\mathcal{L}\approx1.029$
{[}m{]}) - Behavior of $\Psi_{RX}$ versus $r_{RX}$ when $\theta_{RX}\in\left\{ 10,30,50\right\} $
{[}deg{]}.
\item \textbf{Figure 17.} \emph{EMS}-\emph{Synthesis Assessment} ($G_{TX}^{\max}=G_{RX}^{\max}=20.4$
{[}dBi{]}, $r_{TX}=50$ {[}m{]}, $\left(\theta_{TX},\varphi_{TX}\right)=\left(30,180\right)$
{[}deg{]}; $\theta_{RX}=10$ {[}deg{]}; $M\times N=84\times84$ $\to$
$\mathcal{L}=0.72$ {[}m{]}) - Comparison between the predicted and
\emph{HFSS}-computed values of $\Delta\Psi_{RX}$ versus $r_{RX}$.
\end{itemize}

\section*{TABLE CAPTIONS}

\begin{itemize}
\item \textbf{Table I.} \emph{Numerical Study -} Values of the descriptors
of the horn antenna in Fig. 1(\emph{d}).
\end{itemize}
~

\newpage
\begin{center}~\vfill\end{center}

\begin{center}\begin{tabular}{cc}
\multicolumn{2}{c}{\includegraphics[%
  width=0.70\columnwidth]{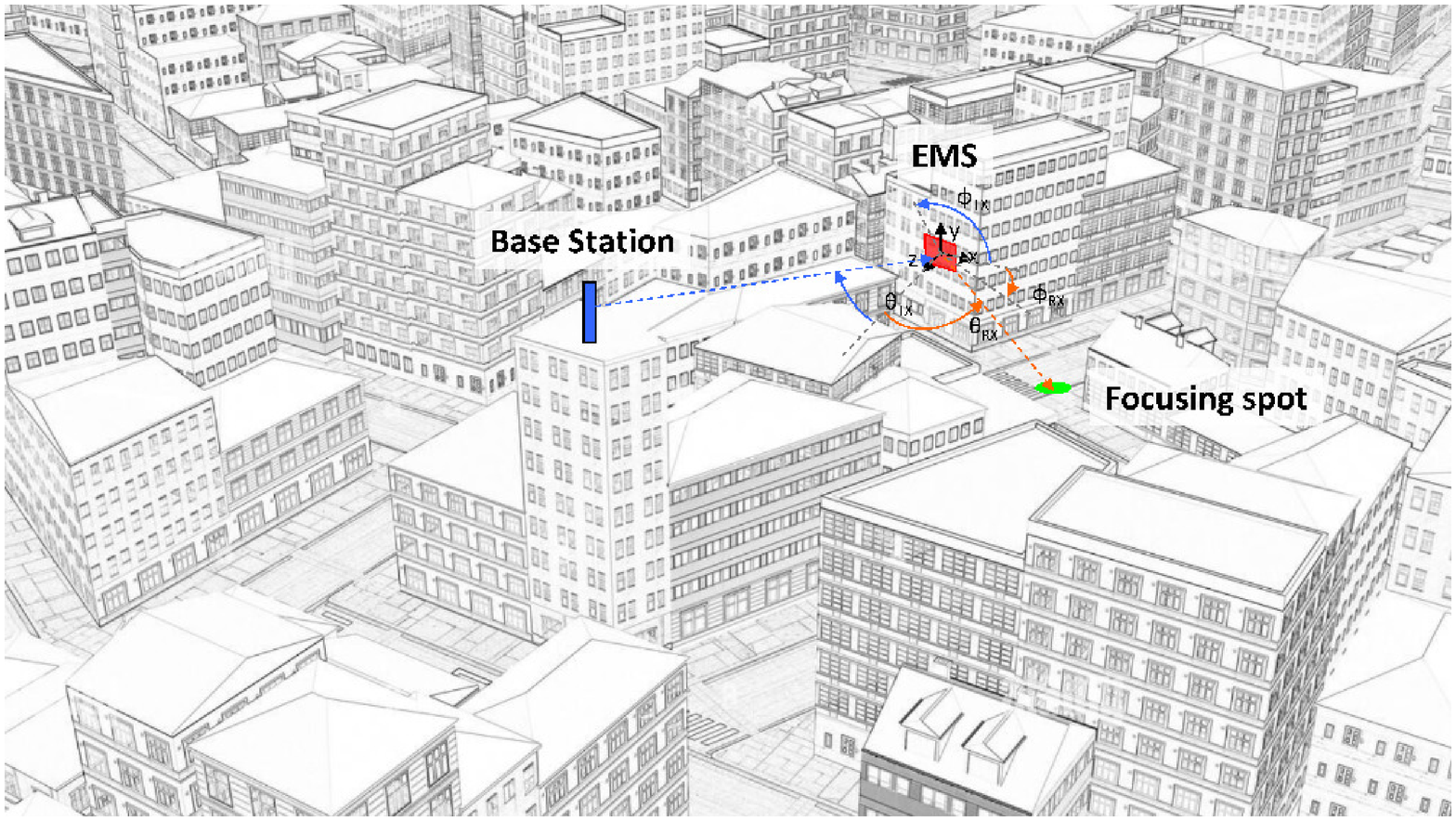}}\tabularnewline
\multicolumn{2}{c}{(\emph{a})}\tabularnewline
\multicolumn{2}{c}{\includegraphics[%
  width=0.50\columnwidth]{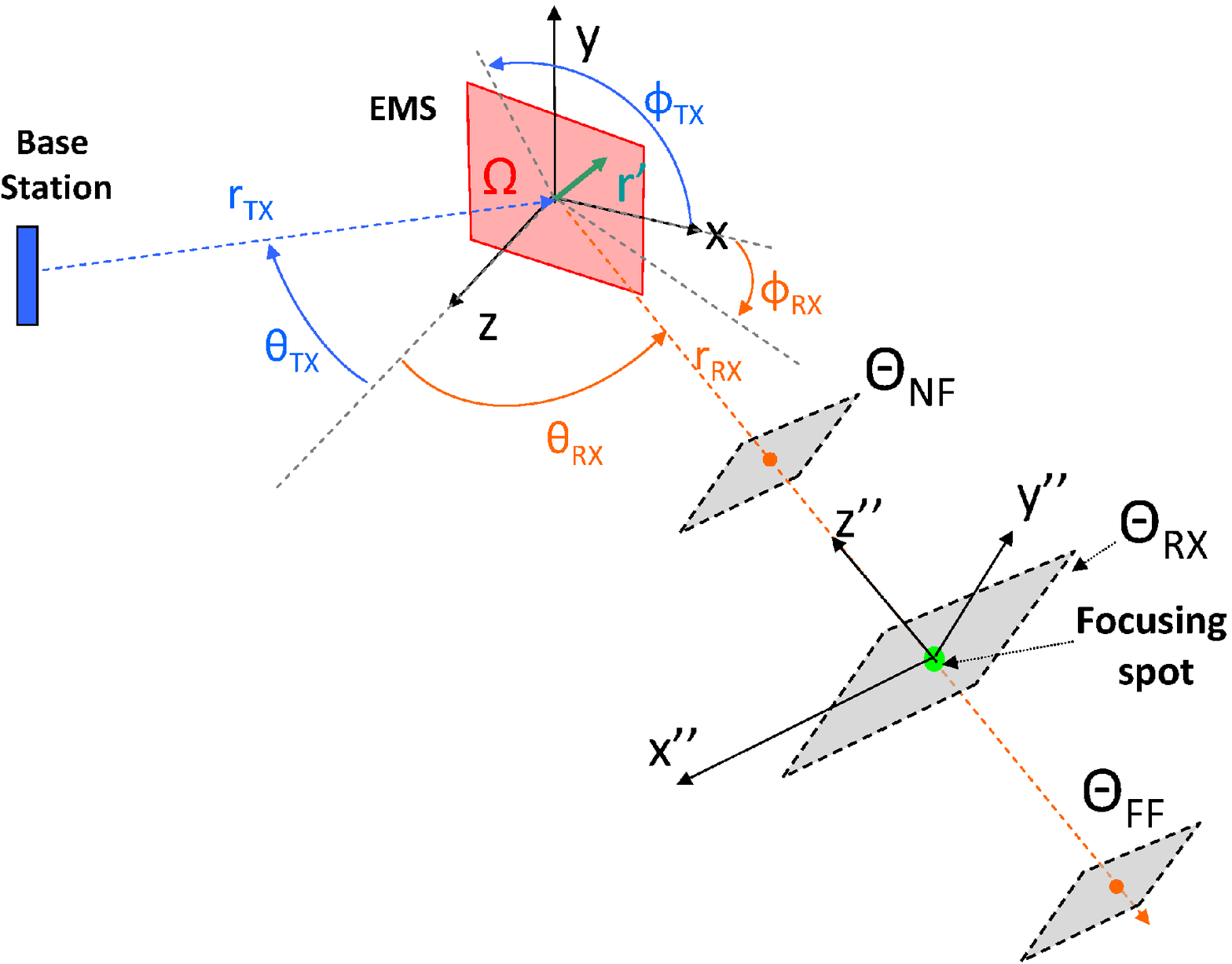}}\tabularnewline
\multicolumn{2}{c}{(\emph{b})}\tabularnewline
\includegraphics[%
  width=0.45\columnwidth]{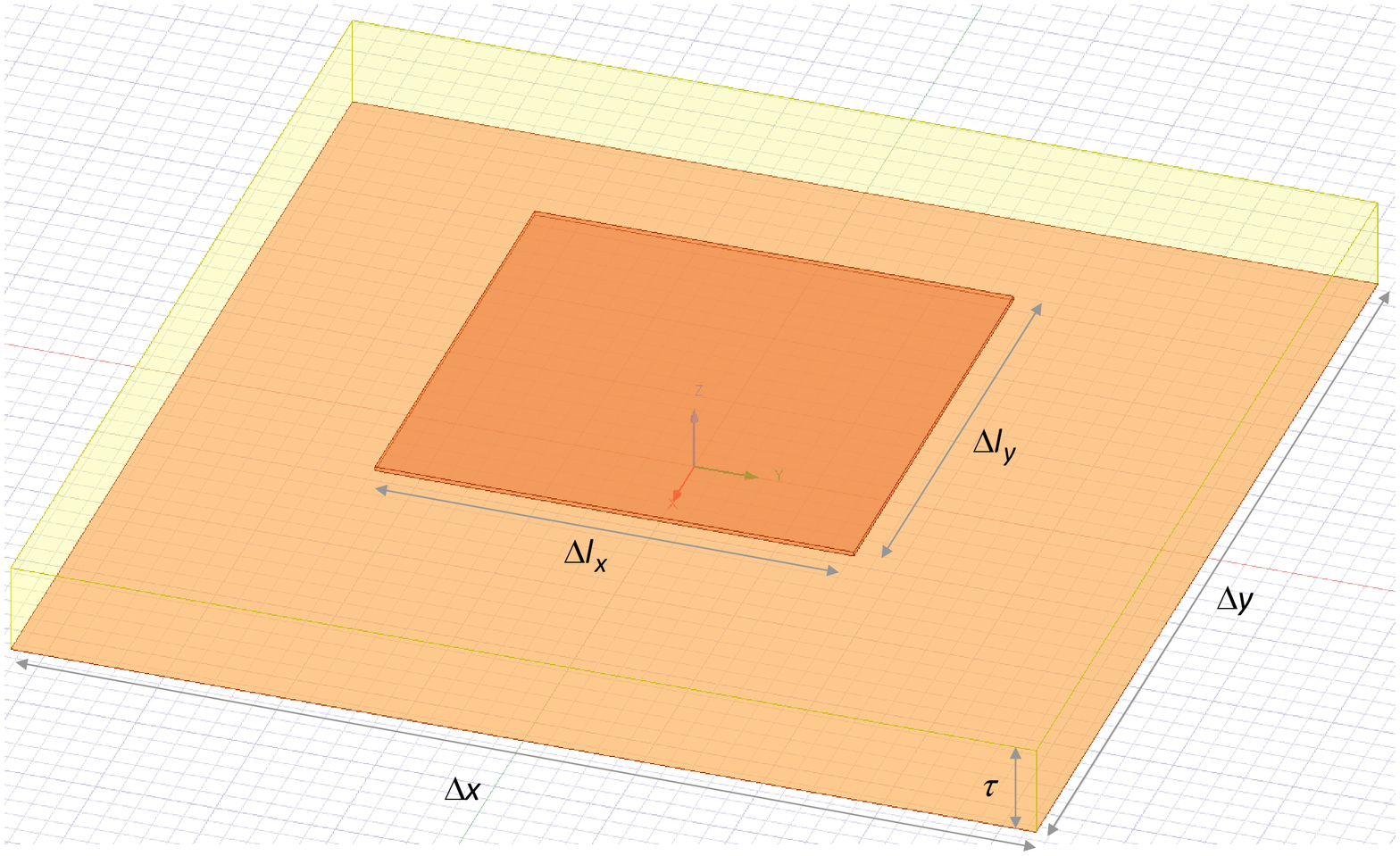}&
\includegraphics[%
  width=0.45\columnwidth]{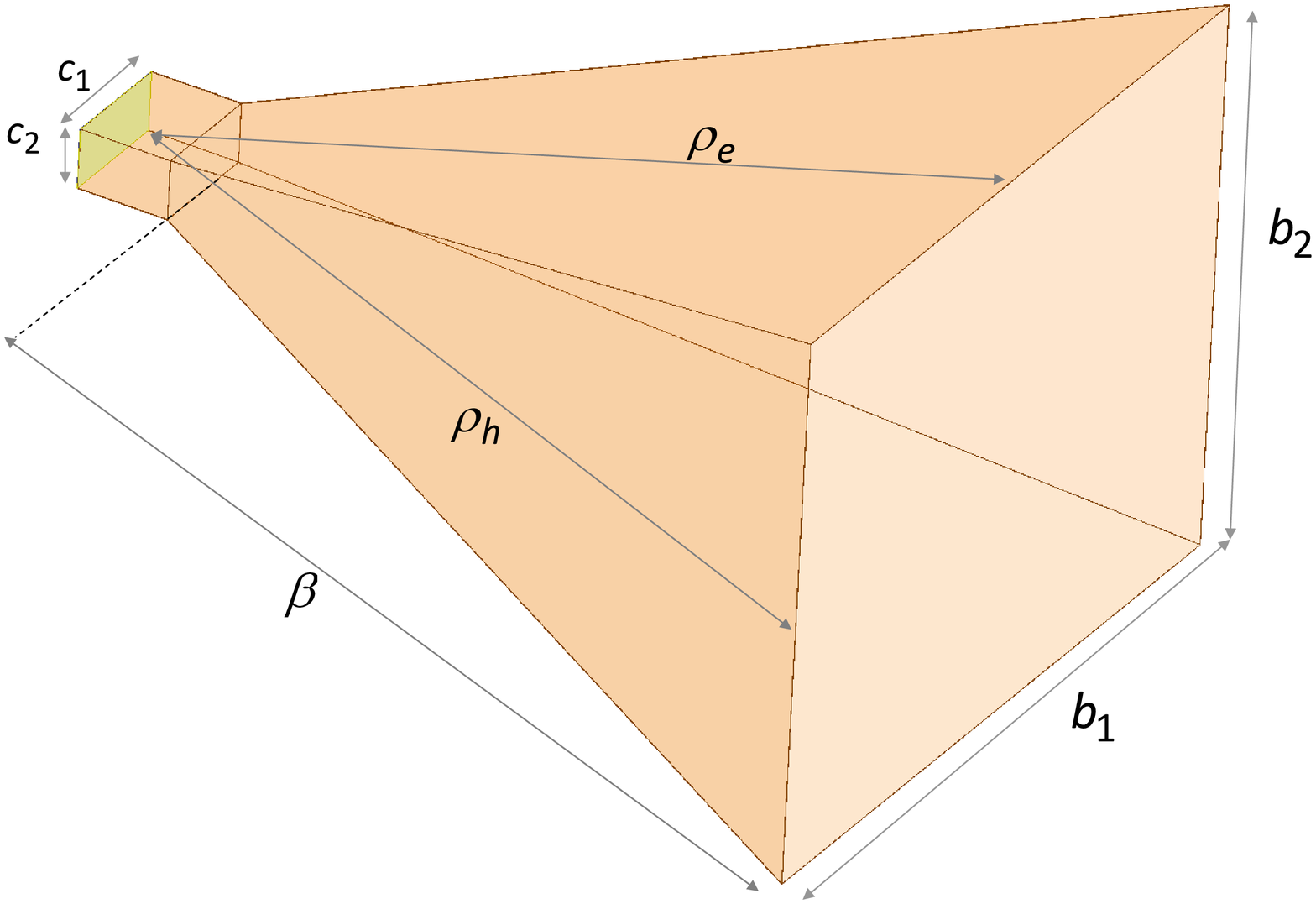}\tabularnewline
(\emph{c})&
(\emph{d})\tabularnewline
\end{tabular}\end{center}

\begin{center}~\vfill\end{center}

\begin{center}\textbf{Fig. 1 - G. Oliveri et} \textbf{\emph{al.}}\textbf{,}
\textbf{\emph{{}``}}Generalized Analysis and Unified Design of \emph{EM}
Skins ...''\end{center}

\newpage
\begin{center}~\vfill\end{center}

\begin{center}\begin{tabular}{cc}
\includegraphics[%
  width=0.45\columnwidth]{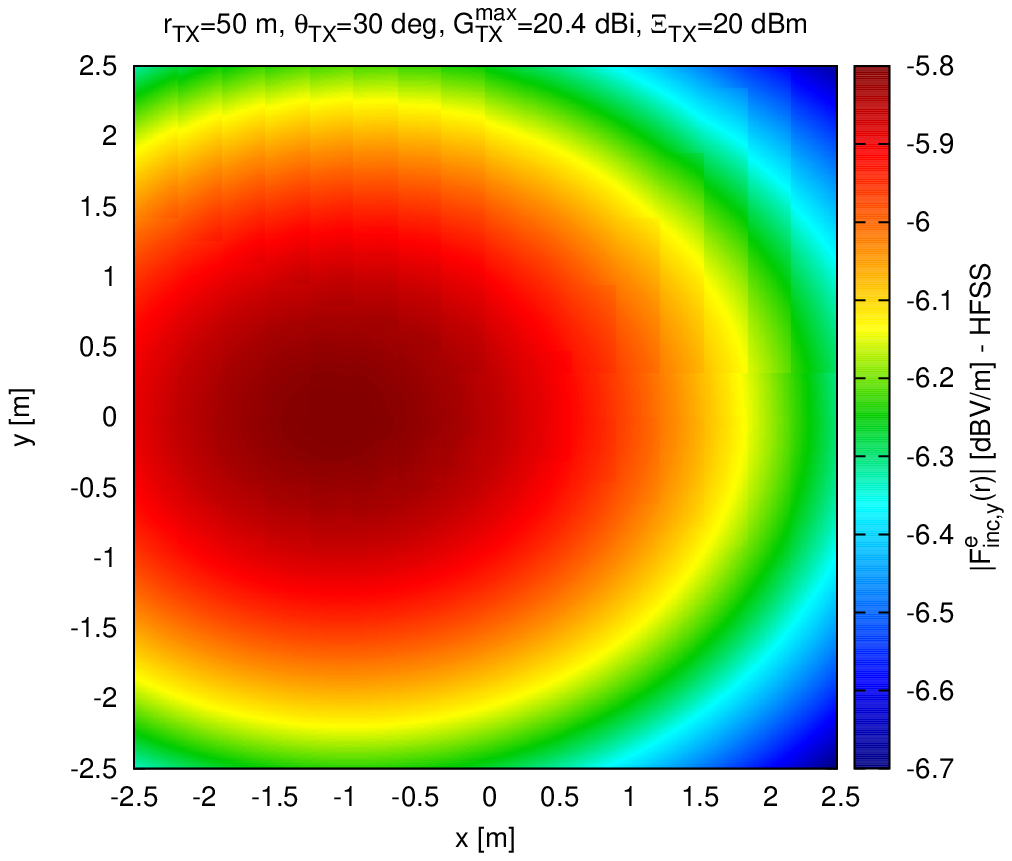}&
\includegraphics[%
  width=0.45\columnwidth]{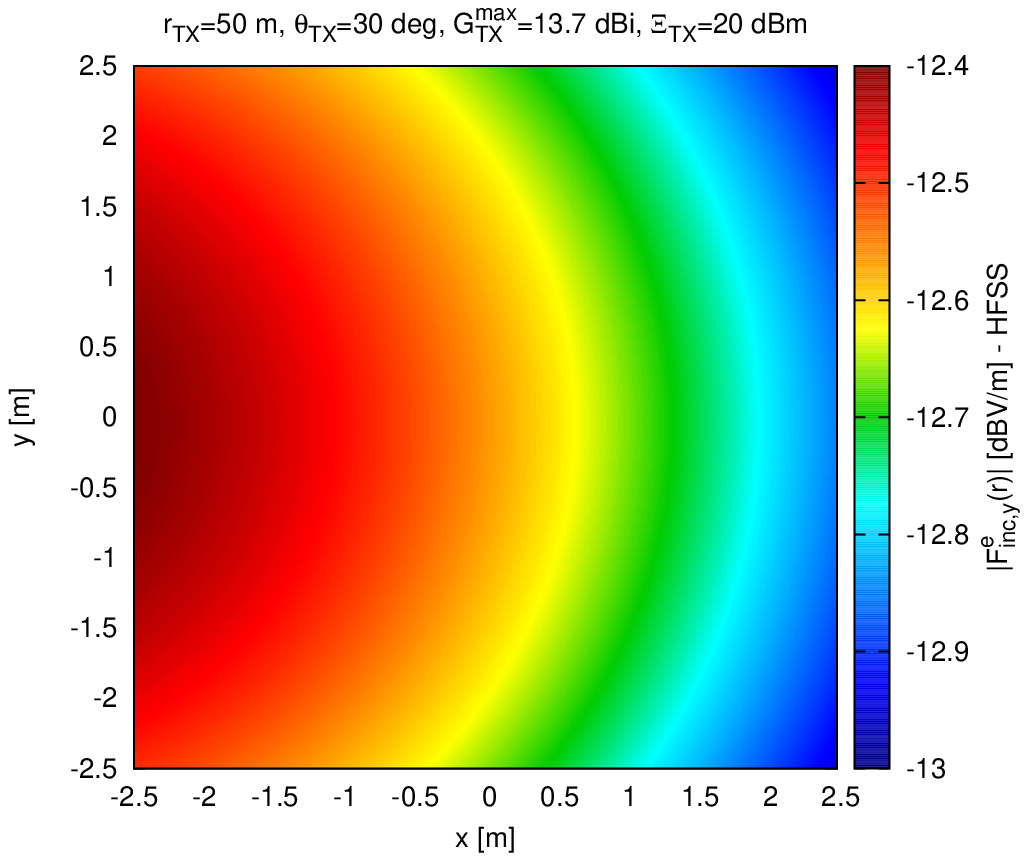}\tabularnewline
(\emph{a})&
(\emph{c})\tabularnewline
\includegraphics[%
  width=0.45\columnwidth]{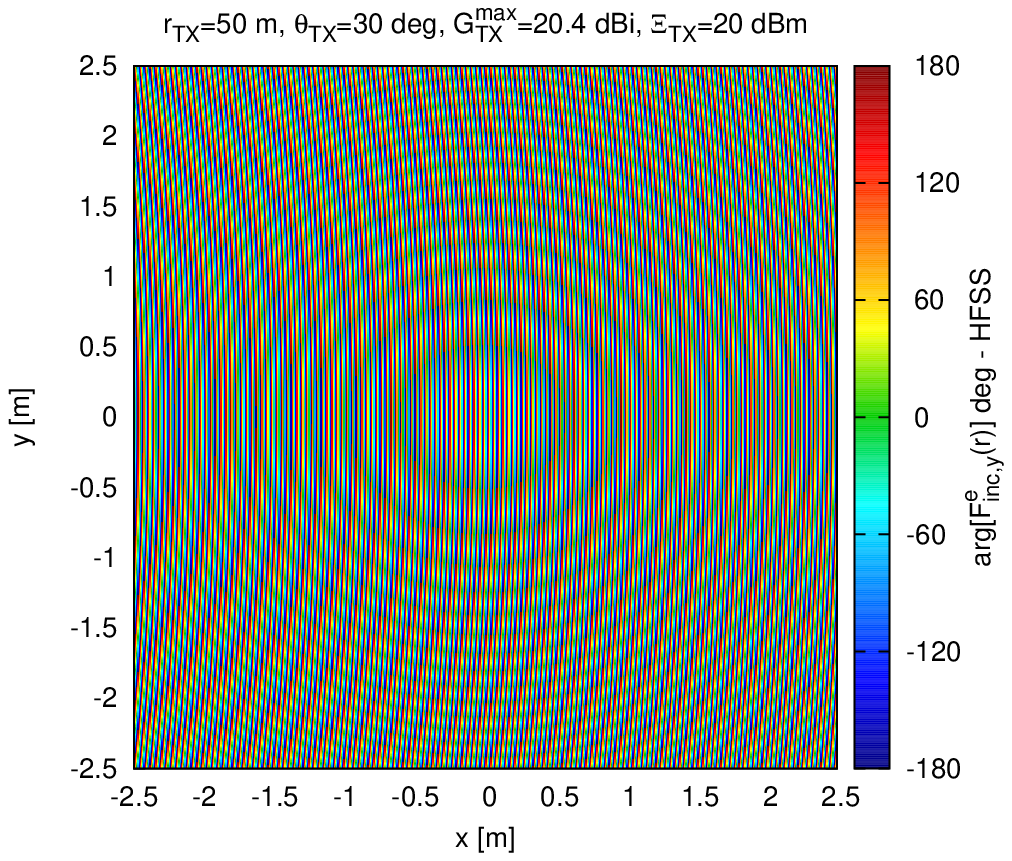}&
\includegraphics[%
  width=0.45\columnwidth]{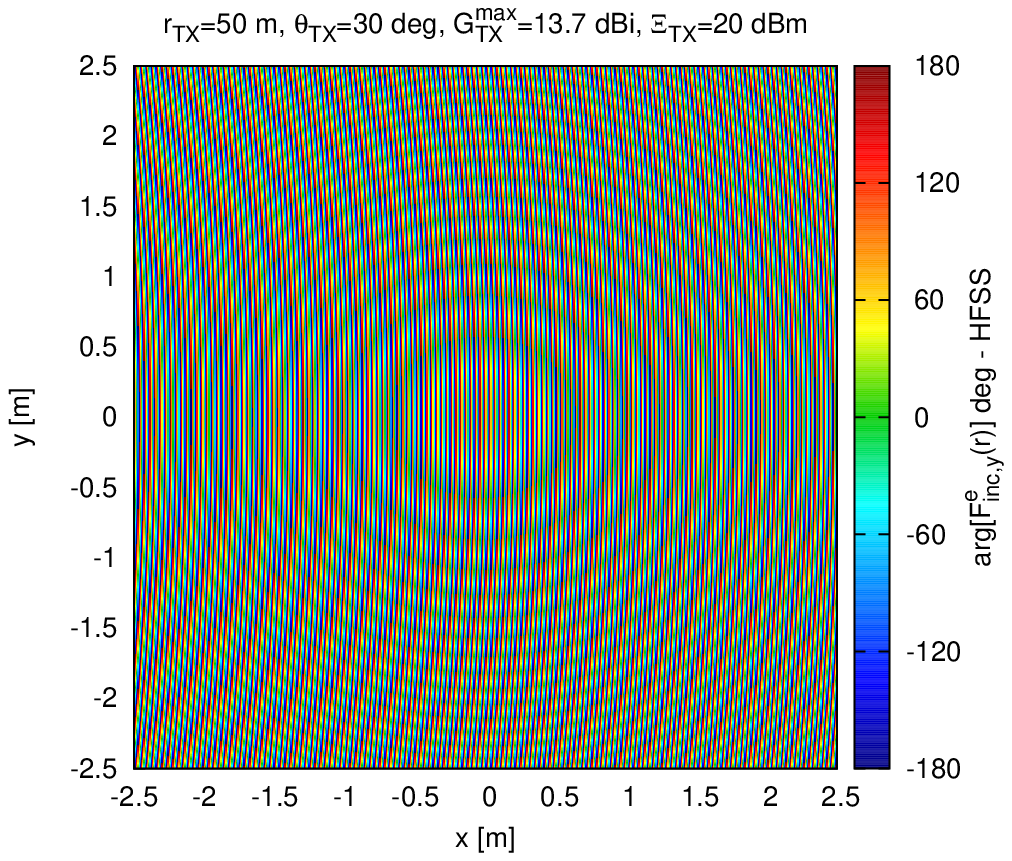}\tabularnewline
(\emph{b})&
(\emph{d})\tabularnewline
\end{tabular}\end{center}

\begin{center}~\vfill\end{center}

\begin{center}\textbf{Fig. 2 - G. Oliveri et} \textbf{\emph{al.}}\textbf{,}
\textbf{\emph{{}``}}Generalized Analysis and Unified Design of \emph{EM}
Skins ...''\end{center}

\newpage
\begin{center}~\vfill\end{center}

\begin{center}\begin{tabular}{cc}
\multicolumn{2}{c}{\includegraphics[%
  width=0.45\columnwidth]{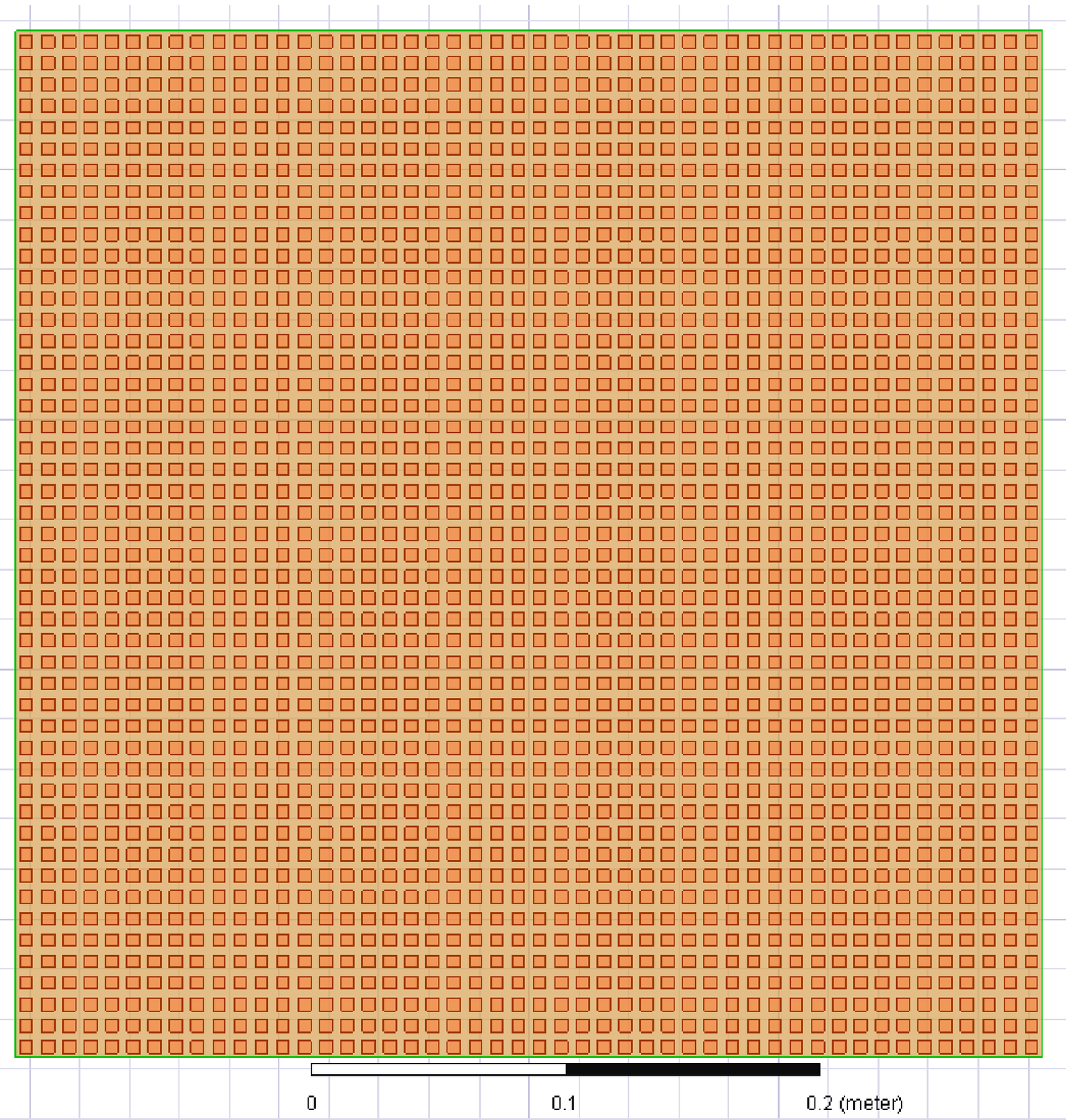}}\tabularnewline
\multicolumn{2}{c}{(\emph{a})}\tabularnewline
\includegraphics[%
  width=0.45\columnwidth]{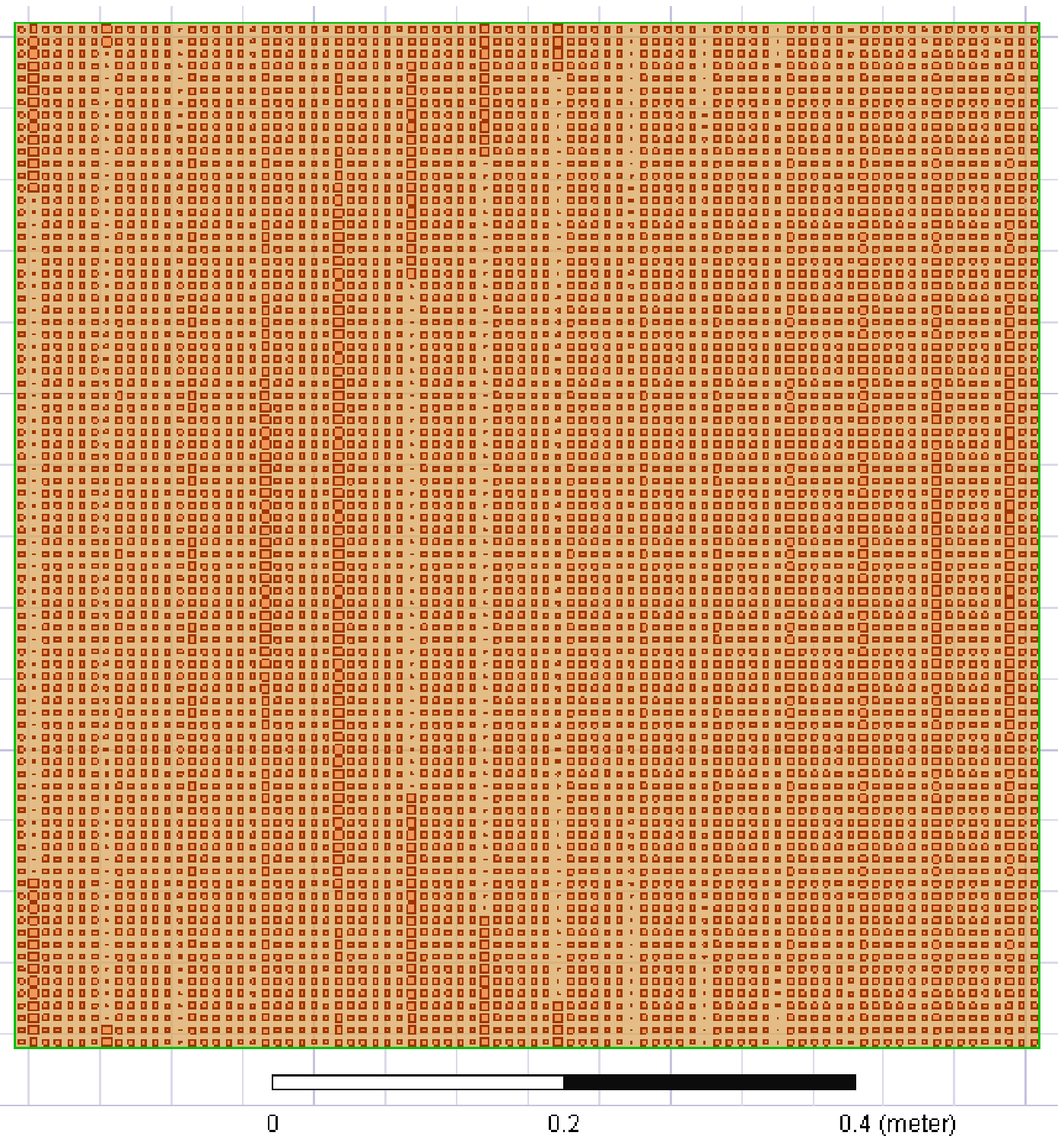}&
\includegraphics[%
  width=0.45\columnwidth]{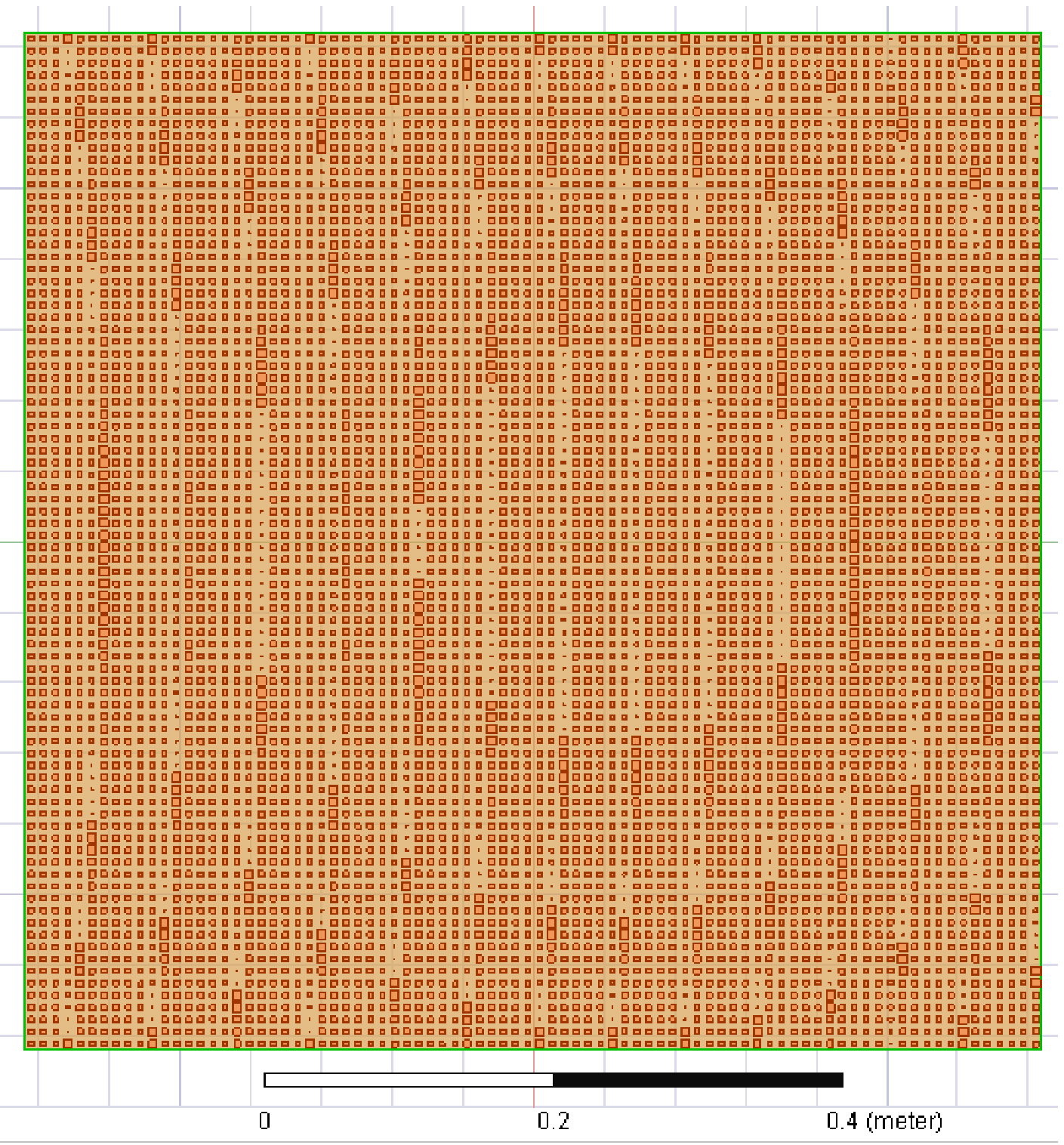}\tabularnewline
(\emph{b})&
(\emph{c})\tabularnewline
\end{tabular}\end{center}

\begin{center}~\vfill\end{center}

\begin{center}\textbf{Fig. 3 - G. Oliveri et} \textbf{\emph{al.}}\textbf{,}
\textbf{\emph{{}``}}Generalized Analysis and Unified Design of \emph{EM}
Skins ...''\end{center}

\newpage
\begin{center}~\vfill\end{center}

\begin{center}\begin{sideways}
\begin{tabular}{cccc}
&
\emph{FF Theory}&
\emph{Generalized Theory}&
\emph{HFSS}\tabularnewline
\begin{sideways}
~~~~~~~~~~~~~~~~~~~~$\Theta=\Theta_{NF}$%
\end{sideways}&
\includegraphics[%
  width=0.40\columnwidth]{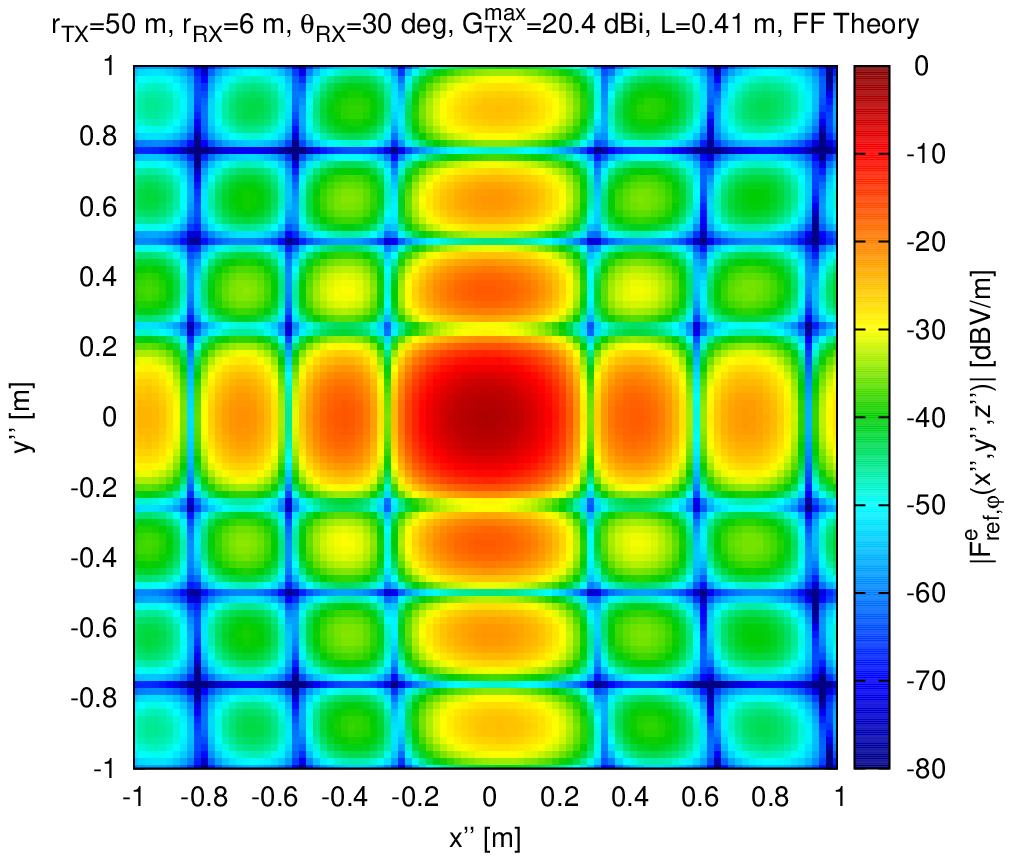}&
\includegraphics[%
  width=0.40\columnwidth]{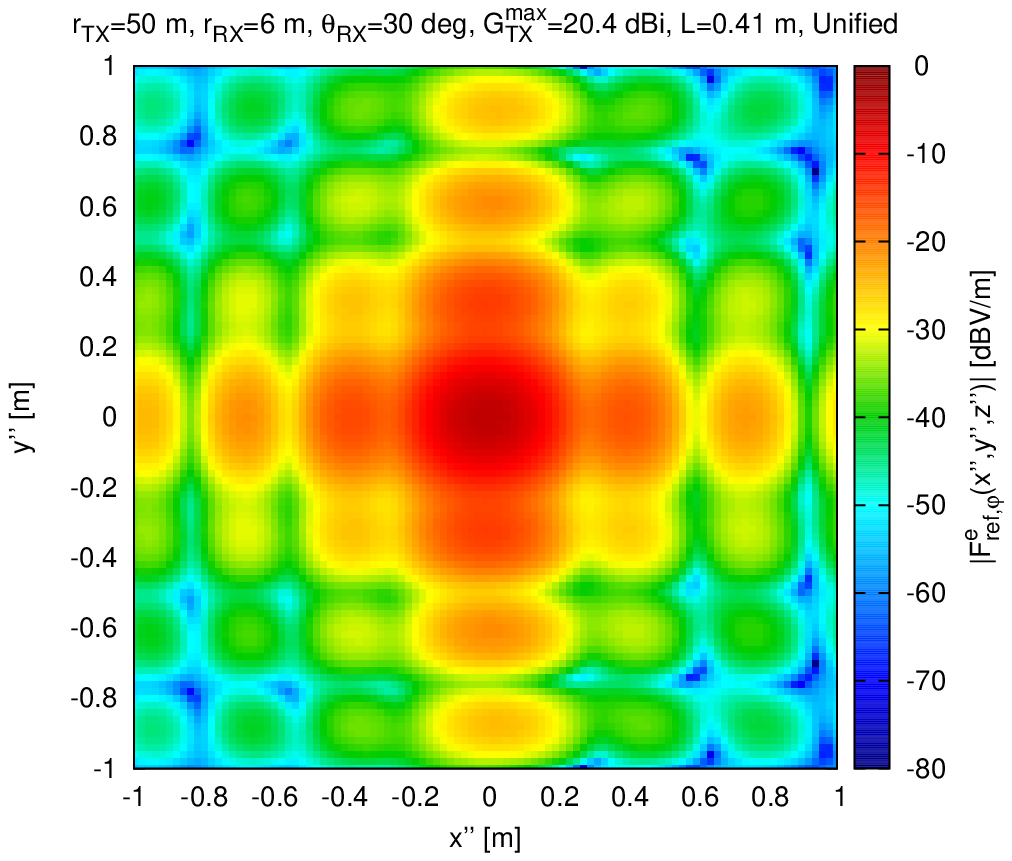}&
\includegraphics[%
  width=0.40\columnwidth]{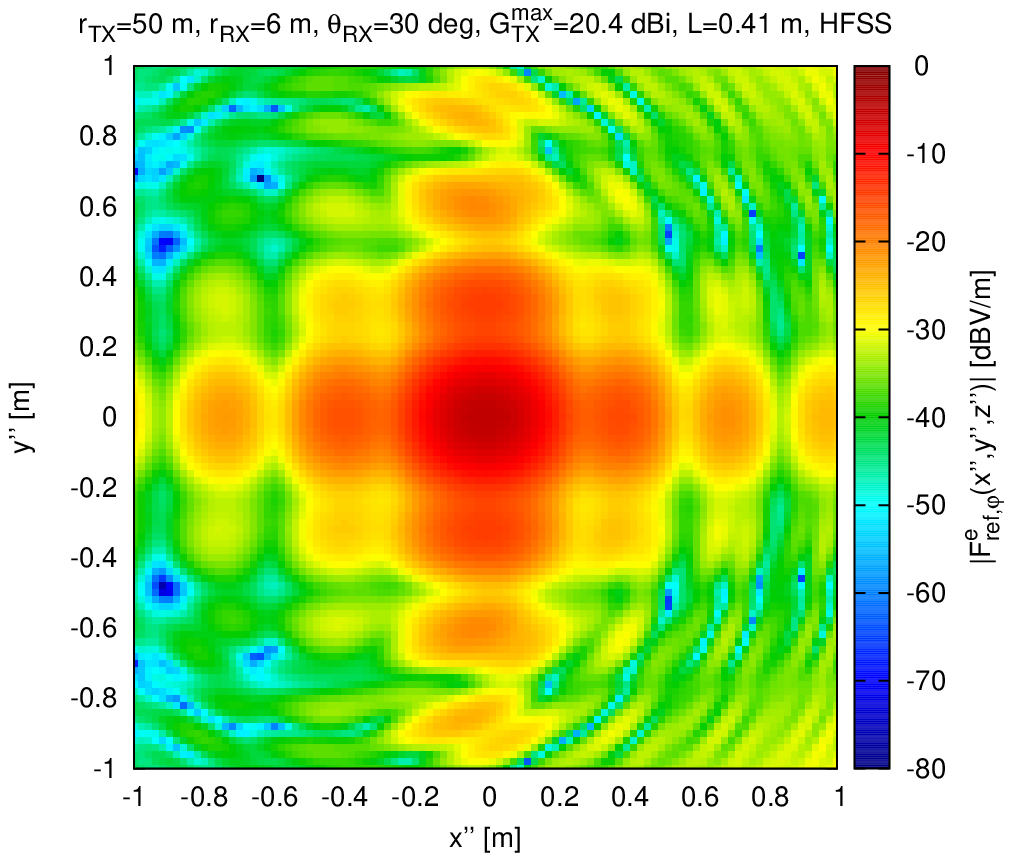}\tabularnewline
&
(\emph{a})&
(\emph{b})&
(\emph{c})\tabularnewline
\begin{sideways}
~~~~~~~~~~~~~~~~~~~~~~$\Theta=\Theta_{FF}$%
\end{sideways}&
\includegraphics[%
  width=0.40\columnwidth]{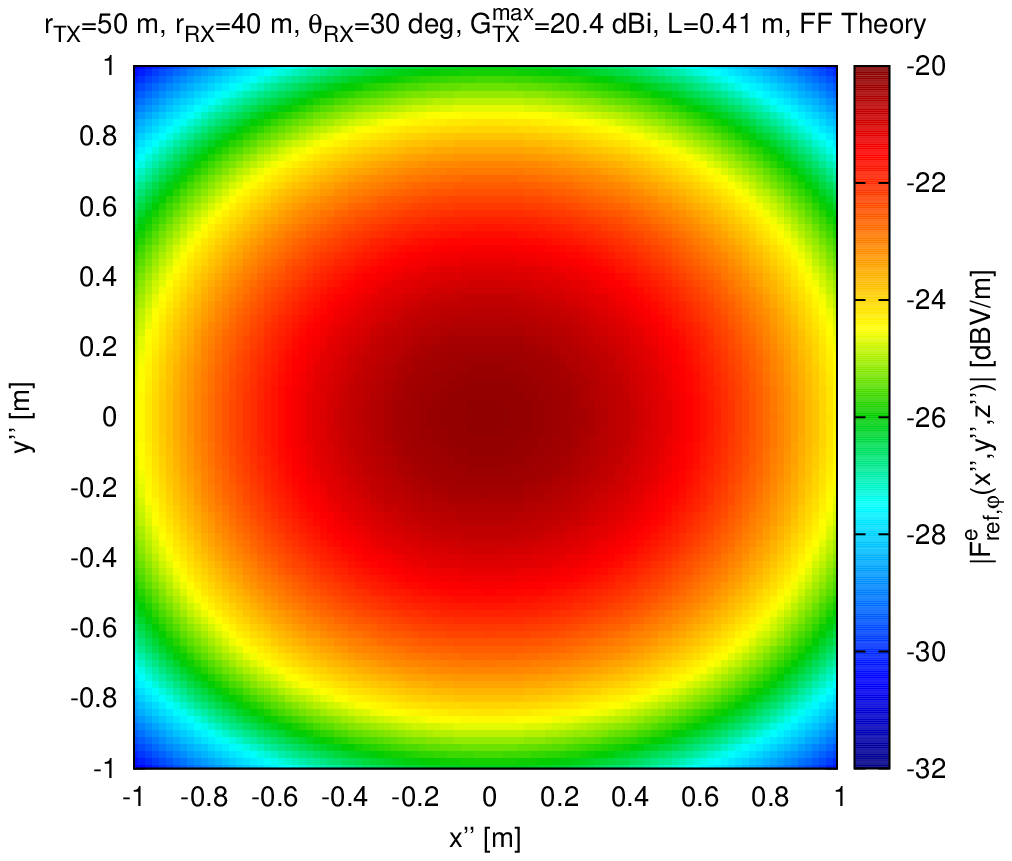}&
\includegraphics[%
  width=0.40\columnwidth]{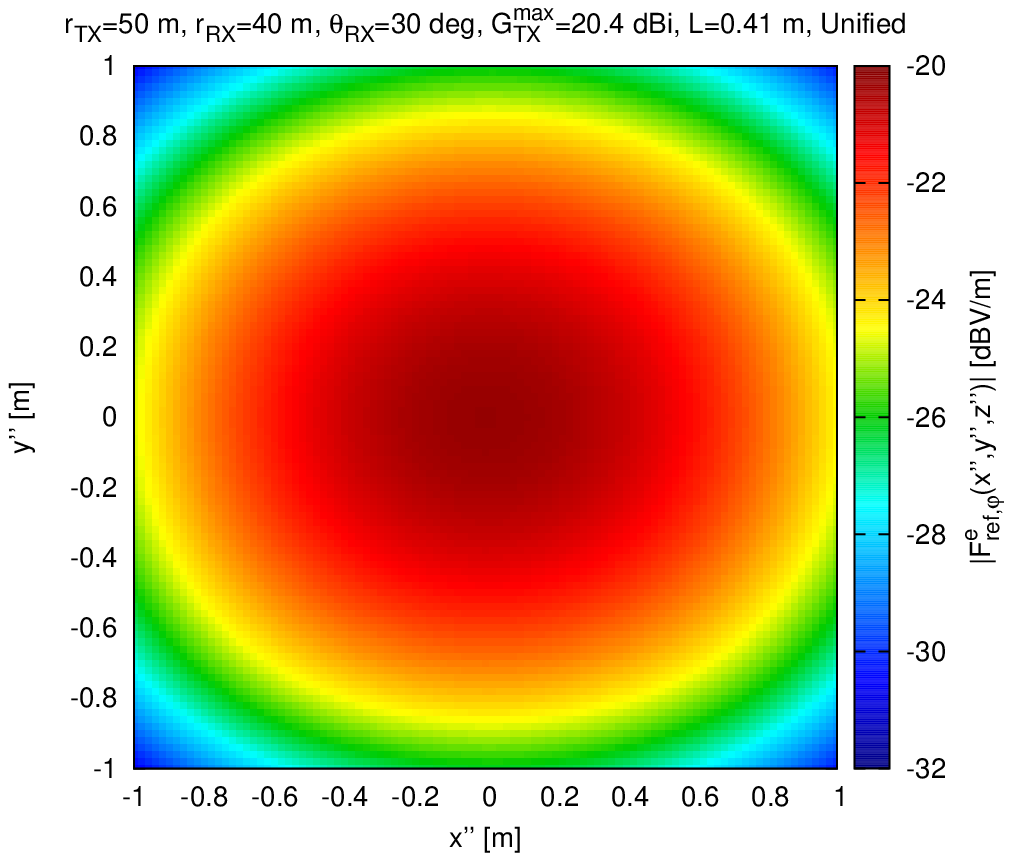}&
\includegraphics[%
  width=0.40\columnwidth]{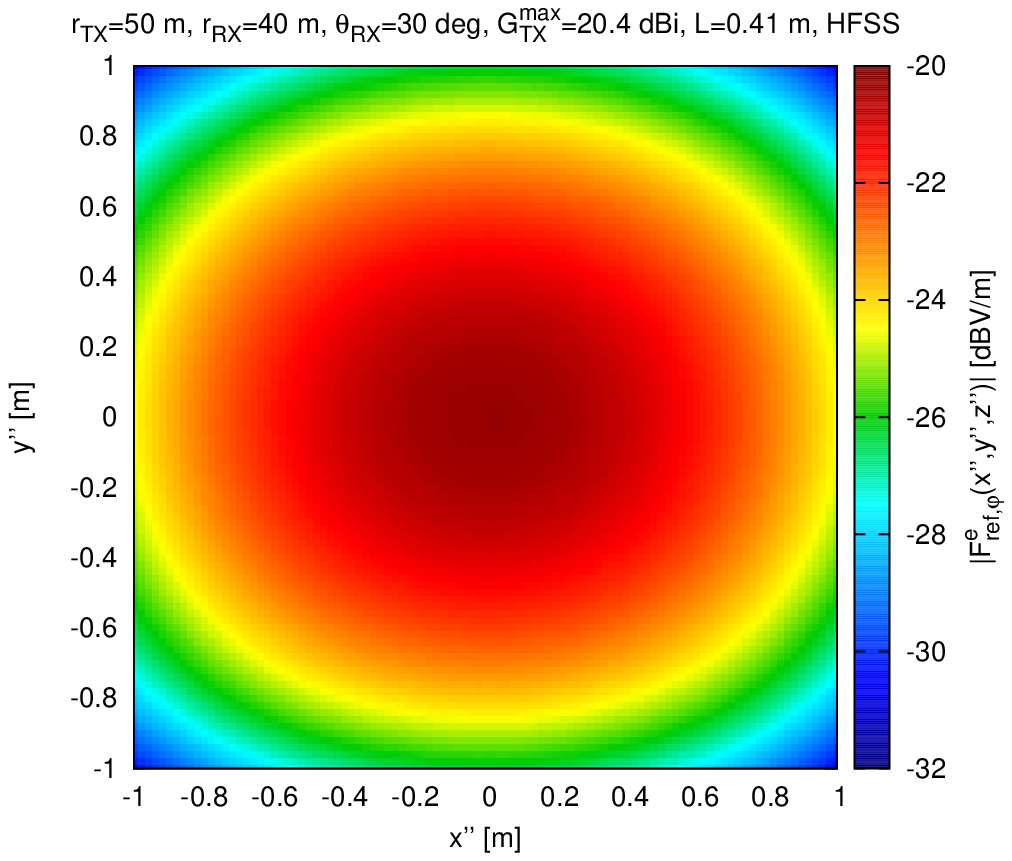}\tabularnewline
&
(\emph{d})&
(\emph{e})&
(\emph{f})\tabularnewline
\end{tabular}
\end{sideways}\end{center}

\begin{center}~\vfill\end{center}

\begin{center}\textbf{Fig. 4 - G. Oliveri et} \textbf{\emph{al.}}\textbf{,}
\textbf{\emph{{}``}}Generalized Analysis and Unified Design of \emph{EM}
Skins ...''\end{center}

\newpage
\begin{center}~\vfill\end{center}

\begin{center}\begin{tabular}{ccc}
&
\emph{FF Theory}&
\emph{Generalized Theory}\tabularnewline
\begin{sideways}
~~~~~~~~~~~~~~~~~~~~$\Theta=\Theta_{NF}$%
\end{sideways}&
\includegraphics[%
  width=0.45\columnwidth]{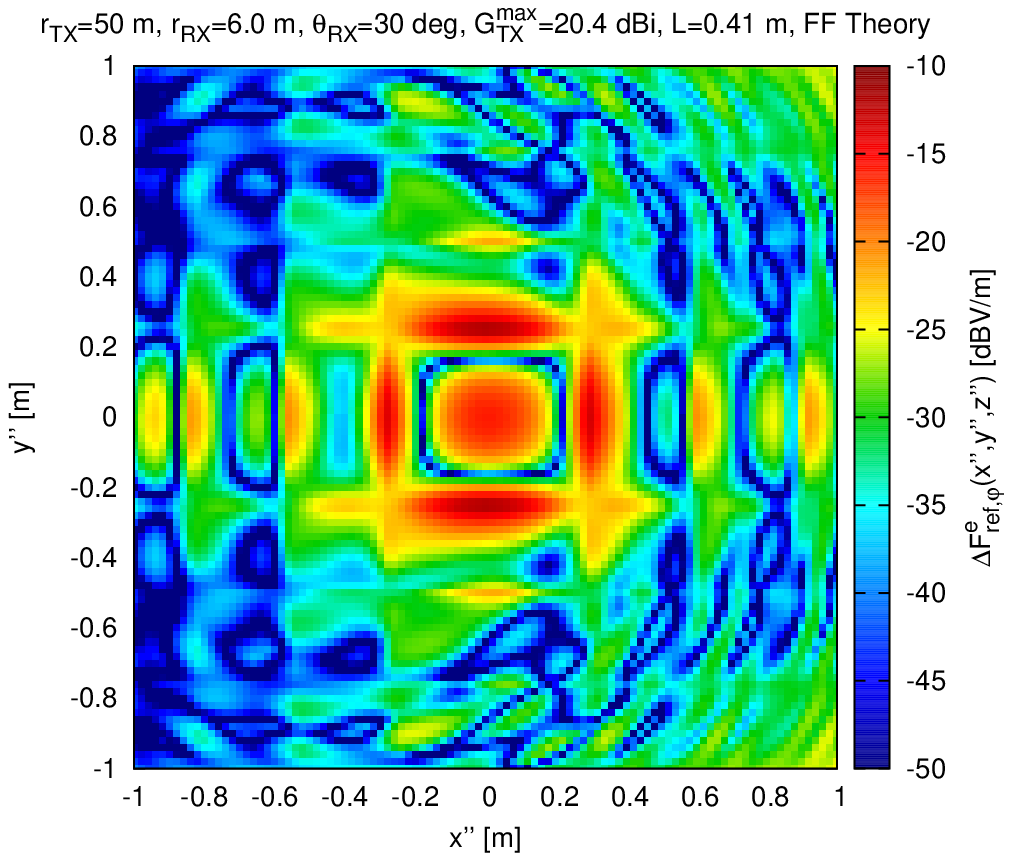}&
\includegraphics[%
  width=0.45\columnwidth]{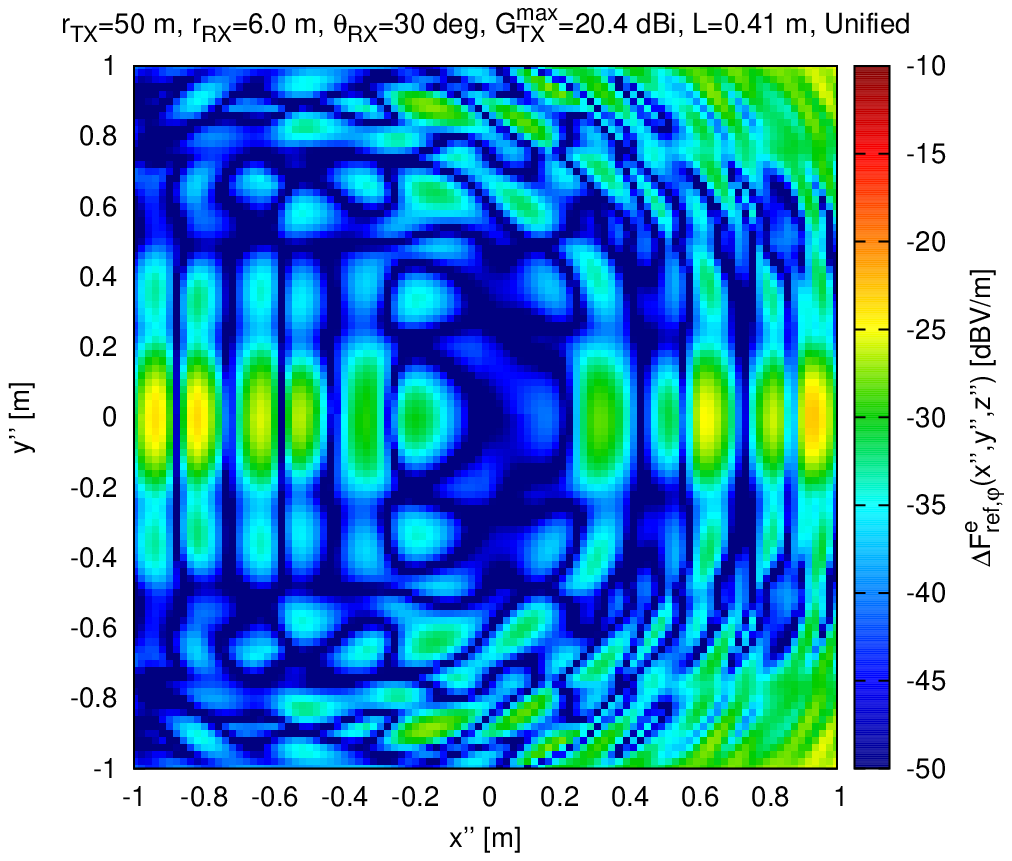}\tabularnewline
&
(\emph{a})&
(\emph{b})\tabularnewline
\begin{sideways}
~~~~~~~~~~~~~~~~~~~~~~$\Theta=\Theta_{FF}$%
\end{sideways}&
\includegraphics[%
  width=0.45\columnwidth]{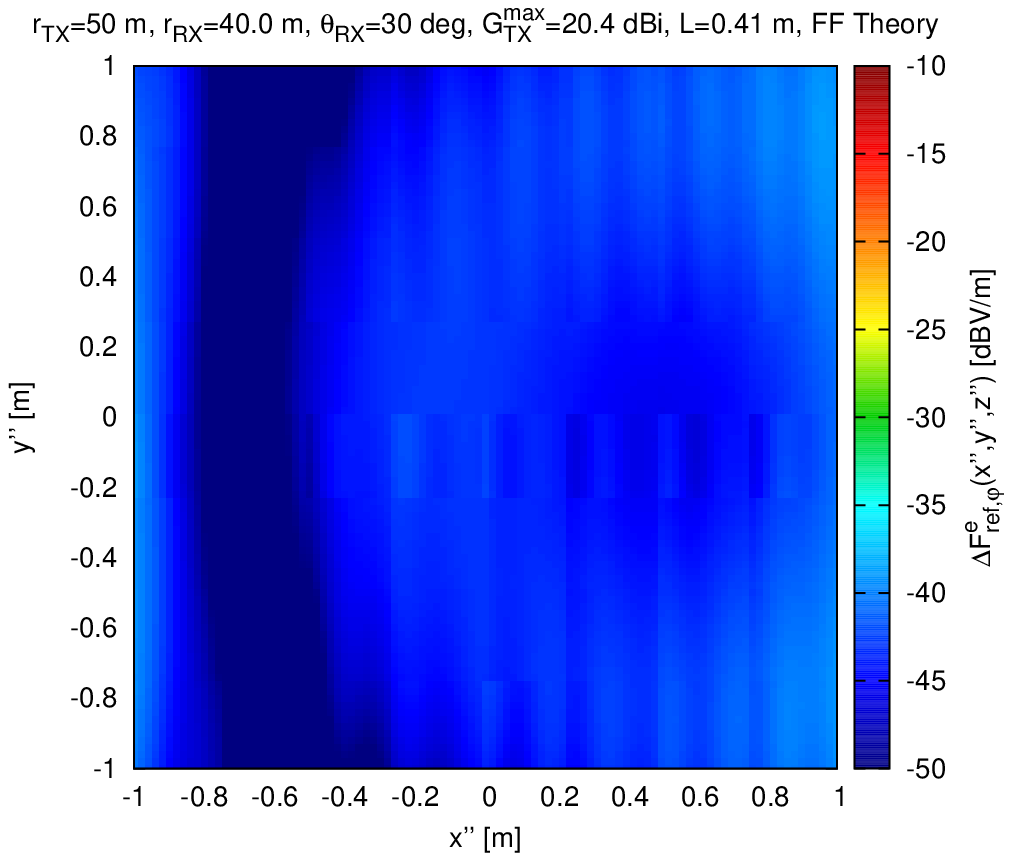}&
\includegraphics[%
  width=0.45\columnwidth]{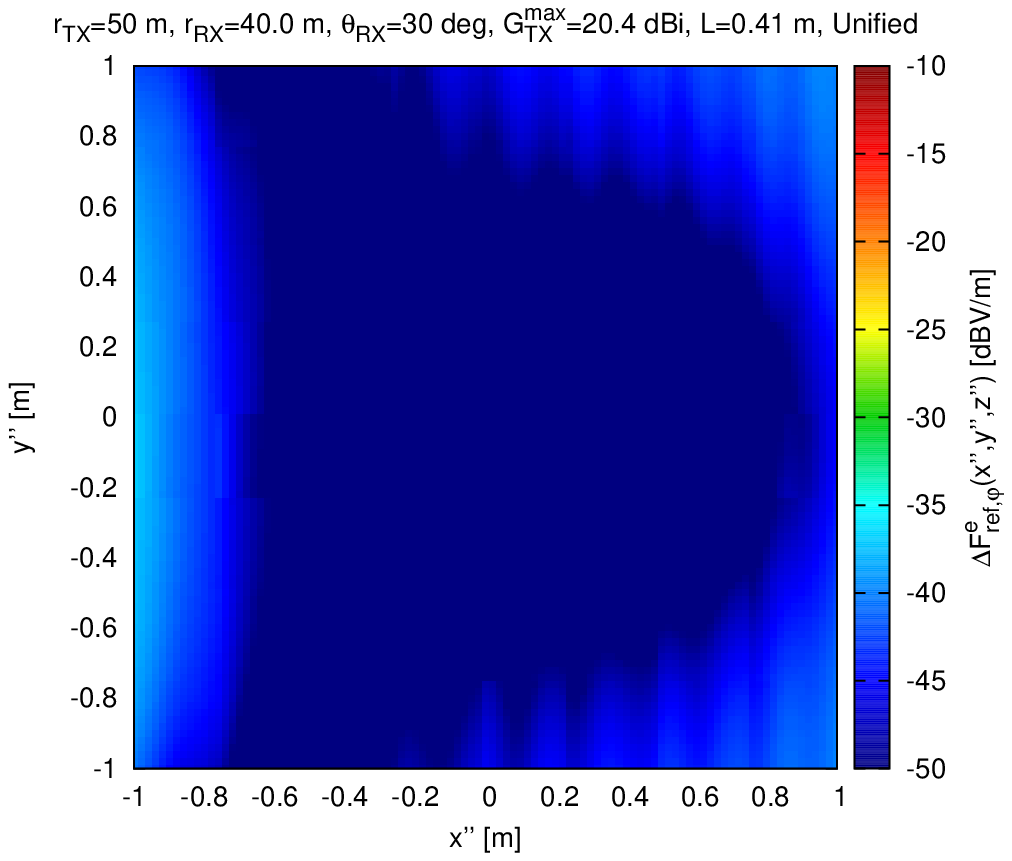}\tabularnewline
&
(\emph{c})&
(\emph{d})\tabularnewline
\end{tabular}\end{center}

\begin{center}~\vfill\end{center}

\begin{center}\textbf{Fig. 5 - G. Oliveri et} \textbf{\emph{al.}}\textbf{,}
\textbf{\emph{{}``}}Generalized Analysis and Unified Design of \emph{EM}
Skins ...''\end{center}

\newpage
\begin{center}~\vfill\end{center}

\begin{center}\begin{tabular}{ccc}
\multicolumn{3}{c}{$\Theta=\Theta_{NF}$}\tabularnewline
&
\emph{FF Theory}&
\emph{Generalized Theory}\tabularnewline
\begin{sideways}
~~~~~~~~~~~~~~~~~~~~Fig. 3(\emph{b})%
\end{sideways}&
\includegraphics[%
  width=0.45\columnwidth]{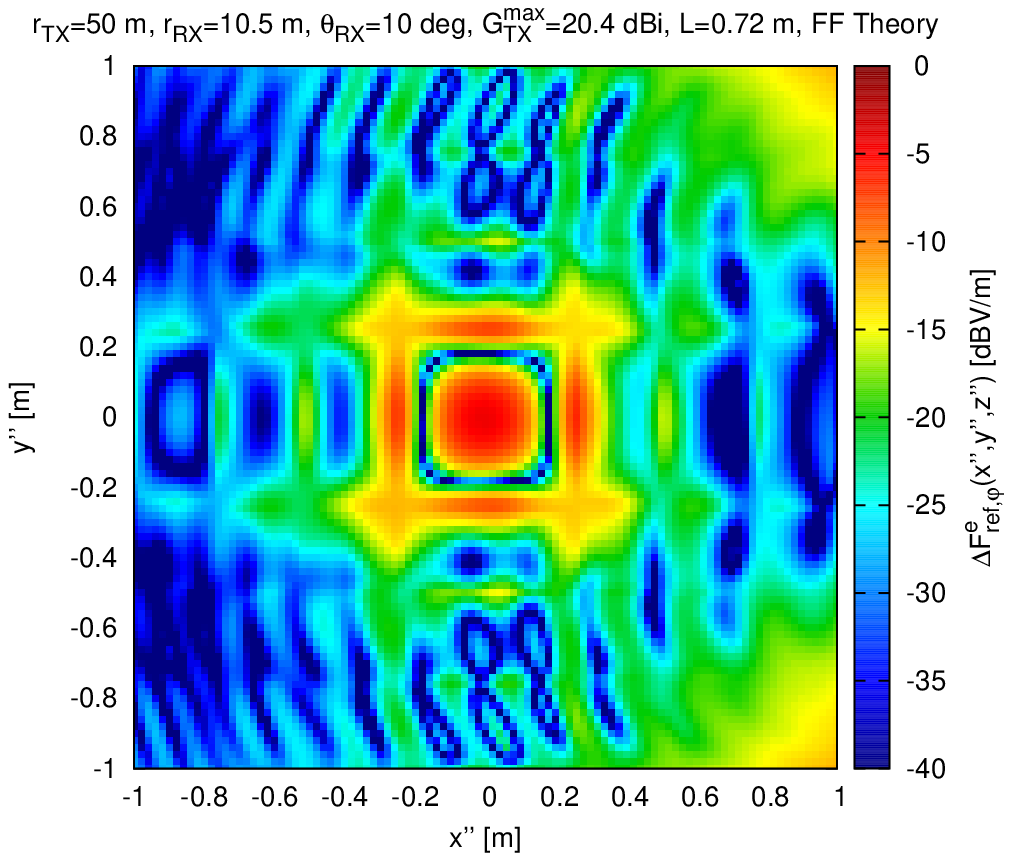}&
\includegraphics[%
  width=0.45\columnwidth]{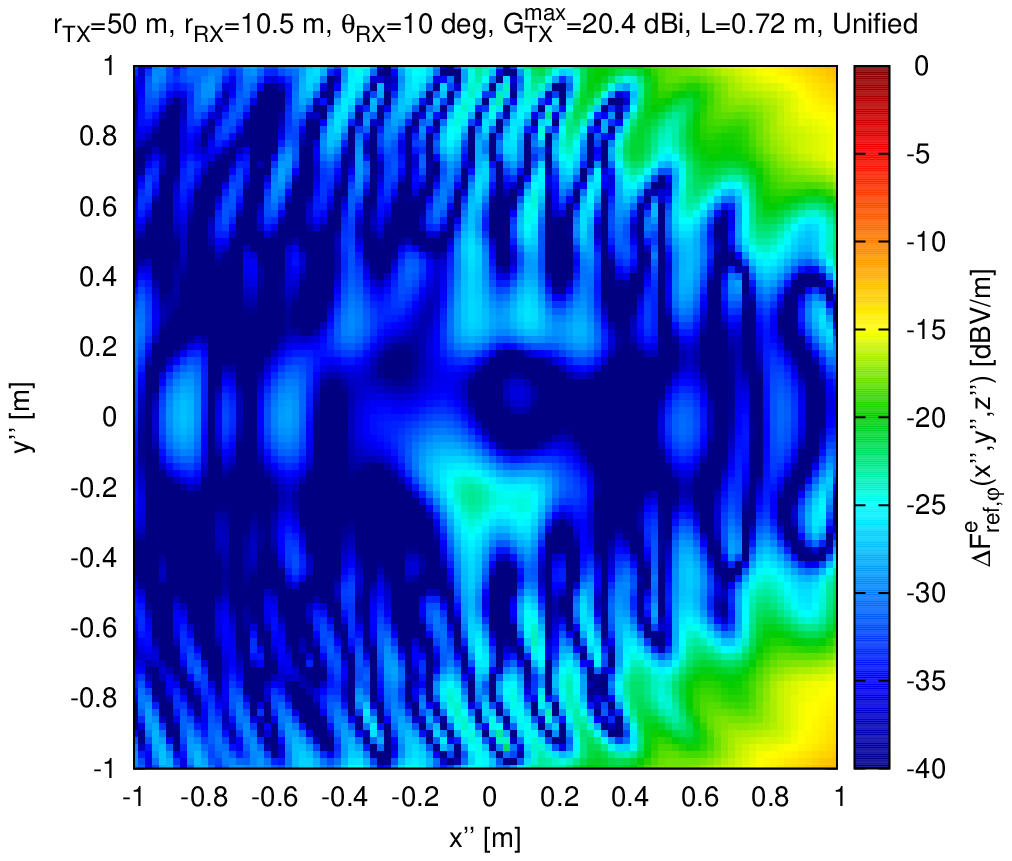}\tabularnewline
&
(\emph{a})&
(\emph{b})\tabularnewline
\begin{sideways}
~~~~~~~~~~~~~~~~~~~~Fig. 3(\emph{c})%
\end{sideways}&
\includegraphics[%
  width=0.45\columnwidth]{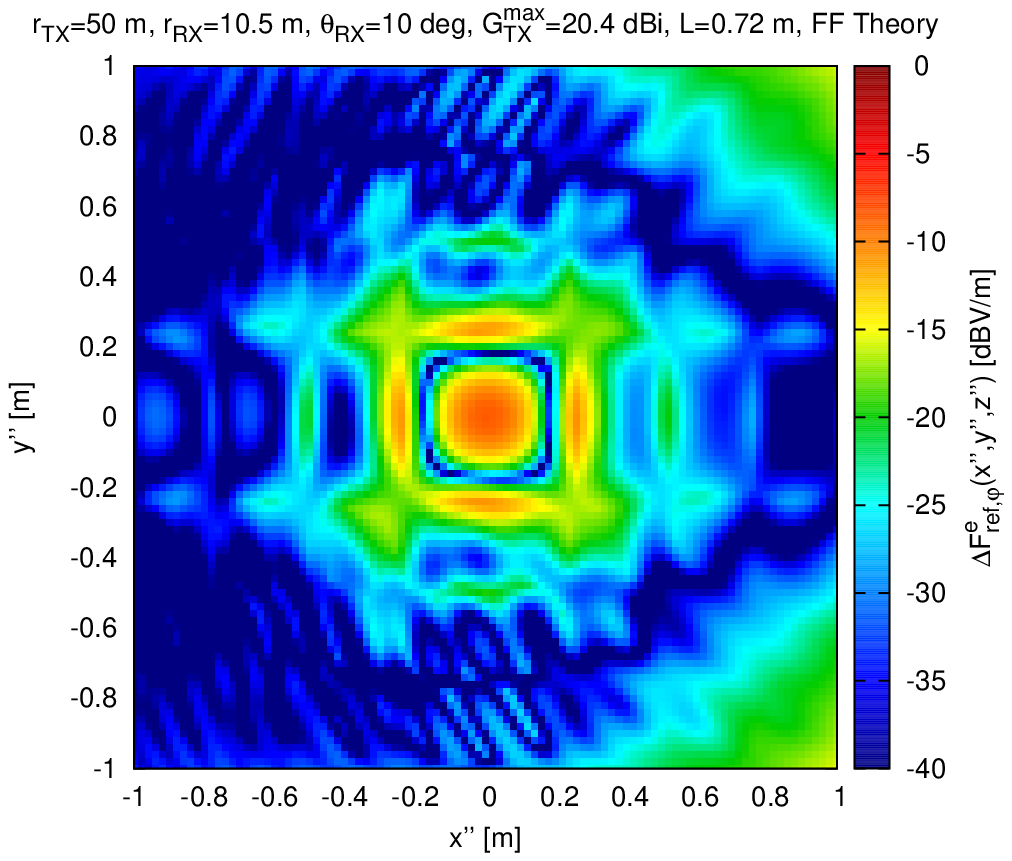}&
\includegraphics[%
  width=0.45\columnwidth]{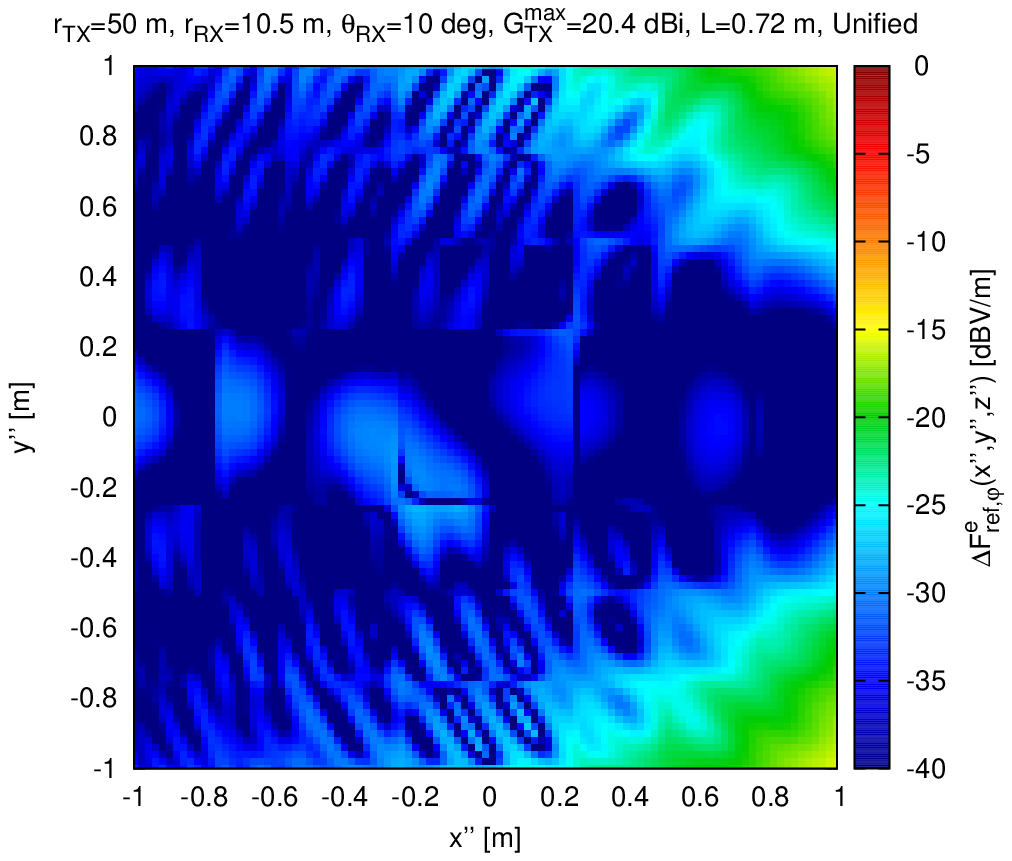}\tabularnewline
&
(\emph{c})&
(\emph{d})\tabularnewline
\end{tabular}\end{center}

\begin{center}~\vfill\end{center}

\begin{center}\textbf{Fig. 6 - G. Oliveri et} \textbf{\emph{al.}}\textbf{,}
\textbf{\emph{{}``}}Generalized Analysis and Unified Design of \emph{EM}
Skins ...''\end{center}

\newpage
\begin{center}~\vfill\end{center}

\begin{center}\begin{tabular}{ccc}
\multicolumn{3}{c}{$\Theta=\Theta_{FF}$}\tabularnewline
&
\emph{FF Theory}&
\emph{Generalized Theory}\tabularnewline
\begin{sideways}
~~~~~~~~~~~~~~~~~~~~Fig. 3(\emph{b})%
\end{sideways}&
\includegraphics[%
  width=0.45\columnwidth]{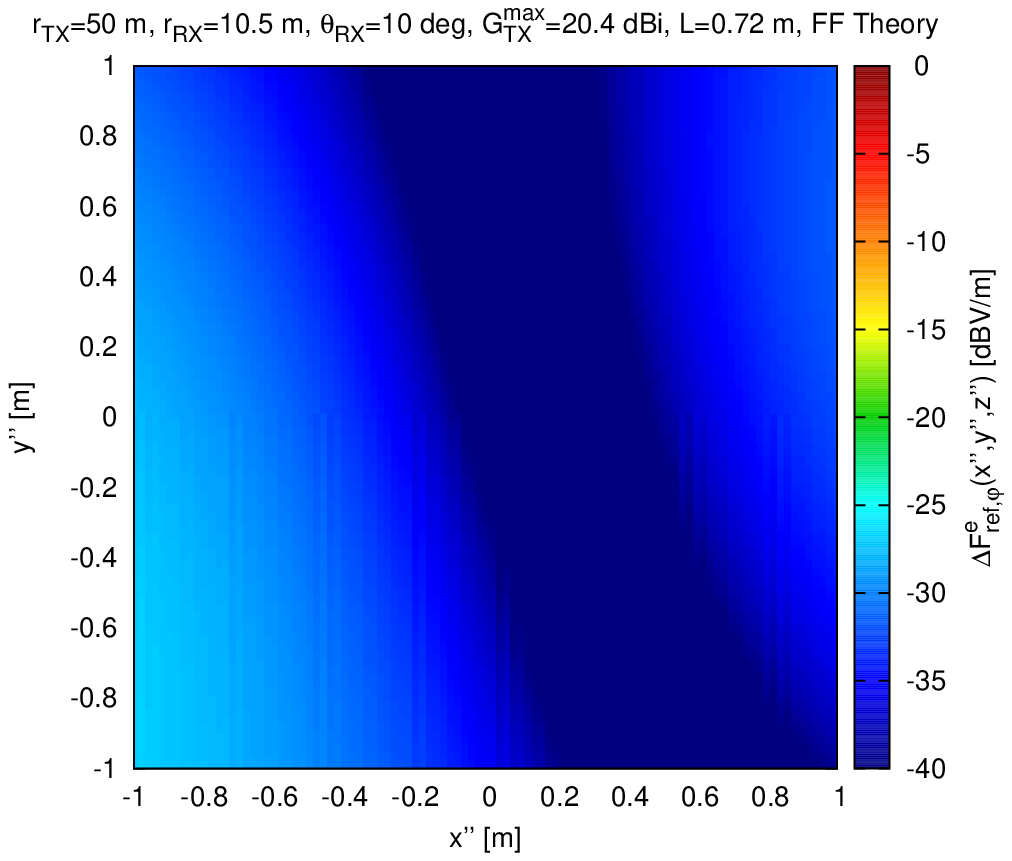}&
\includegraphics[%
  width=0.45\columnwidth]{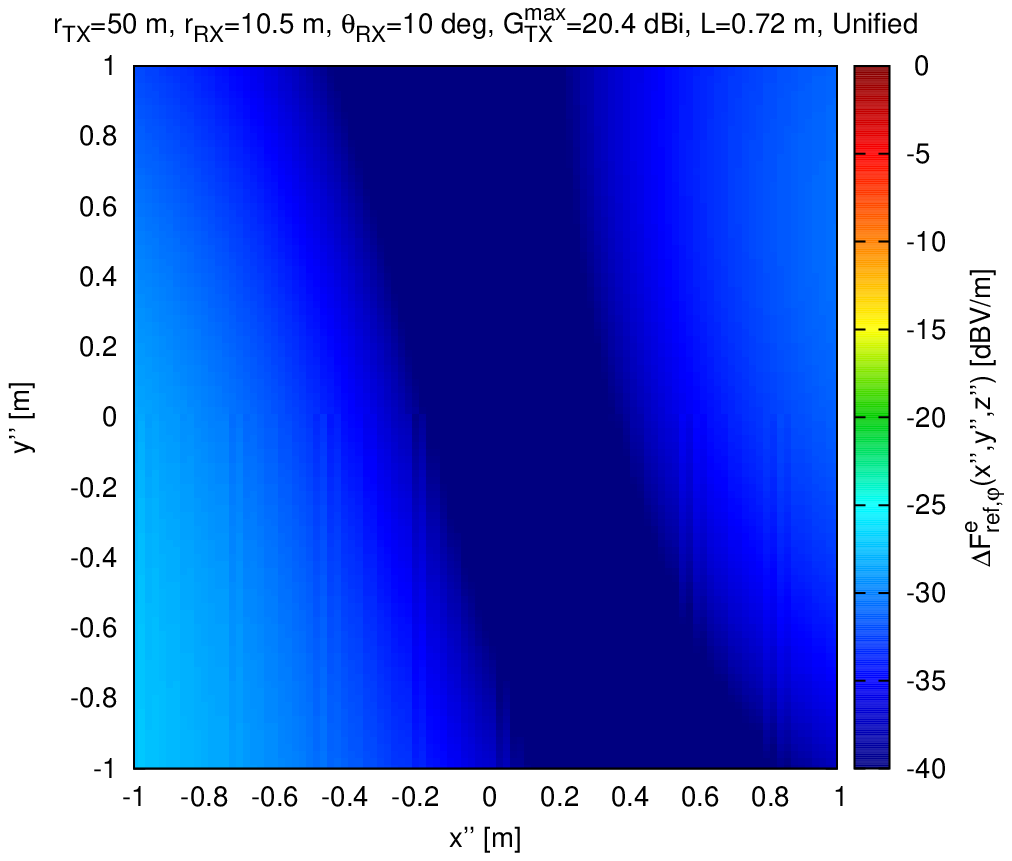}\tabularnewline
&
(\emph{a})&
(\emph{b})\tabularnewline
\begin{sideways}
~~~~~~~~~~~~~~~~~~~~Fig. 3(\emph{c})%
\end{sideways}&
\includegraphics[%
  width=0.45\columnwidth]{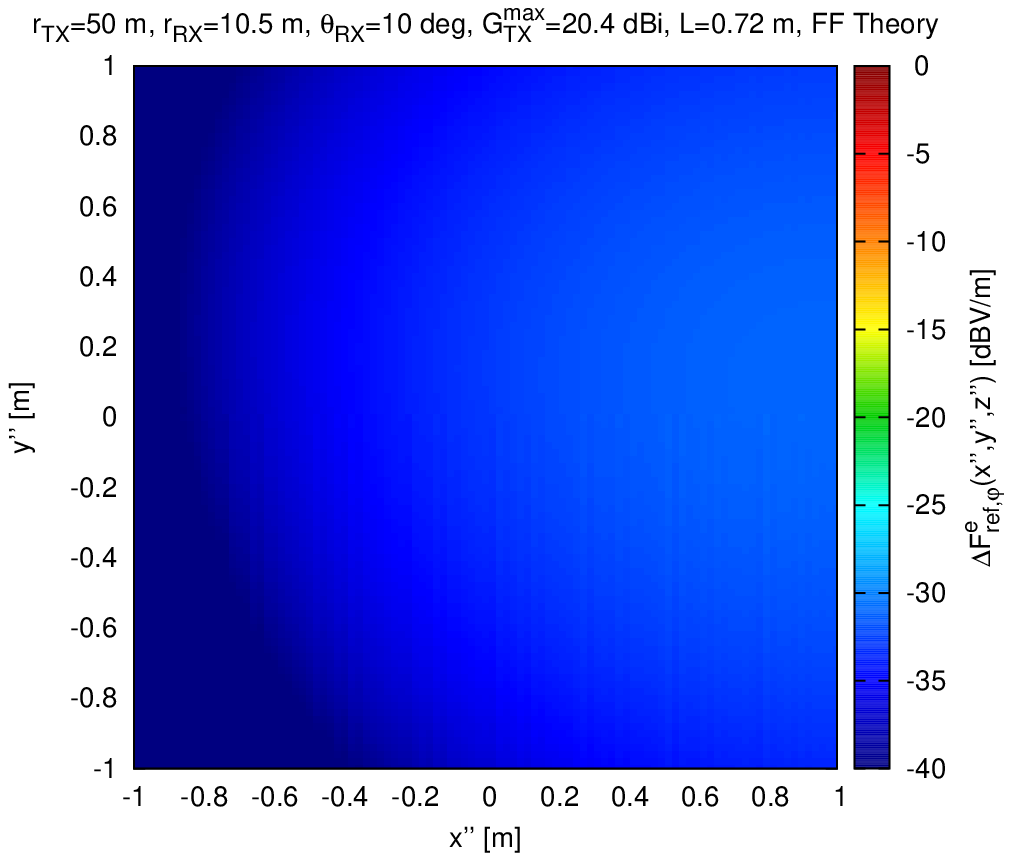}&
\includegraphics[%
  width=0.45\columnwidth]{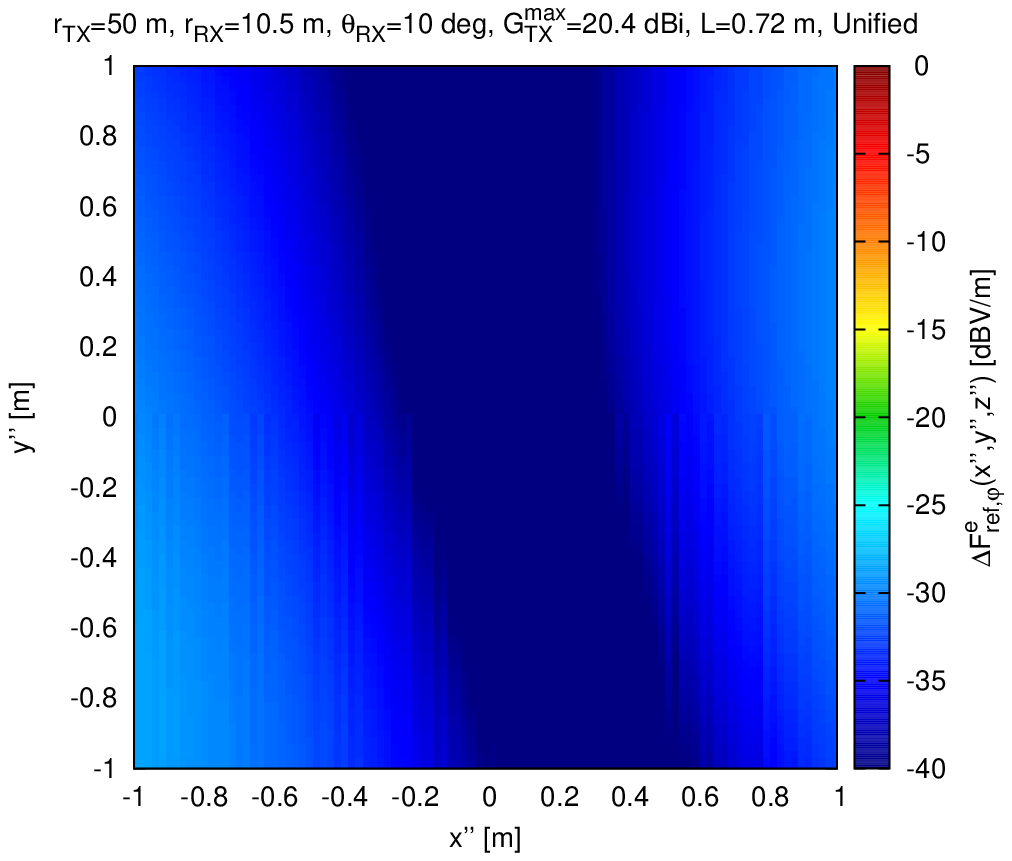}\tabularnewline
&
(\emph{c})&
(\emph{d})\tabularnewline
\end{tabular}\end{center}

\begin{center}~\vfill\end{center}

\begin{center}\textbf{Fig. 7 - G. Oliveri et} \textbf{\emph{al.}}\textbf{,}
\textbf{\emph{{}``}}Generalized Analysis and Unified Design of \emph{EM}
Skins ...''\end{center}
\newpage

\begin{center}\begin{tabular}{cc}
\emph{FFM}&
\emph{USM}\tabularnewline
\includegraphics[%
  width=0.45\columnwidth]{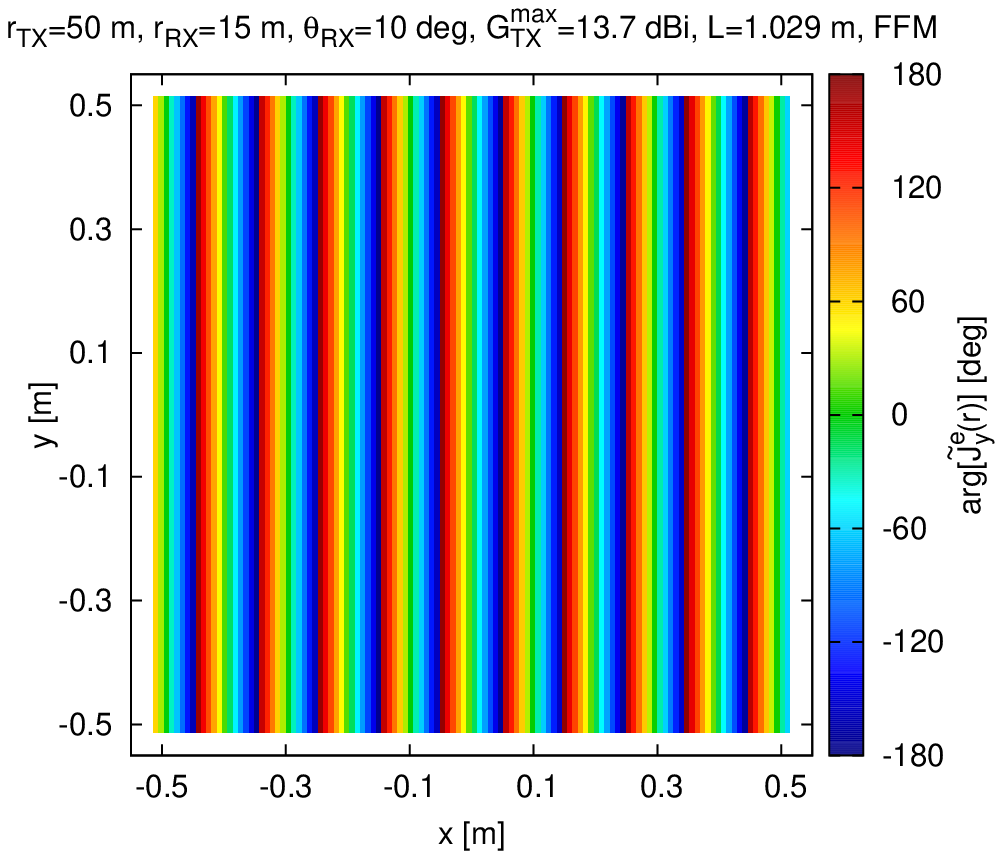}&
\includegraphics[%
  width=0.45\columnwidth]{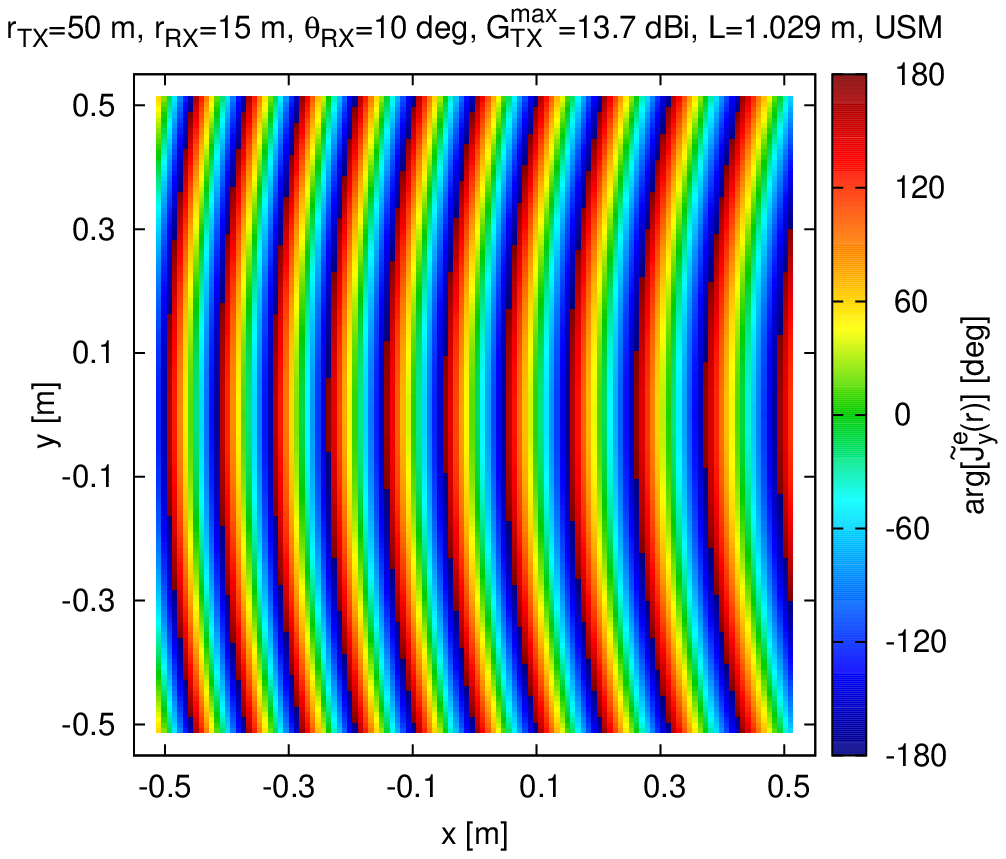}\tabularnewline
(\emph{a})&
(\emph{b})\tabularnewline
\includegraphics[%
  width=0.45\columnwidth]{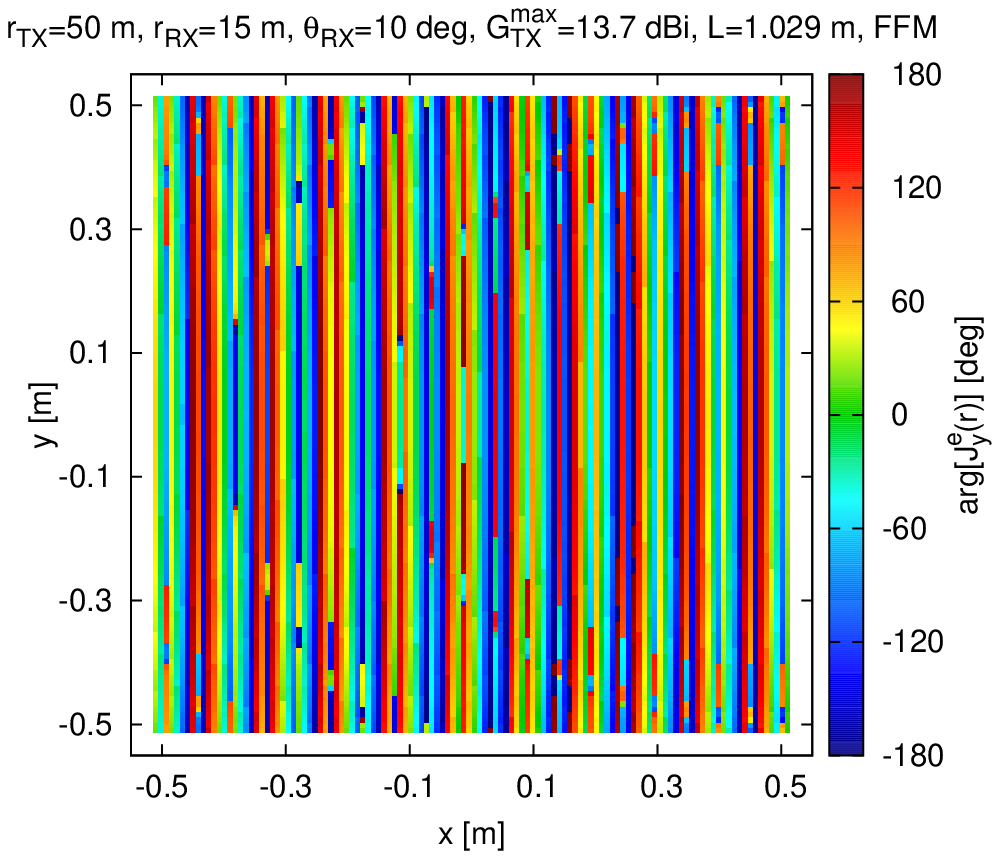}&
\includegraphics[%
  width=0.45\columnwidth]{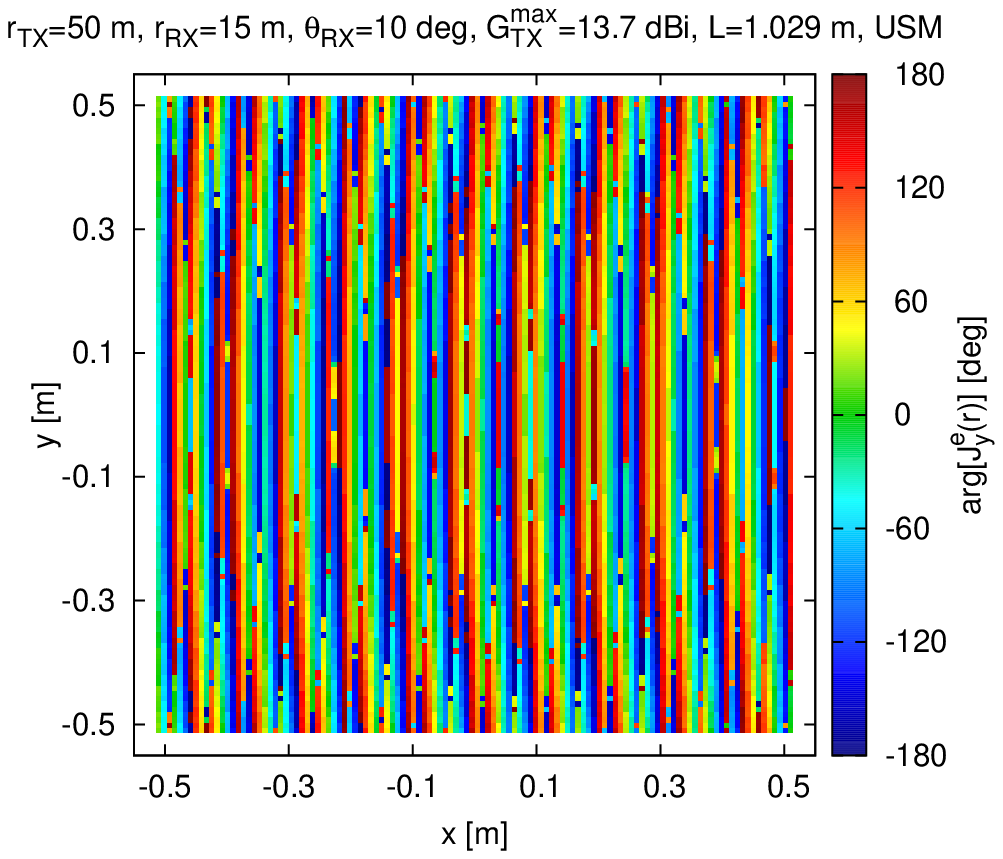}\tabularnewline
(\emph{c})&
(\emph{d})\tabularnewline
\multicolumn{2}{c}{\includegraphics[%
  width=0.70\columnwidth]{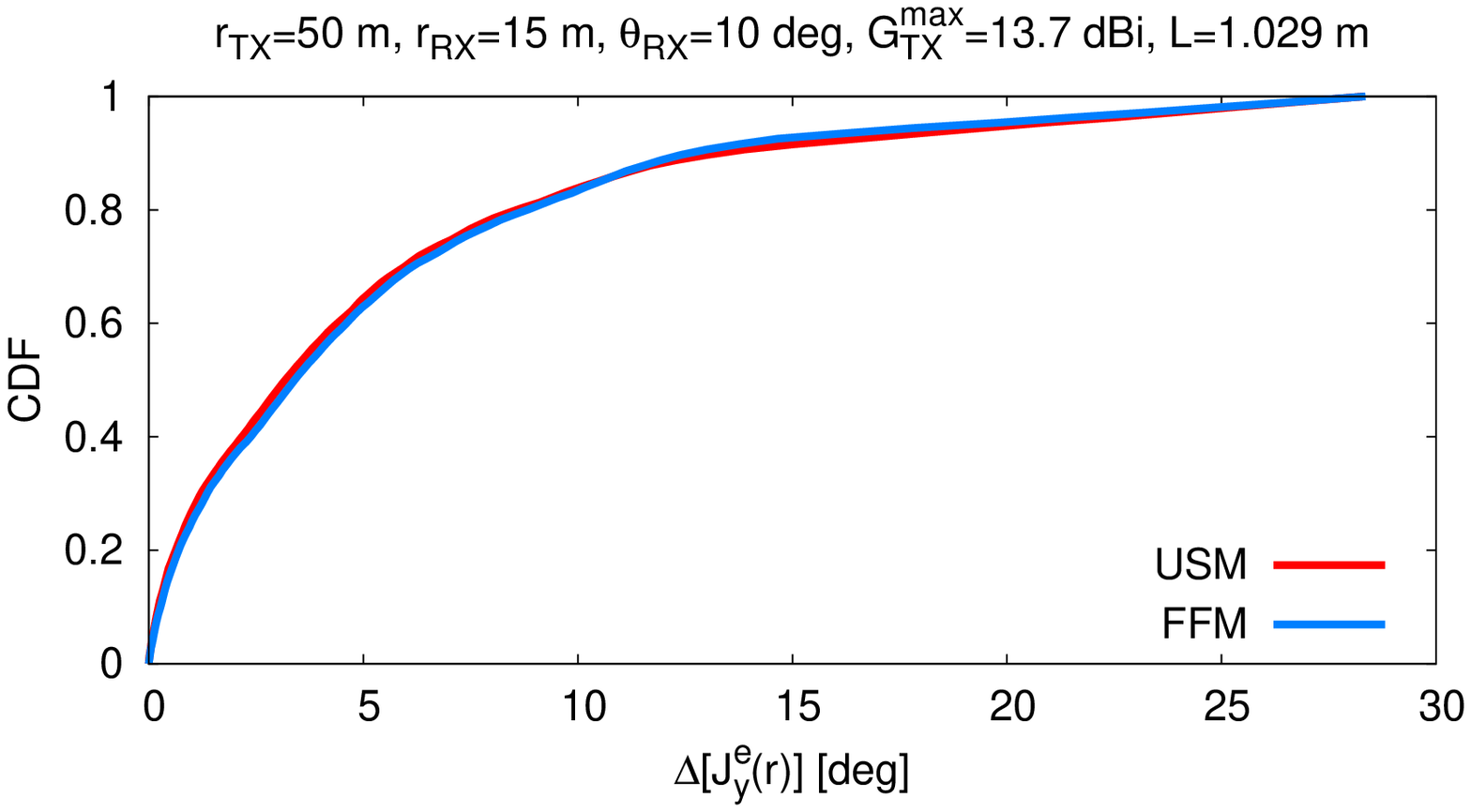}}\tabularnewline
\multicolumn{2}{c}{(\emph{e})}\tabularnewline
\end{tabular}\end{center}

\begin{center}~\vfill\end{center}

\begin{center}\textbf{Fig. 8 - G. Oliveri et} \textbf{\emph{al.}}\textbf{,}
\textbf{\emph{{}``}}Generalized Analysis and Unified Design of \emph{EM}
Skins ...''\end{center}

\newpage
\begin{center}~\vfill\end{center}

\begin{center}\begin{tabular}{cc}
\emph{FFM}&
\emph{USM}\tabularnewline
\includegraphics[%
  width=0.45\columnwidth]{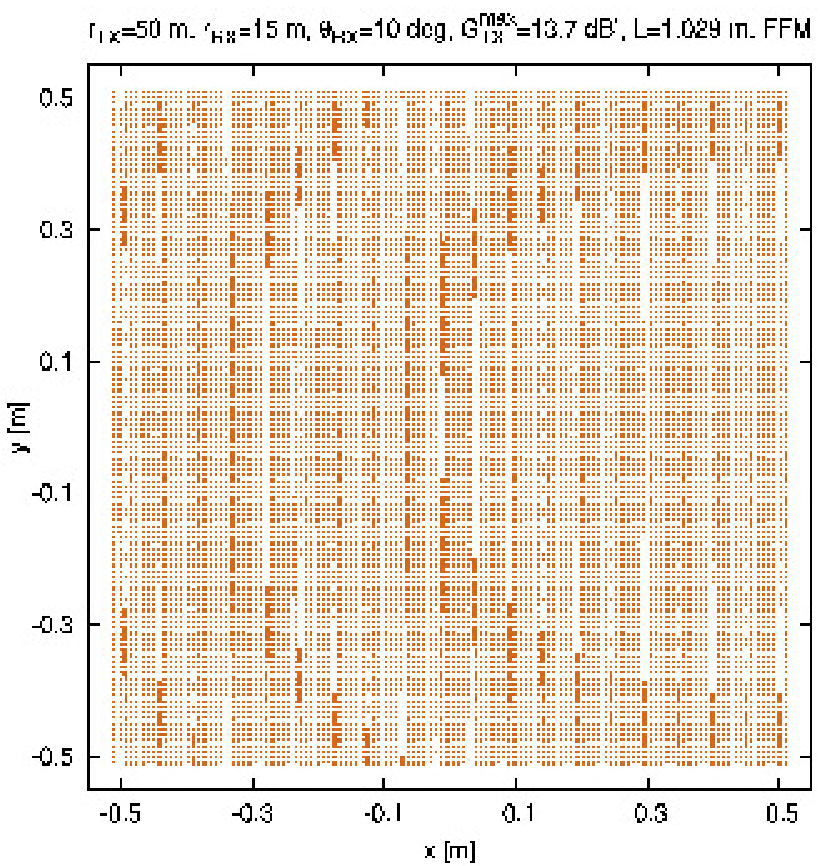}&
\includegraphics[%
  width=0.45\columnwidth]{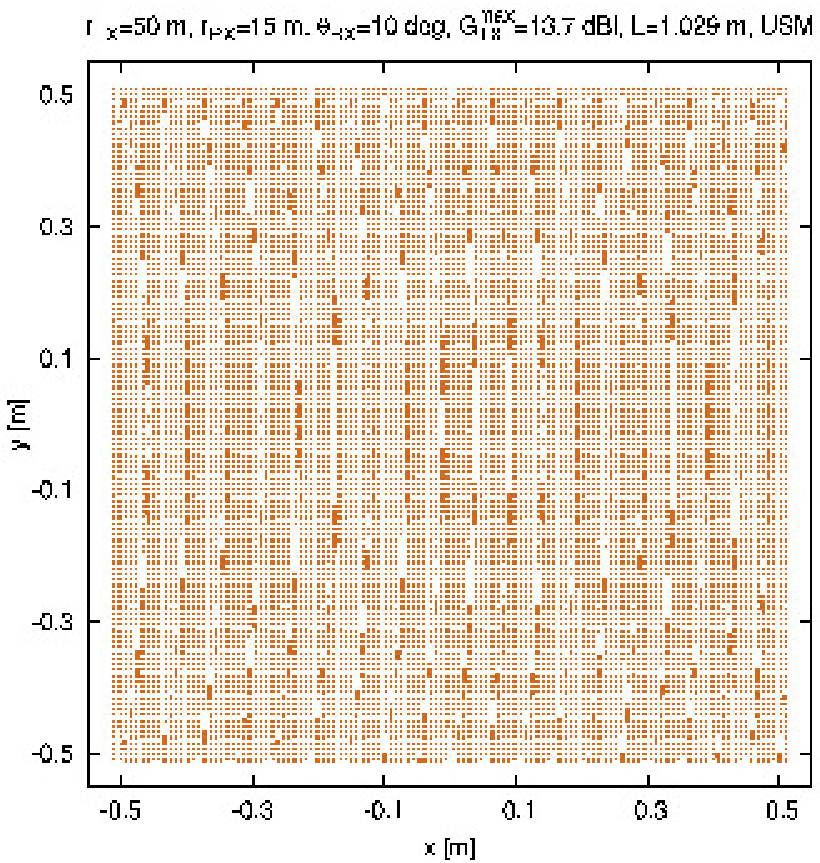}\tabularnewline
(\emph{a})&
(\emph{b})\tabularnewline
\includegraphics[%
  width=0.45\columnwidth]{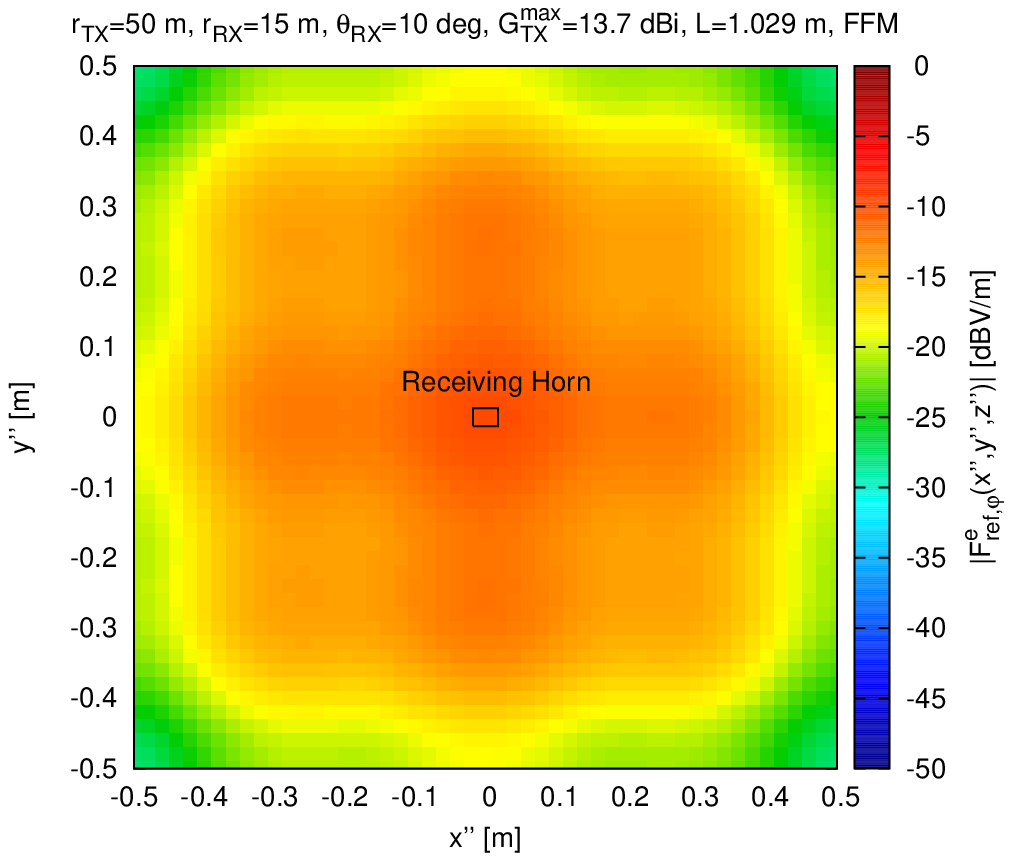}&
\includegraphics[%
  width=0.45\columnwidth]{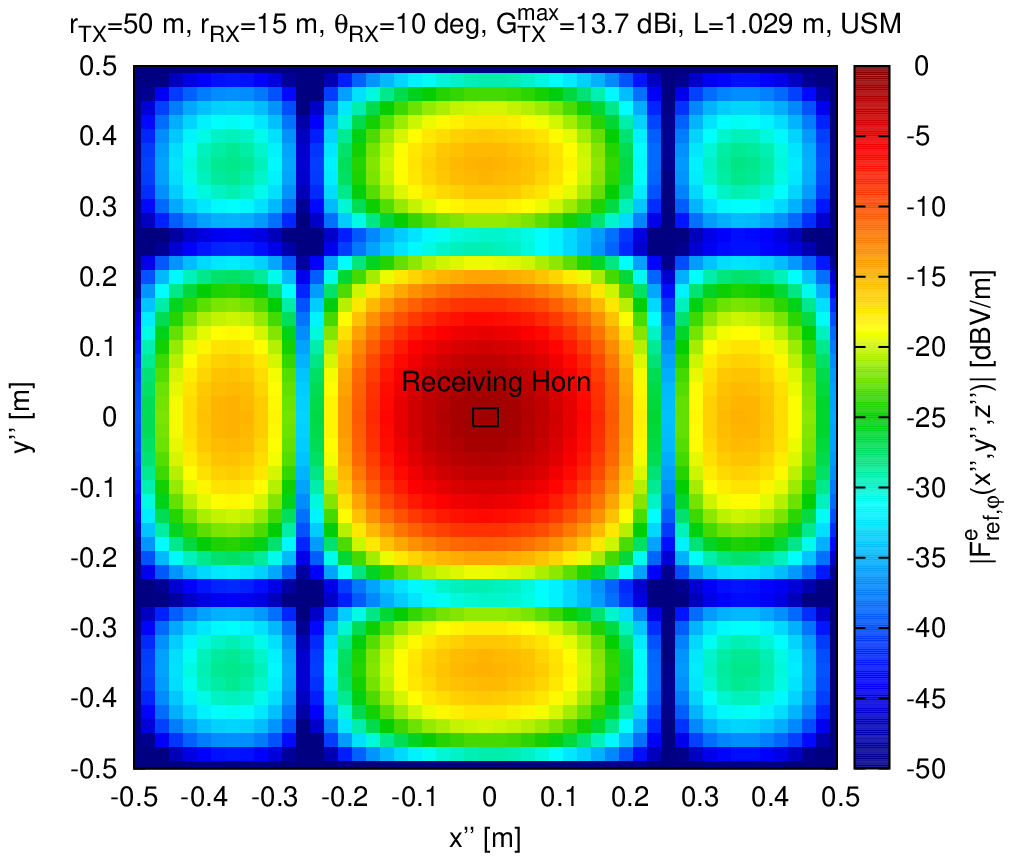}\tabularnewline
(\emph{c})&
(\emph{d})\tabularnewline
\end{tabular}~\vfill\end{center}

\begin{center}\textbf{Fig. 9 - G. Oliveri et} \textbf{\emph{al.}}\textbf{,}
\textbf{\emph{{}``}}Generalized Analysis and Unified Design of \emph{EM}
Skins ...''\end{center}

\newpage
\begin{center}~\vfill\end{center}

\begin{center}\includegraphics[%
  width=0.95\columnwidth]{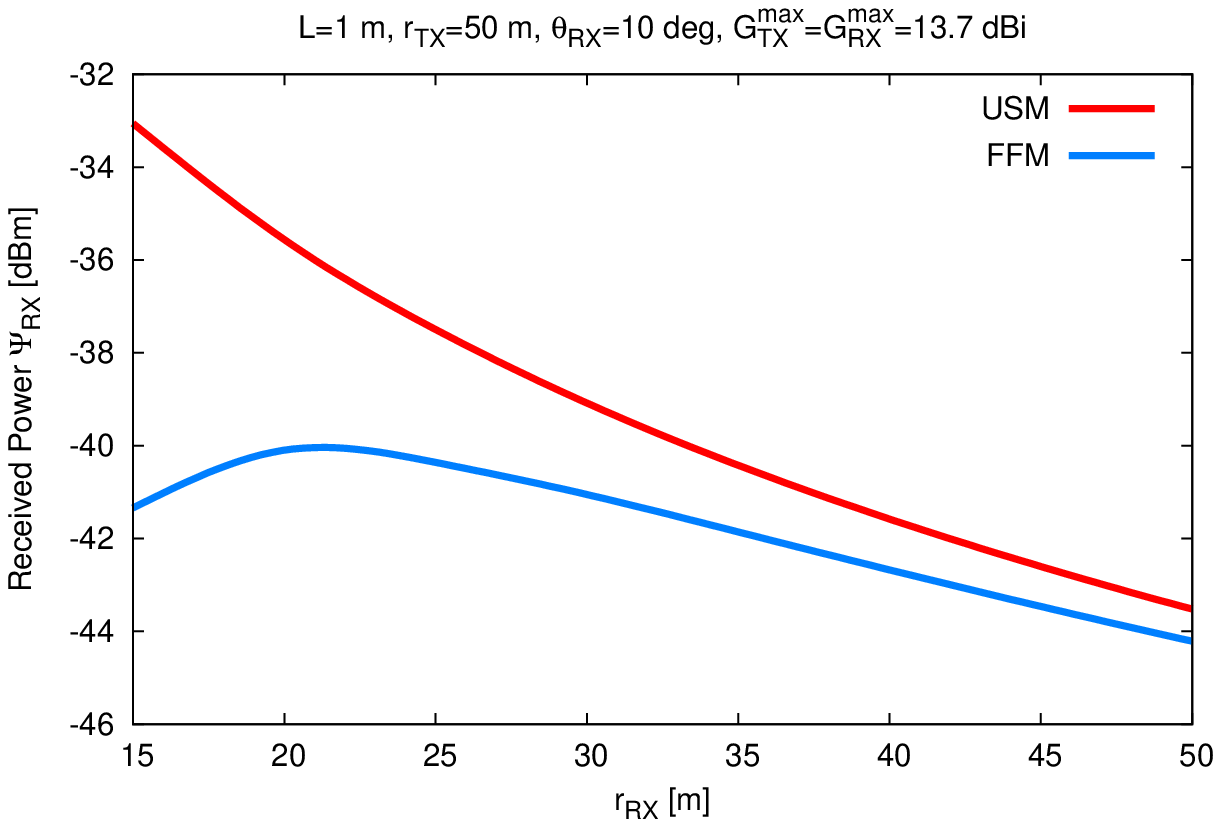}\end{center}

\begin{center}~\vfill\end{center}

\begin{center}\textbf{Fig. 10 - G. Oliveri et} \textbf{\emph{al.}}\textbf{,}
\textbf{\emph{{}``}}Generalized Analysis and Unified Design of \emph{EM}
Skins ...''\end{center}

\newpage
\begin{center}\begin{tabular}{c}
\includegraphics[%
  width=0.60\columnwidth]{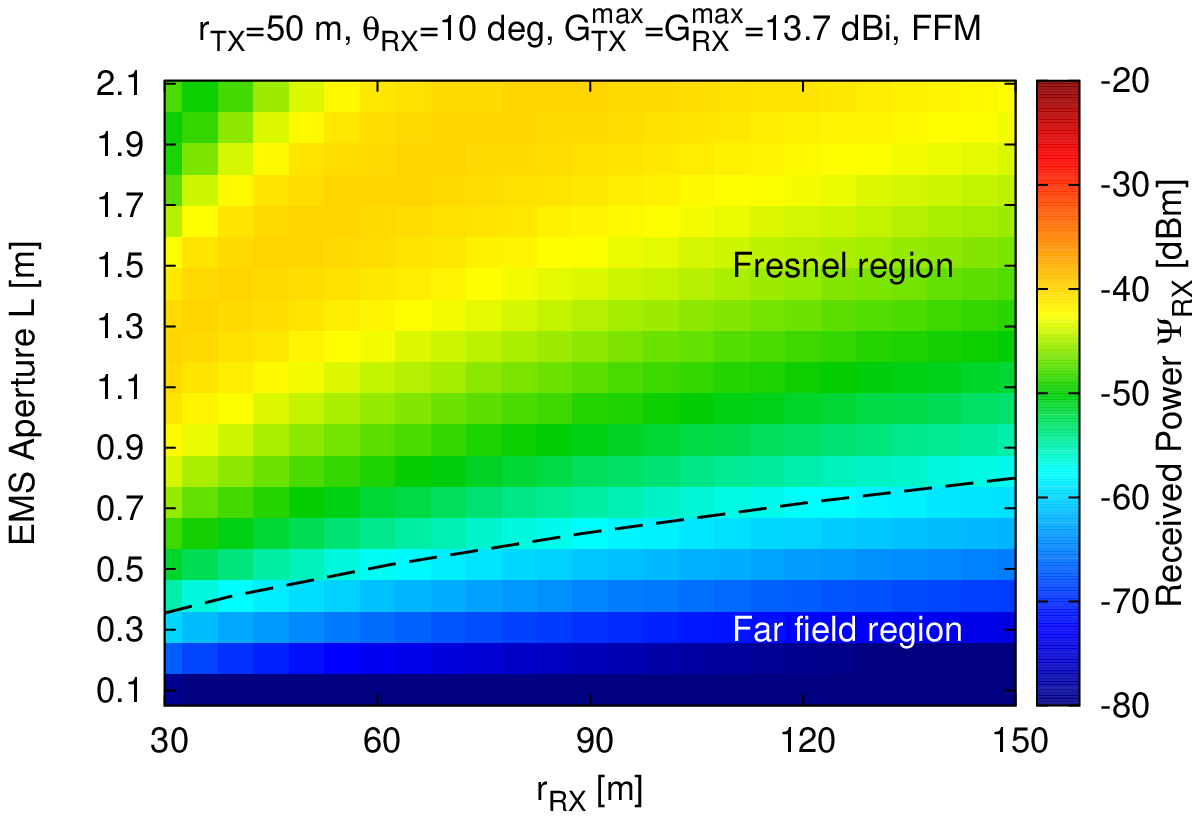}\tabularnewline
(\emph{a})\tabularnewline
\includegraphics[%
  width=0.60\columnwidth]{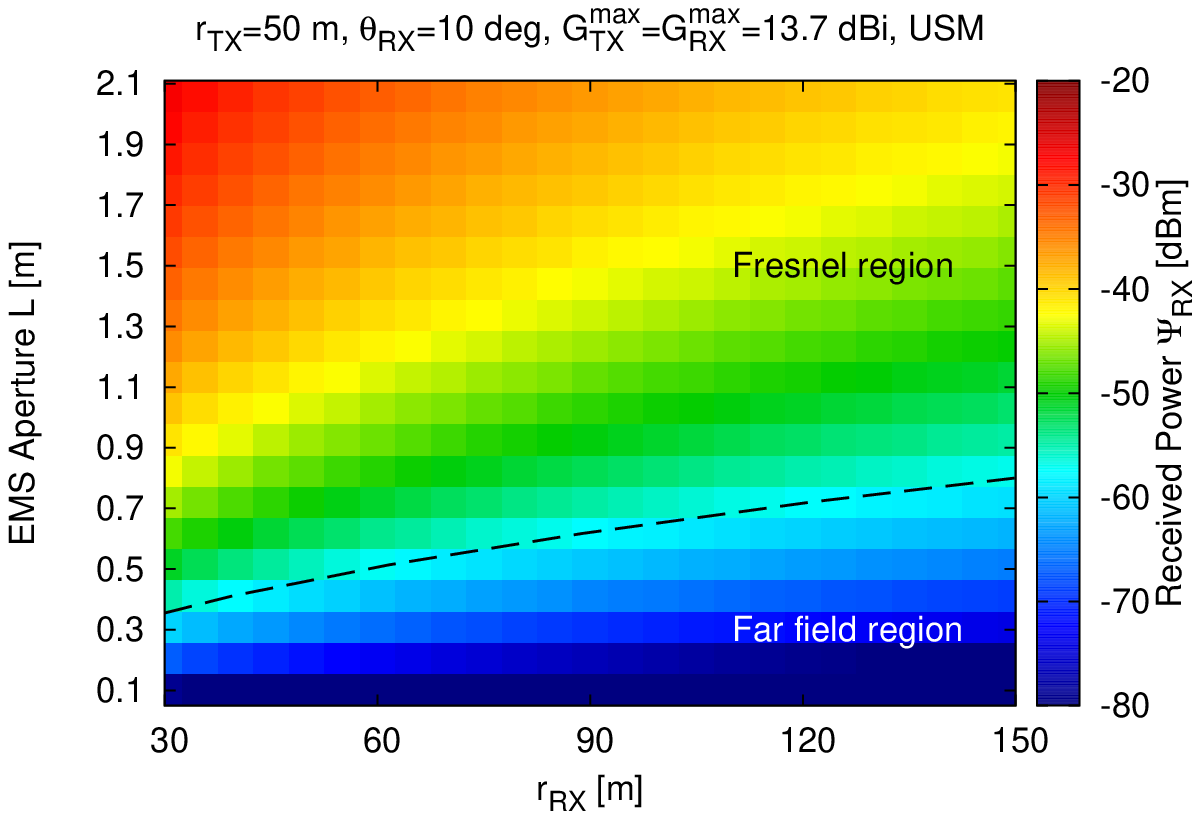}\tabularnewline
(\emph{b})\tabularnewline
\includegraphics[%
  width=0.60\columnwidth]{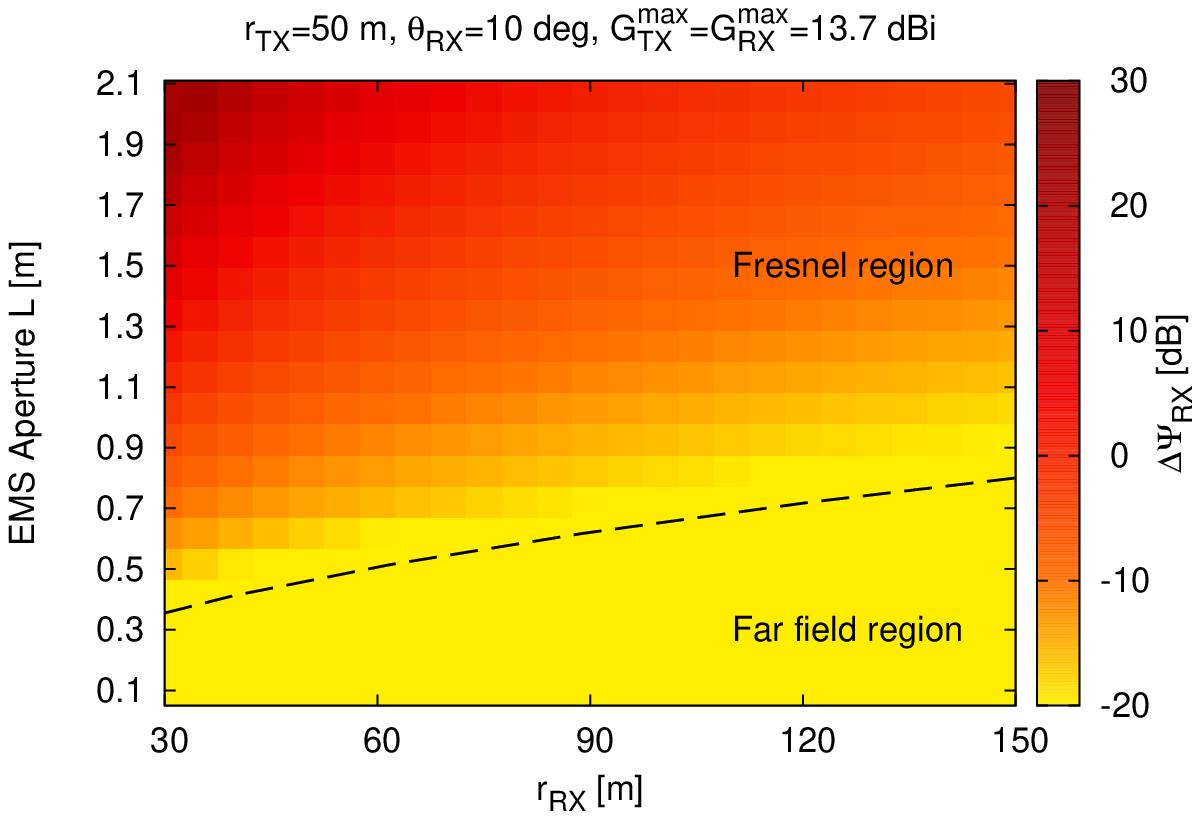}\tabularnewline
(\emph{c})\tabularnewline
\end{tabular}\end{center}

\begin{center}~\vfill\end{center}

\begin{center}\textbf{Fig. 11 - G. Oliveri et} \textbf{\emph{al.}}\textbf{,}
\textbf{\emph{{}``}}Generalized Analysis and Unified Design of \emph{EM}
Skins ...''\end{center}

\newpage
\begin{center}~\vfill\end{center}

\begin{center}\begin{tabular}{cc}
\emph{FFM}&
\emph{USM}\tabularnewline
\includegraphics[%
  width=0.45\columnwidth]{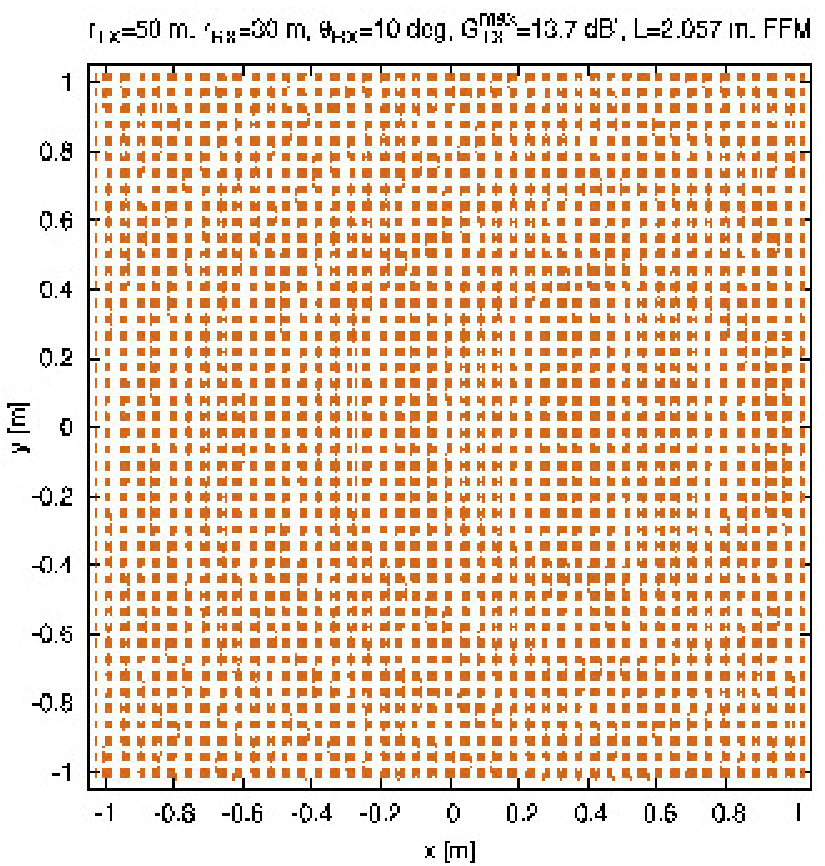}&
\includegraphics[%
  width=0.45\columnwidth]{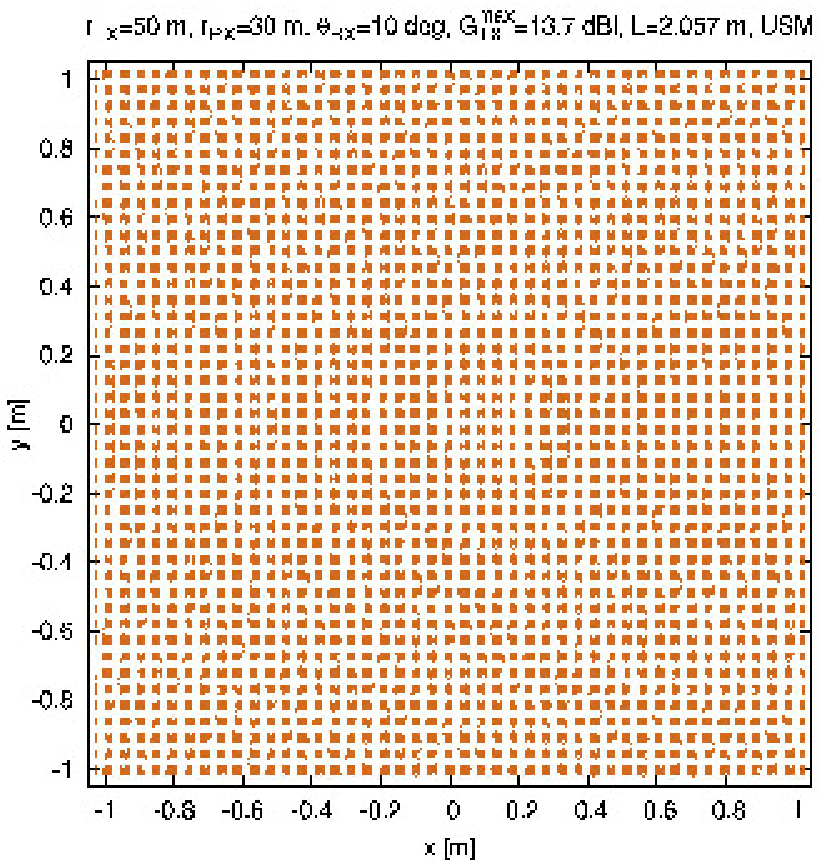}\tabularnewline
(\emph{a})&
(\emph{b})\tabularnewline
\includegraphics[%
  width=0.45\columnwidth]{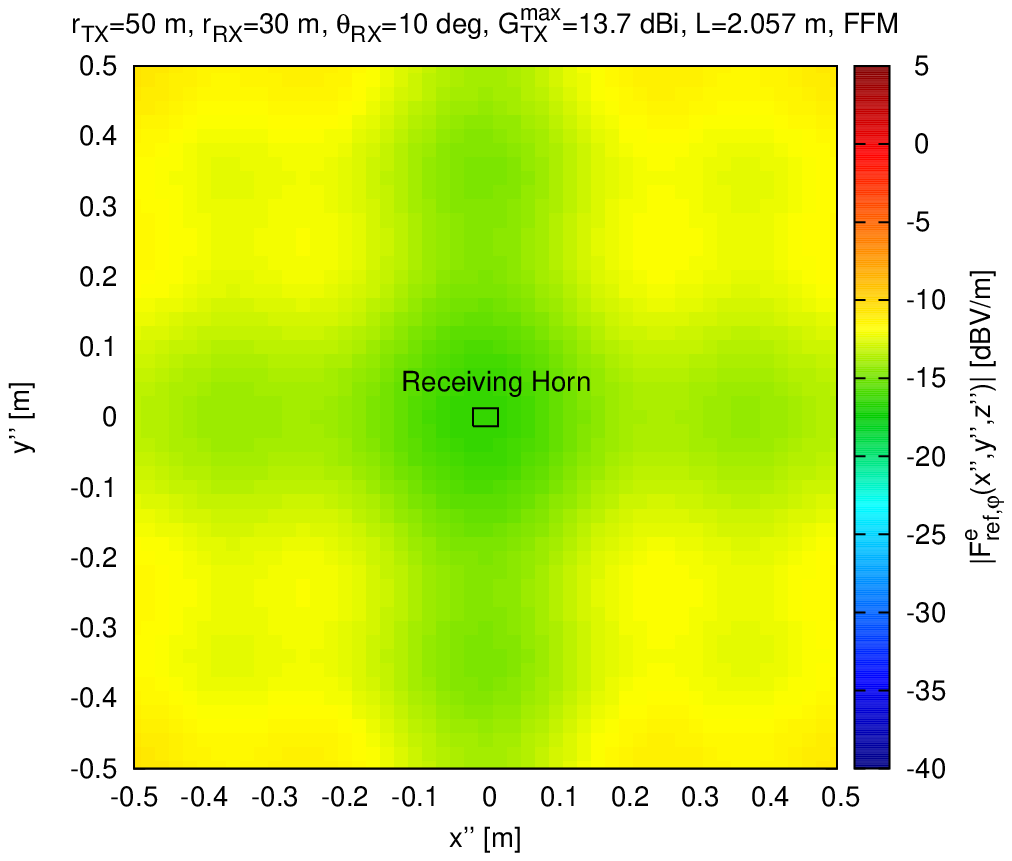}&
\includegraphics[%
  width=0.45\columnwidth]{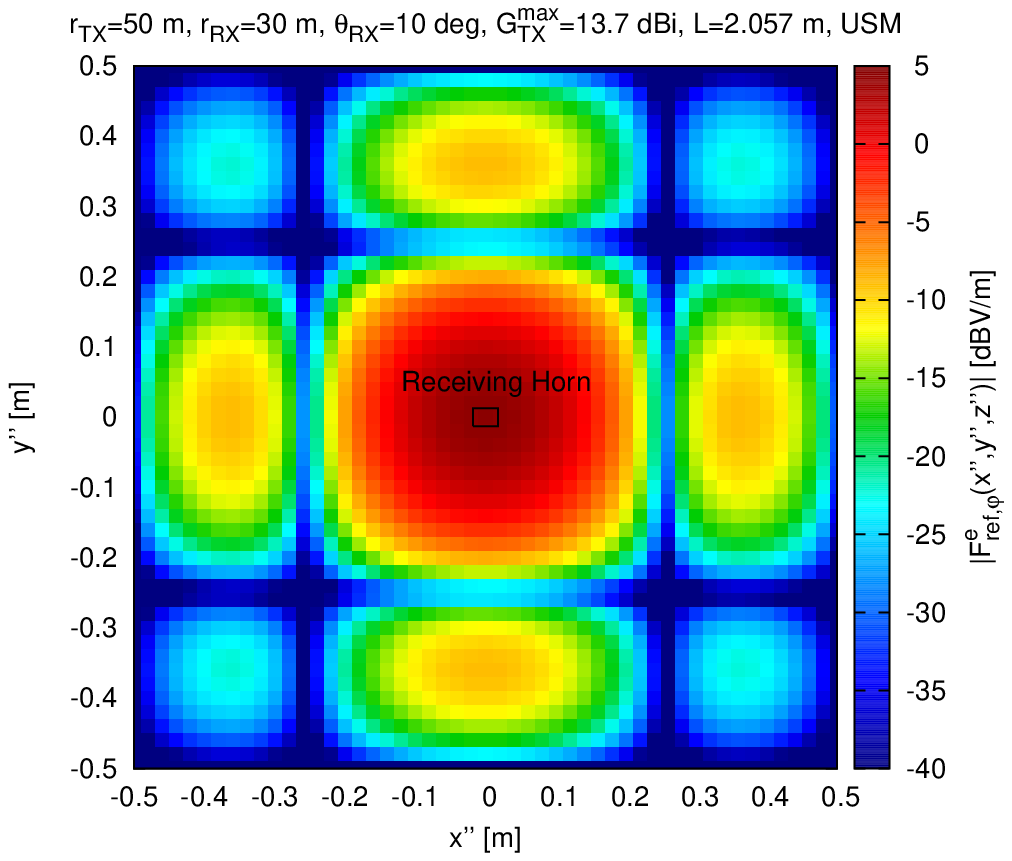}\tabularnewline
(\emph{c})&
(\emph{d})\tabularnewline
\end{tabular}\end{center}

\begin{center}~\vfill\end{center}

\begin{center}\textbf{Fig. 12 - G. Oliveri et} \textbf{\emph{al.}}\textbf{,}
\textbf{\emph{{}``}}Generalized Analysis and Unified Design of \emph{EM}
Skins ...''\end{center}

\newpage
\begin{center}\begin{tabular}{ccc}
\multicolumn{3}{c}{\includegraphics[%
  width=0.50\columnwidth]{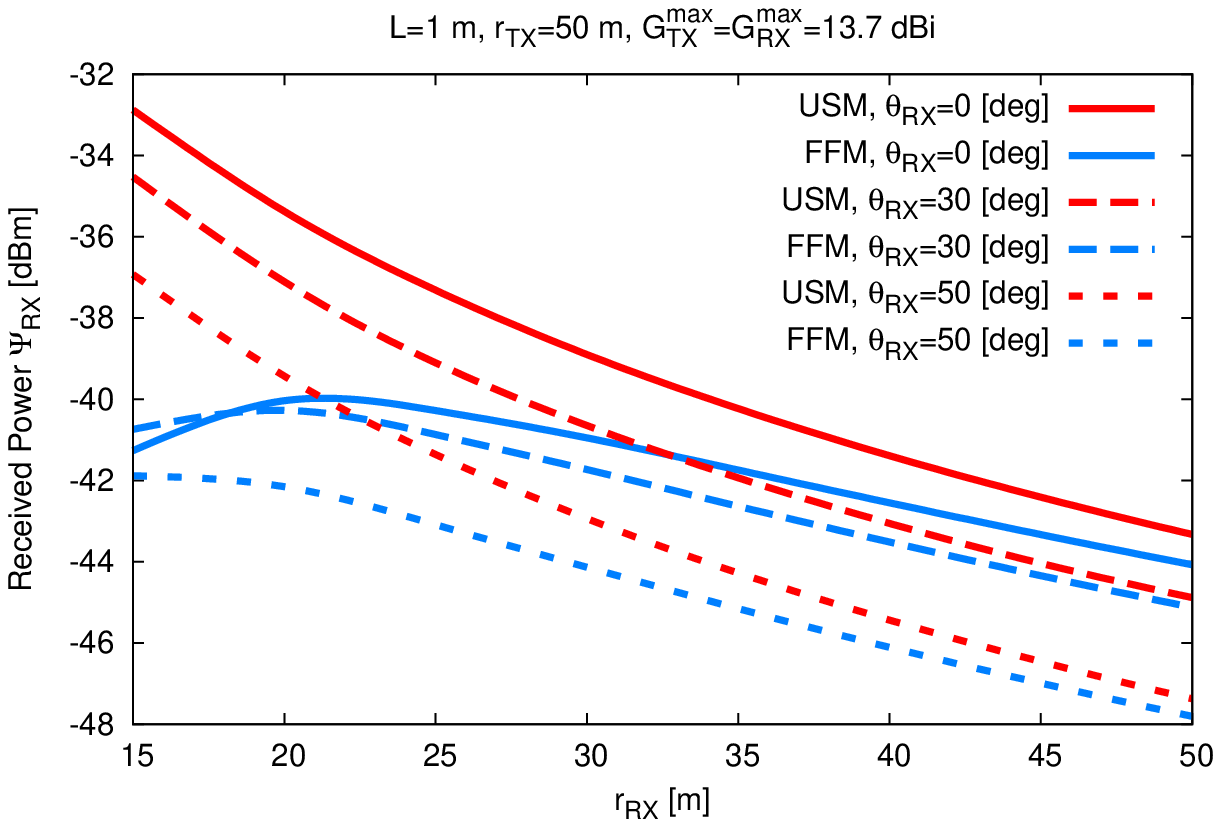}}\tabularnewline
\multicolumn{3}{c}{(\emph{a})}\tabularnewline
&
$\theta_{RX}=30$ {[}deg{]}&
$\theta_{RX}=50$ {[}deg{]}\tabularnewline
\begin{sideways}
~~~~~~~~~~~~~~~~~~\emph{FFM}%
\end{sideways}&
\includegraphics[%
  width=0.40\columnwidth]{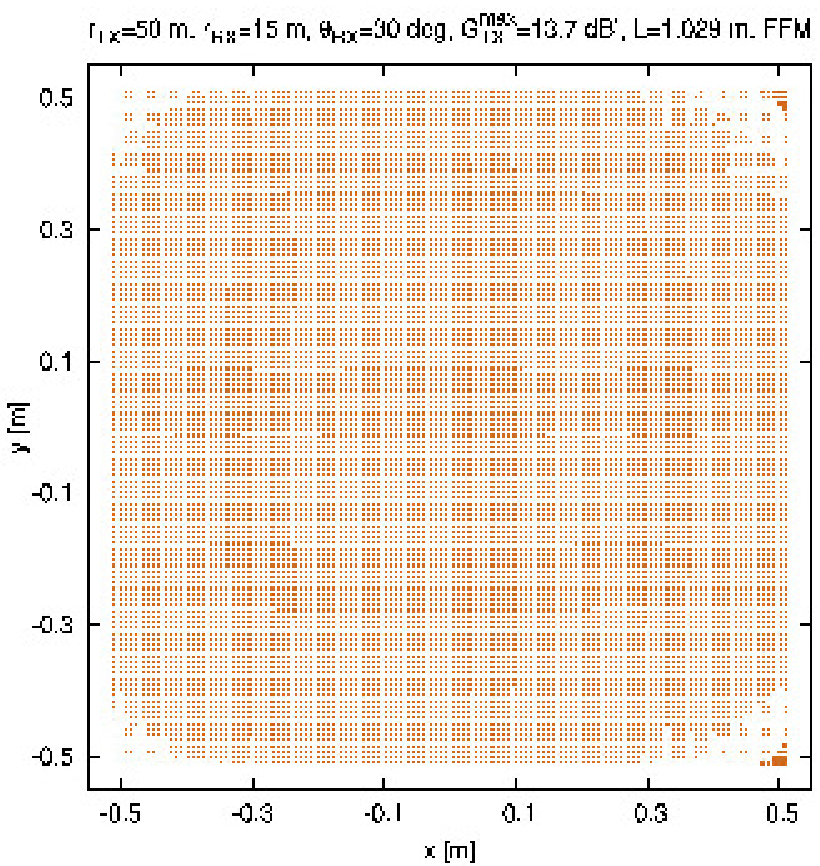}&
\includegraphics[%
  width=0.40\columnwidth]{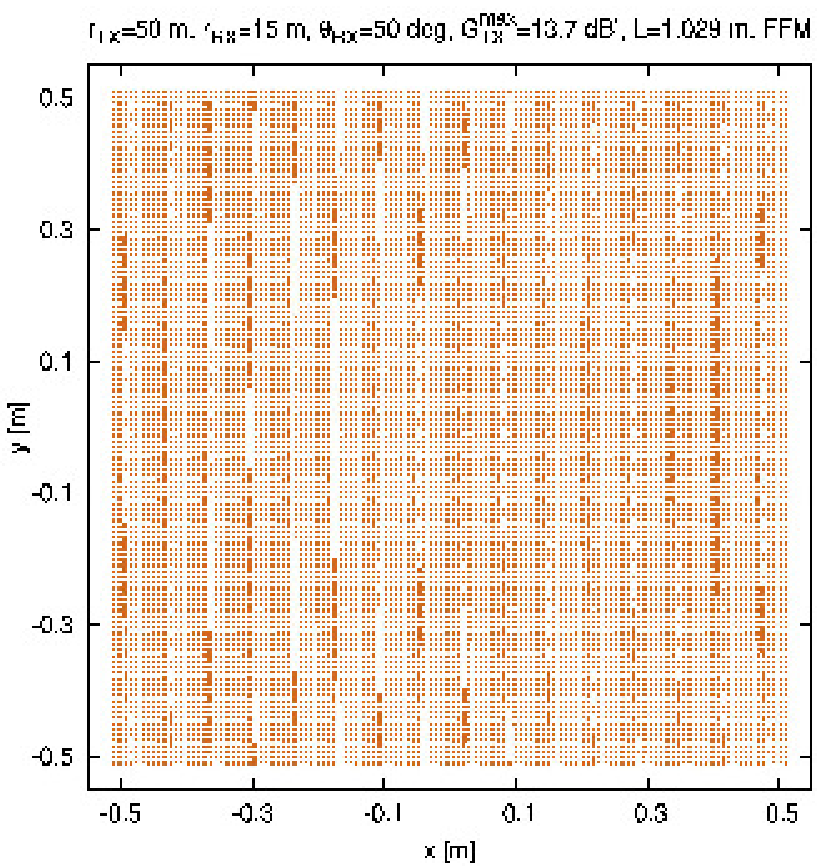}\tabularnewline
&
(\emph{b})&
(\emph{c})\tabularnewline
\begin{sideways}
~~~~~~~~~~~~~~~~~~\emph{USM}%
\end{sideways}&
\includegraphics[%
  width=0.40\columnwidth]{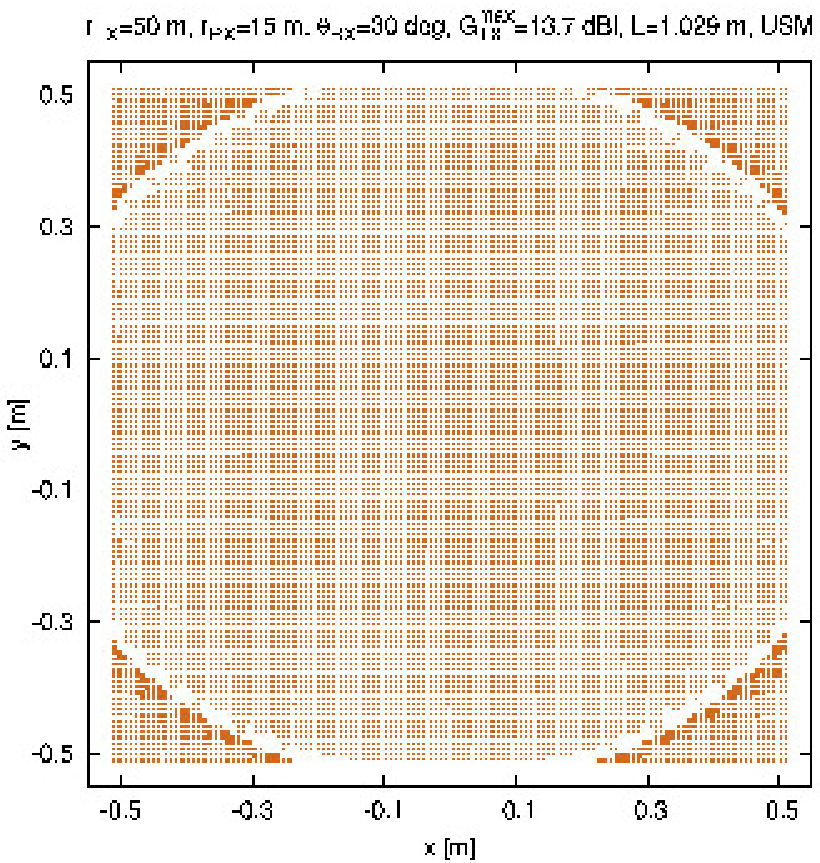}&
\includegraphics[%
  width=0.40\columnwidth]{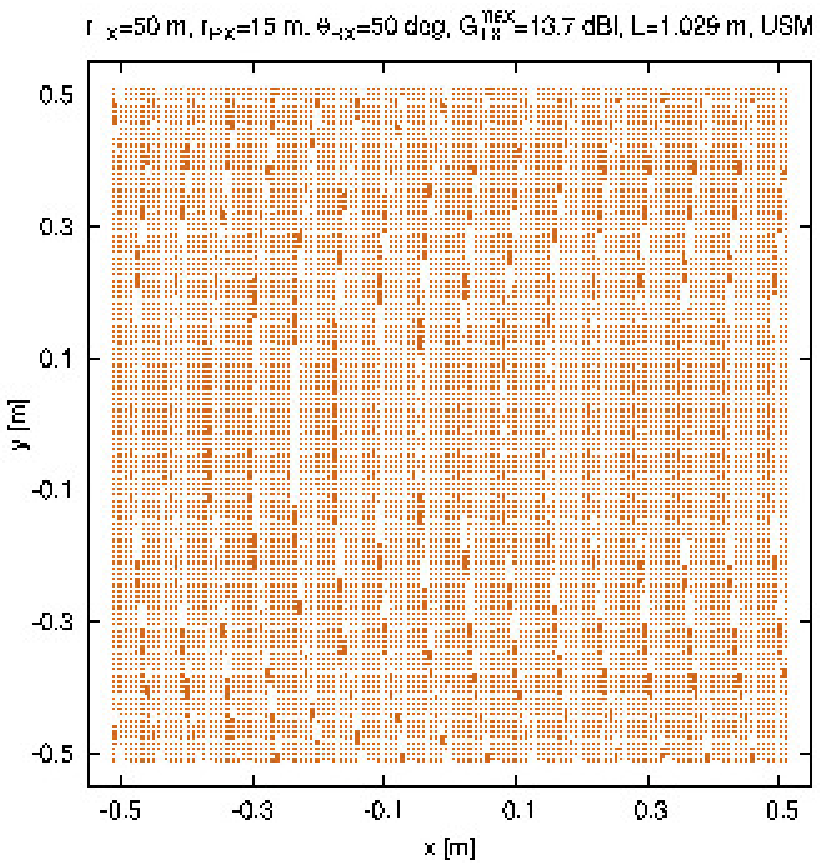}\tabularnewline
&
(\emph{d})&
(\emph{e})\tabularnewline
\end{tabular}\end{center}

\begin{center}~\vfill\end{center}

\begin{center}\textbf{Fig. 13 - G. Oliveri et} \textbf{\emph{al.}}\textbf{,}
\textbf{\emph{{}``}}Generalized Analysis and Unified Design of \emph{EM}
Skins ...''\end{center}

\newpage
\begin{center}~\vfill\end{center}

\begin{center}\begin{tabular}{ccc}
&
$\theta_{RX}=30$ {[}deg{]}&
$\theta_{RX}=50$ {[}deg{]}\tabularnewline
\begin{sideways}
~~~~~~~~~~~~~~~~~~\emph{FFM}%
\end{sideways}&
\includegraphics[%
  width=0.45\columnwidth]{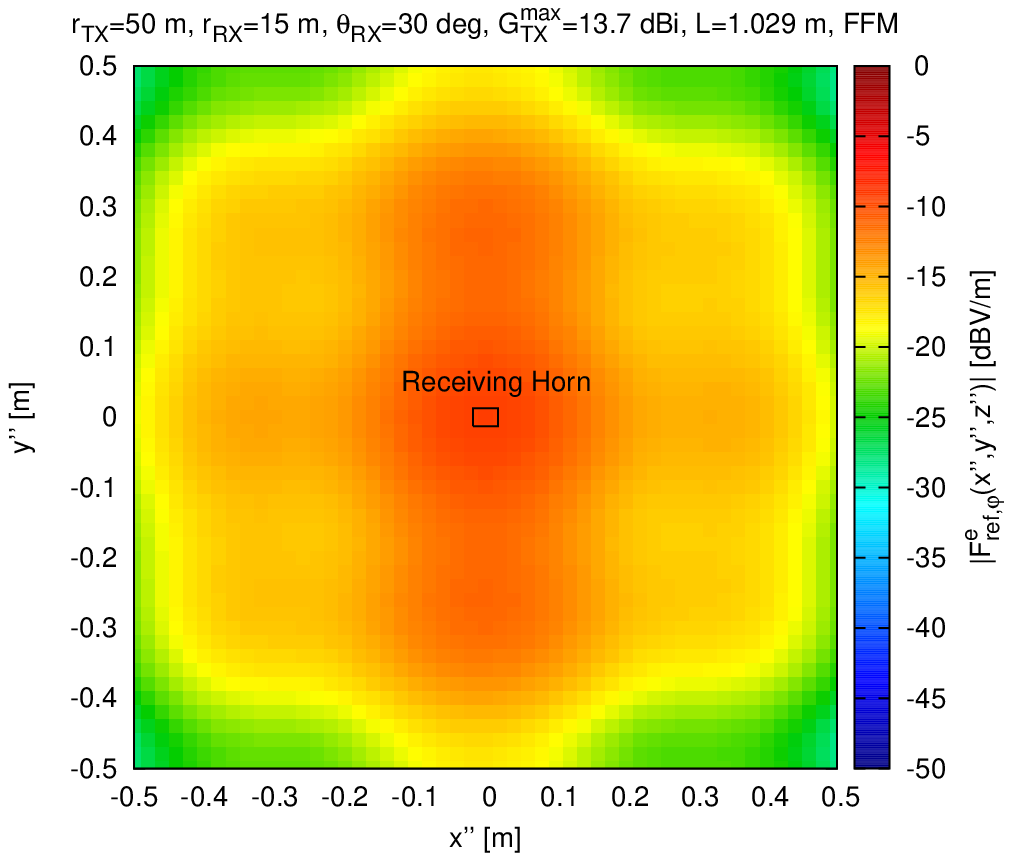}&
\includegraphics[%
  width=0.45\columnwidth]{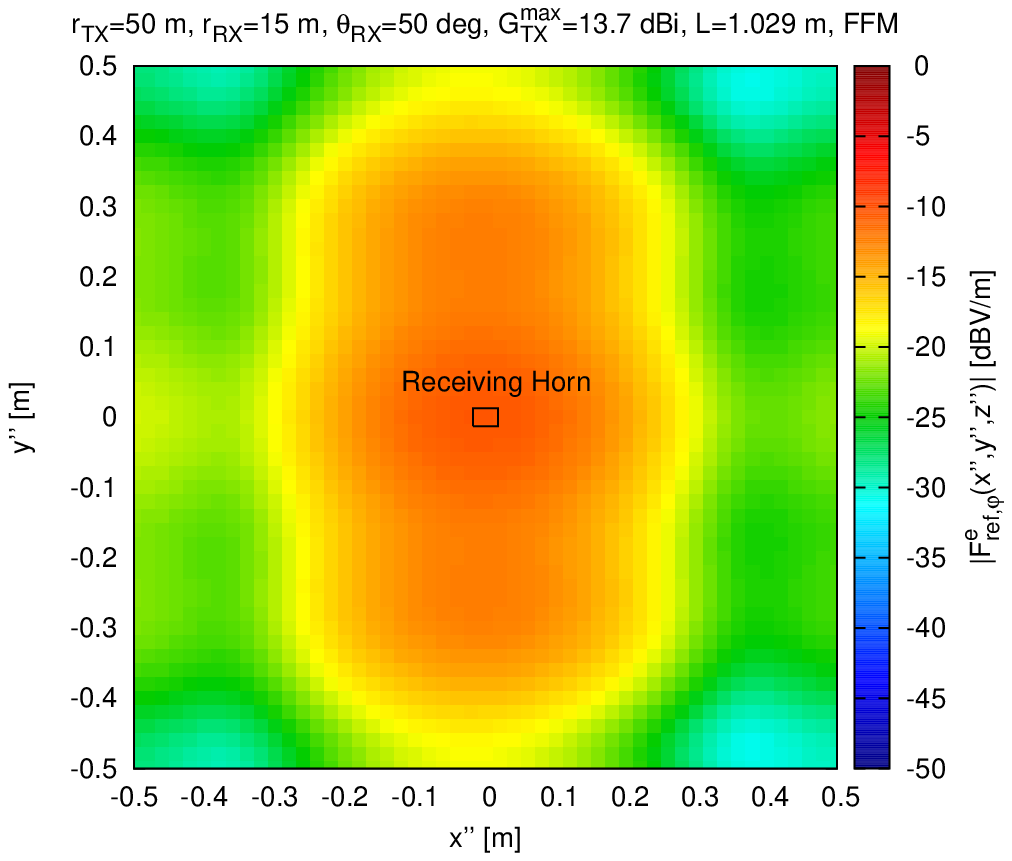}\tabularnewline
&
(\emph{a})&
(\emph{b})\tabularnewline
\begin{sideways}
~~~~~~~~~~~~~~~~~~\emph{USM}%
\end{sideways}&
\includegraphics[%
  width=0.45\columnwidth]{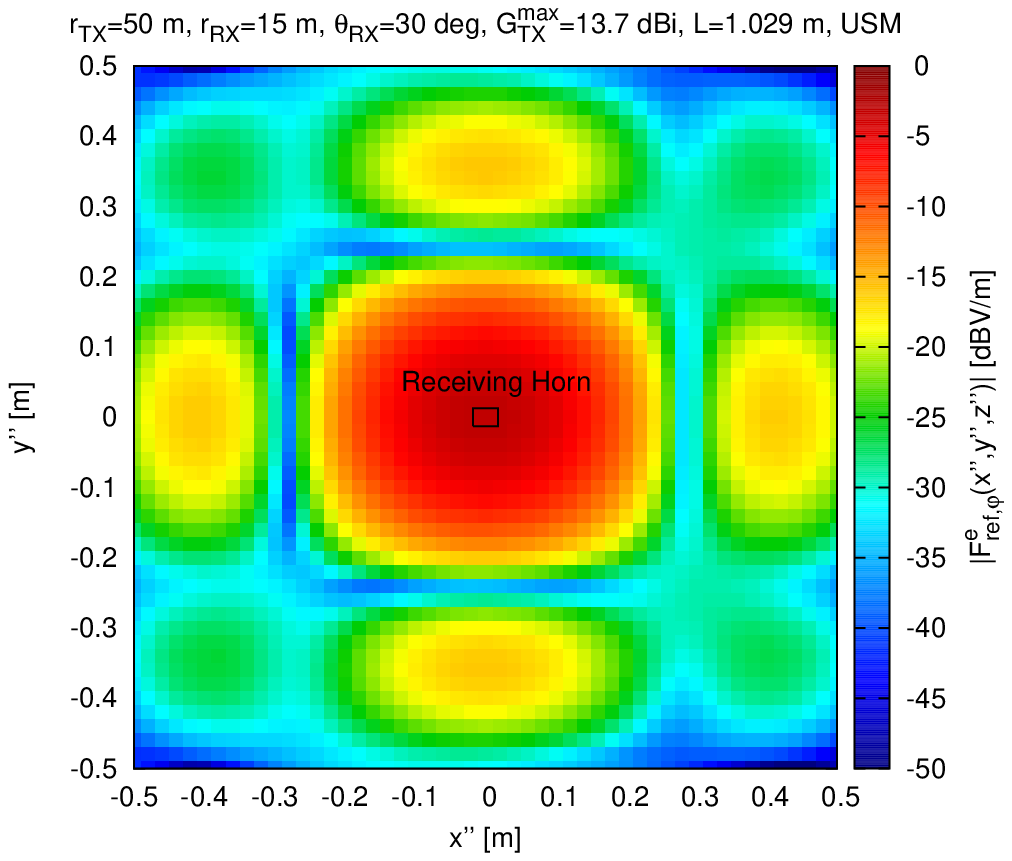}&
\includegraphics[%
  width=0.45\columnwidth]{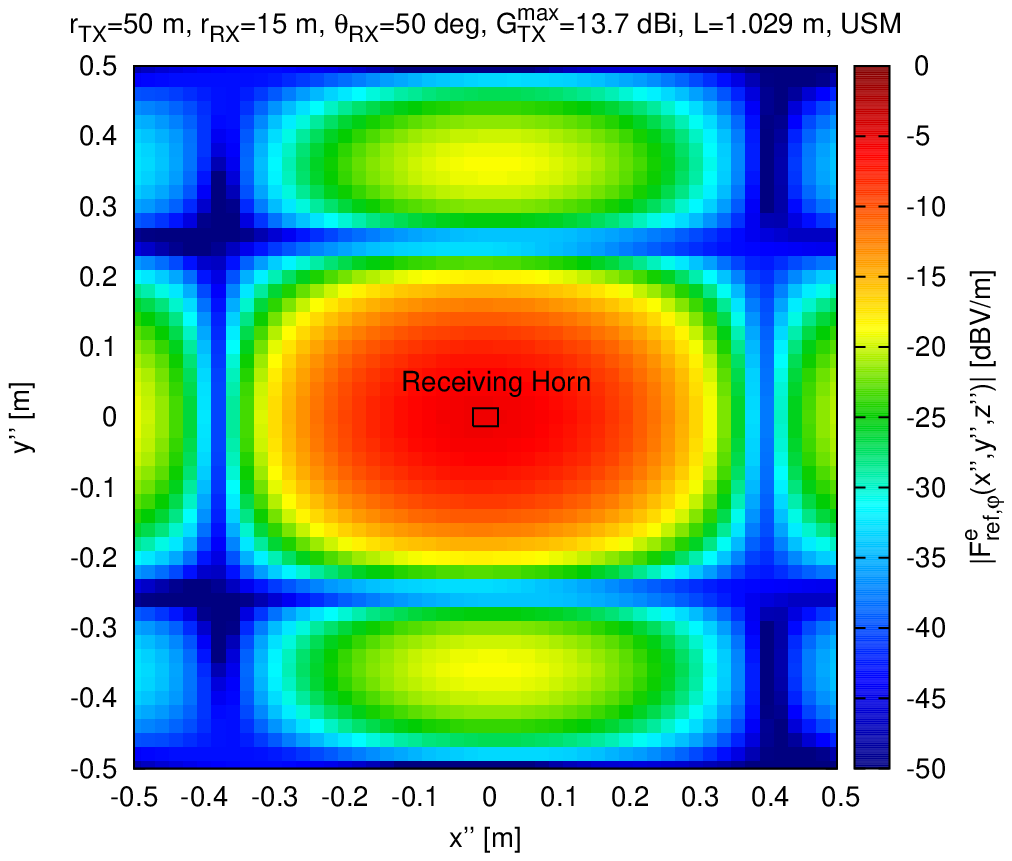}\tabularnewline
&
(\emph{c})&
(\emph{d})\tabularnewline
\end{tabular}\end{center}

\begin{center}~\vfill\end{center}

\begin{center}\textbf{Fig. 14 - G. Oliveri et} \textbf{\emph{al.}}\textbf{,}
\textbf{\emph{{}``}}Generalized Analysis and Unified Design of \emph{EM}
Skins ...''\end{center}

\newpage
\begin{center}\begin{tabular}{cc}
\multicolumn{2}{c}{\includegraphics[%
  width=0.50\columnwidth]{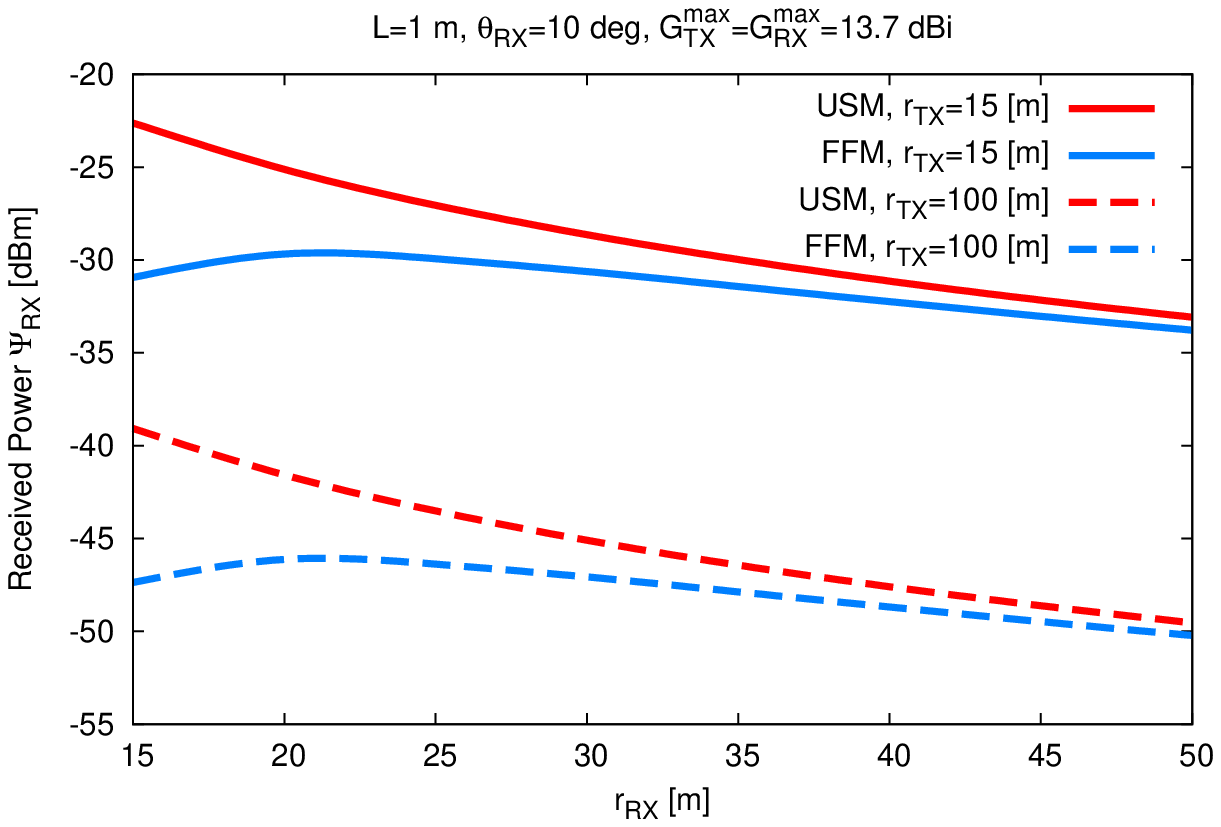}}\tabularnewline
\multicolumn{2}{c}{(\emph{a})}\tabularnewline
\emph{FFM}&
\emph{USM}\tabularnewline
\includegraphics[%
  width=0.42\columnwidth]{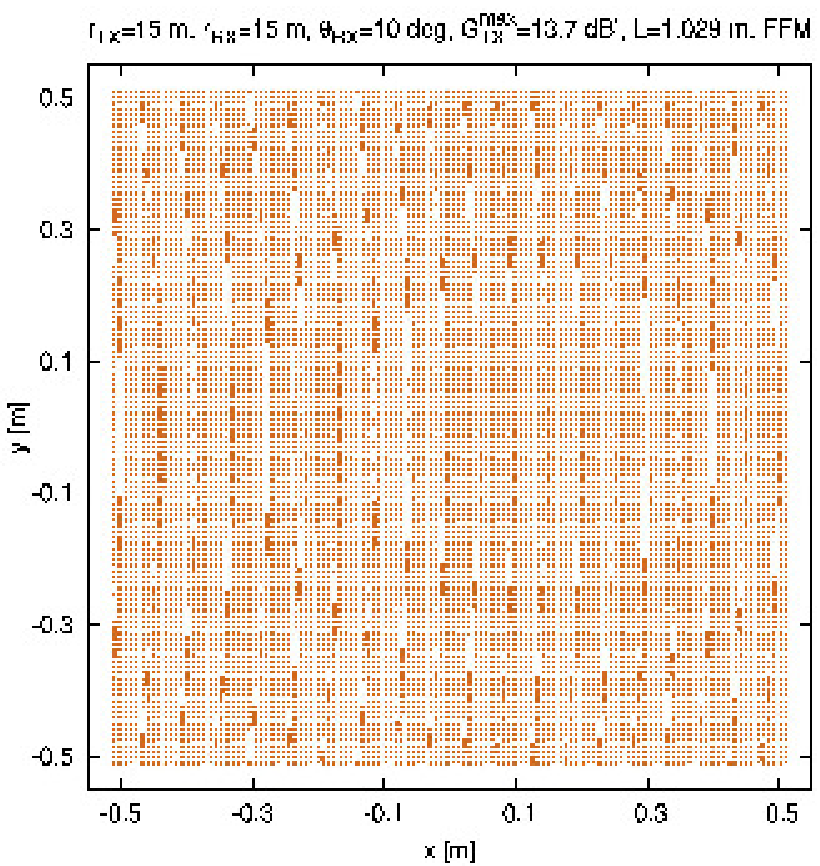}&
\includegraphics[%
  width=0.42\columnwidth]{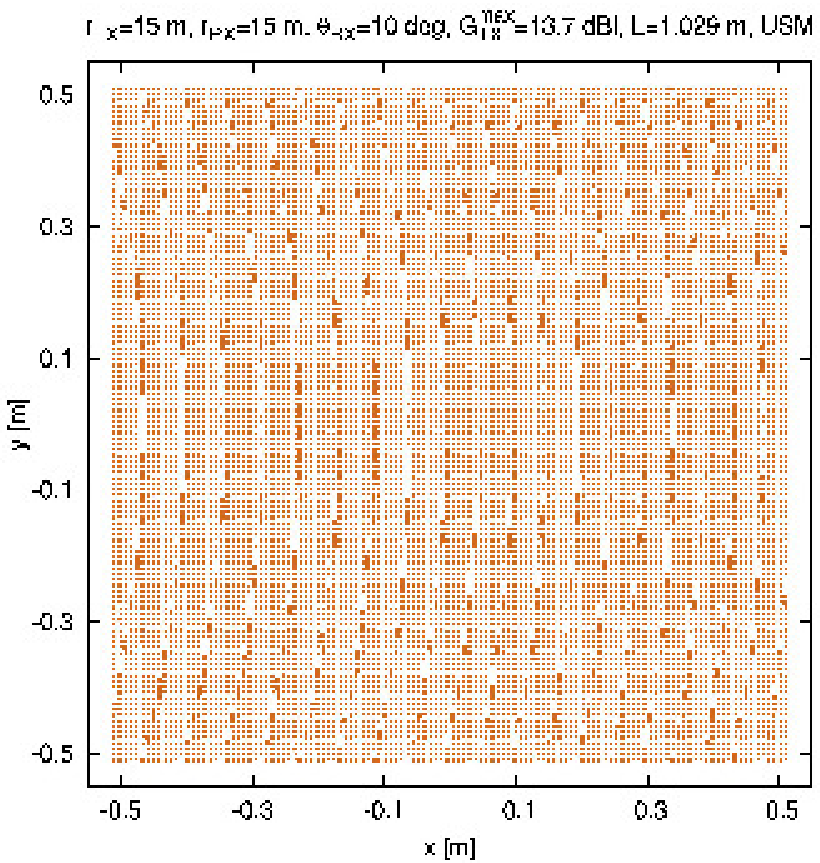}\tabularnewline
(\emph{b})&
(\emph{c})\tabularnewline
\includegraphics[%
  width=0.45\columnwidth]{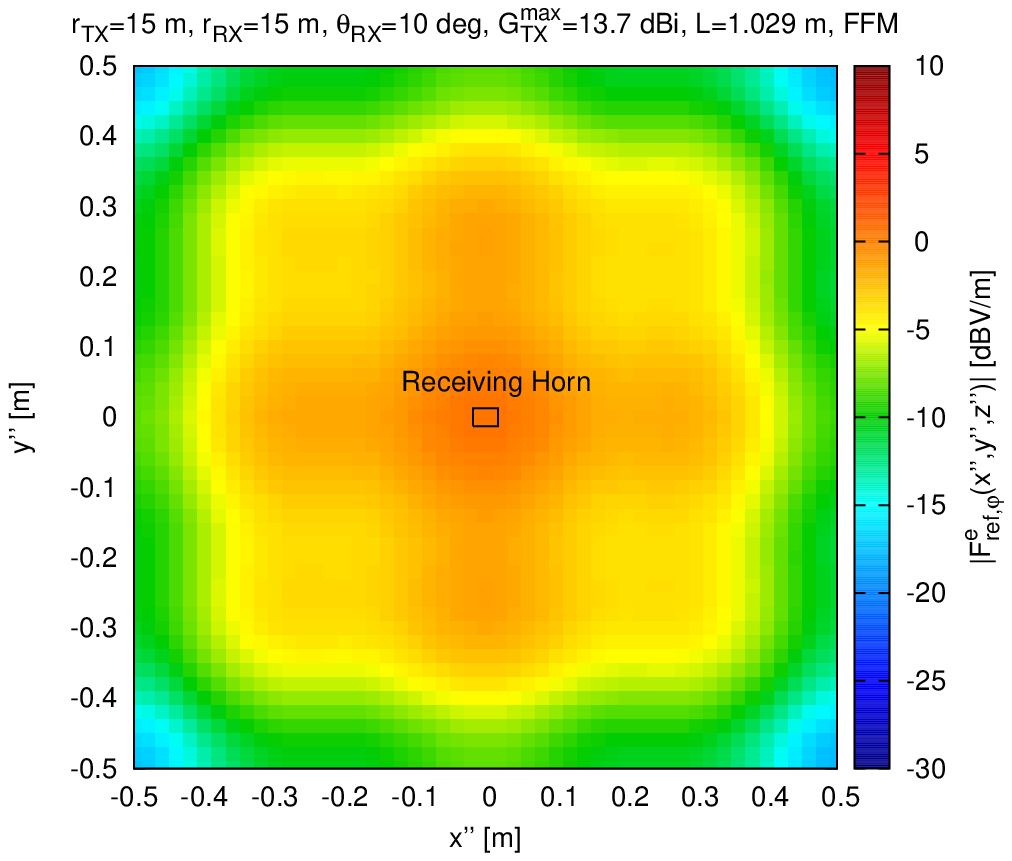}&
\includegraphics[%
  width=0.45\columnwidth]{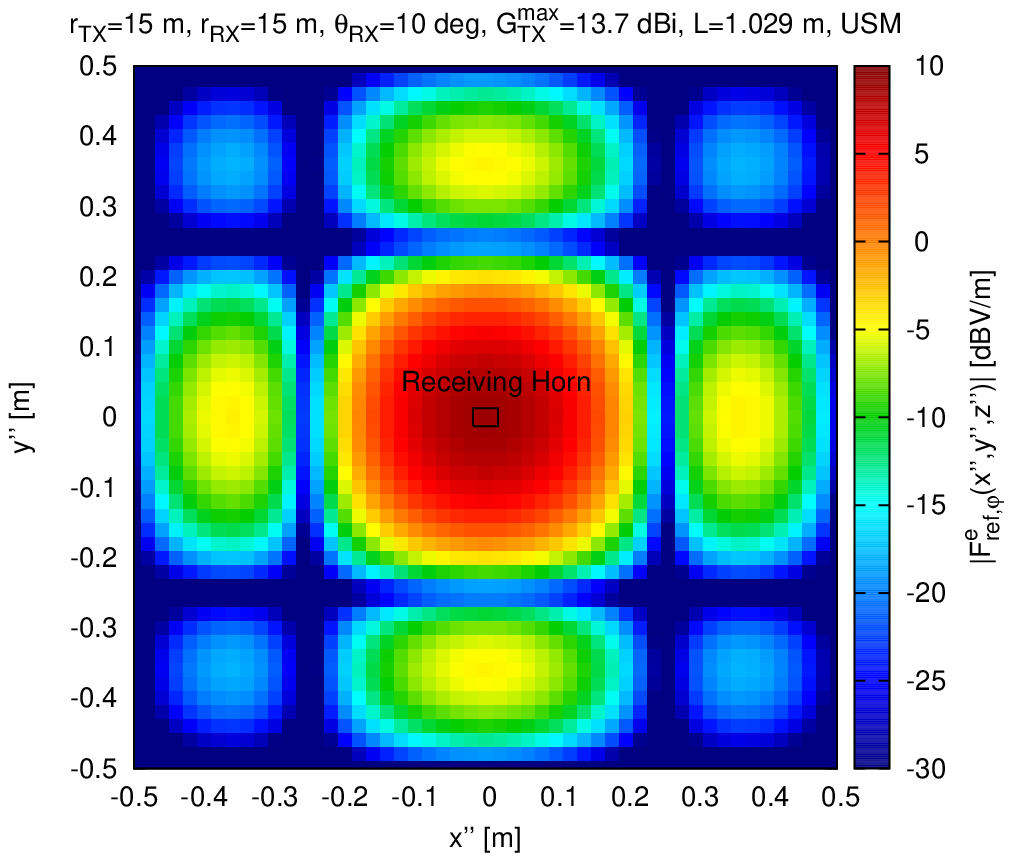}\tabularnewline
(\emph{d})&
(\emph{e})\tabularnewline
\end{tabular}\end{center}

\begin{center}~\vfill\end{center}

\begin{center}\textbf{Fig. 15 - G. Oliveri et} \textbf{\emph{al.}}\textbf{,}
\textbf{\emph{{}``}}Generalized Analysis and Unified Design of \emph{EM}
Skins ...''\end{center}

\newpage
\begin{center}~\vfill\end{center}

\begin{center}\begin{tabular}{cc}
\multicolumn{2}{c}{\includegraphics[%
  width=0.90\columnwidth]{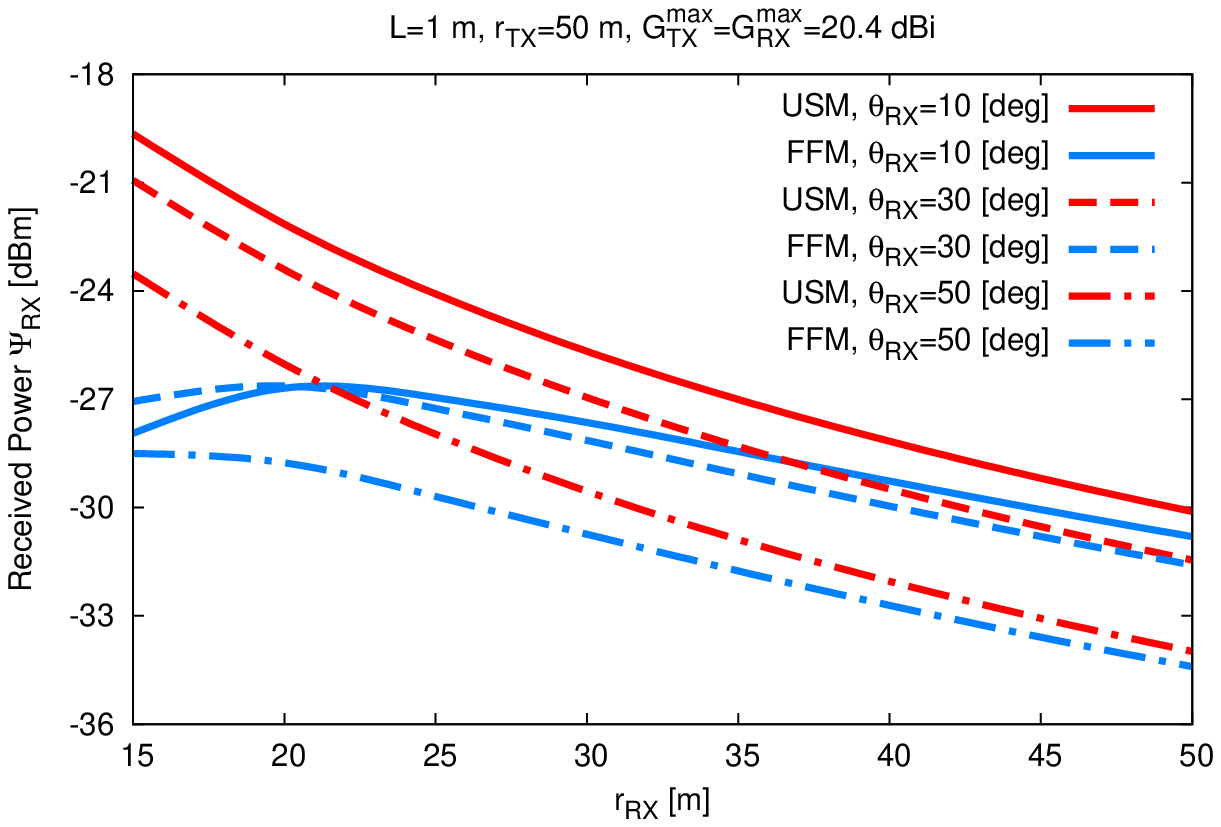}}\tabularnewline
\end{tabular}\end{center}

\begin{center}~\vfill\end{center}

\begin{center}\textbf{Fig. 16 - G. Oliveri et} \textbf{\emph{al.}}\textbf{,}
\textbf{\emph{{}``}}Generalized Analysis and Unified Design of \emph{EM}
Skins ...''\end{center}

\newpage
\begin{center}~\vfill\end{center}

\begin{center}\begin{tabular}{cc}
\multicolumn{2}{c}{\includegraphics[%
  width=1.0\columnwidth]{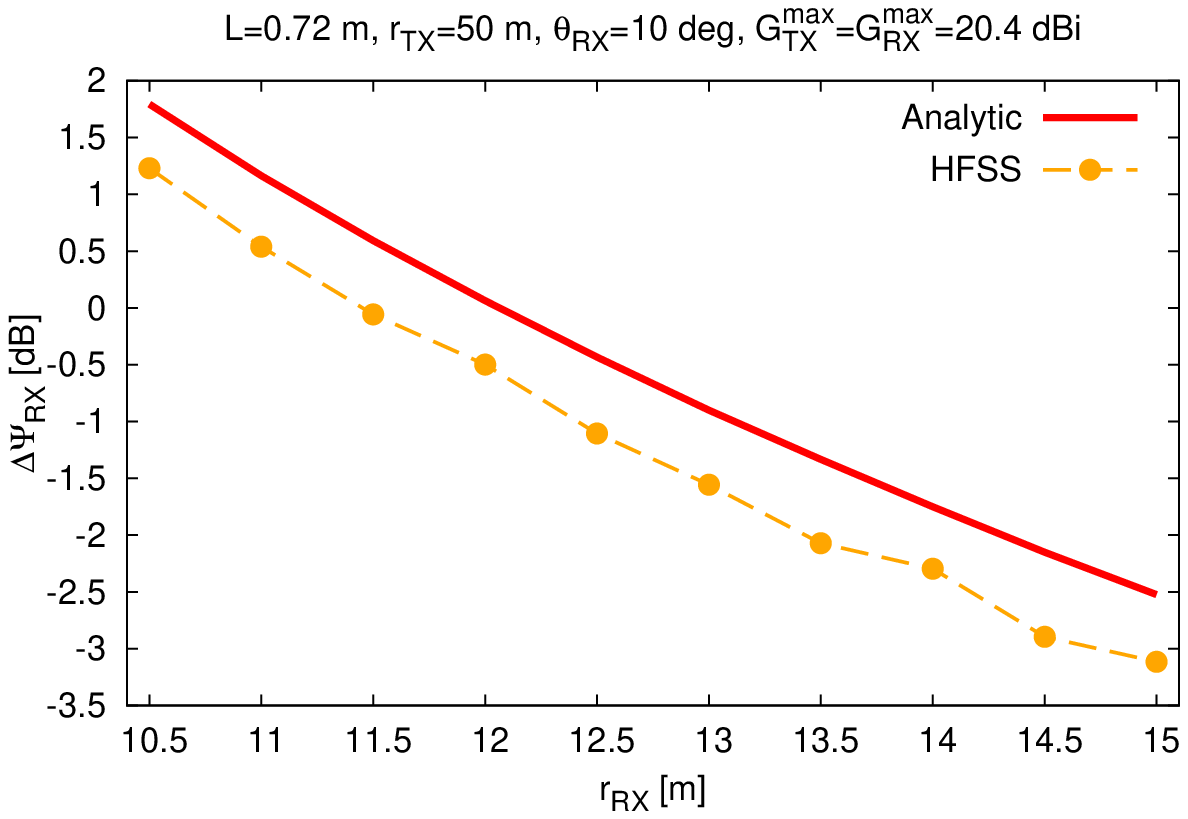}}\tabularnewline
\end{tabular}\end{center}

\begin{center}\vfill~\end{center}

\begin{center}\textbf{Fig. 17 - G. Oliveri et} \textbf{\emph{al.}}\textbf{,}
\textbf{\emph{{}``}}Generalized Analysis and Unified Design of \emph{EM}
Skins ...''\end{center}

\newpage
\begin{center}~\vfill\end{center}

\begin{center}\begin{tabular}{|c|c|c|}
\hline 
&
\textbf{Low-Gain Horn}&
\textbf{High-Gain Horn}\tabularnewline
\hline
\hline 
$f$ {[}GHz{]}&
$17.5$&
$17.5$\tabularnewline
\hline 
$c_{1}$ {[}m{]}&
$1.295\times10^{-2}$&
$1.295\times10^{-2}$\tabularnewline
\hline 
$c_{2}$ {[}m{]}&
$6.477\times10^{-3}$&
$6.477\times10^{-3}$\tabularnewline
\hline 
$\beta$ {[}m{]}&
$1.072\times10^{-3}$&
$8.897\times10^{-2}$\tabularnewline
\hline 
$\rho_{e}$ {[}m{]}&
$1.925\times10^{-2}$&
$1.041\times10^{-1}$\tabularnewline
\hline 
$\rho_{h}$ {[}m{]}&
$2.449\times10^{-2}$&
$1.137\times10^{-1}$\tabularnewline
\hline 
$b_{1}$ {[}m{]}&
$3.549\times10^{-2}$&
$7.648\times10^{-2}$\tabularnewline
\hline 
$b_{2}$ {[}m{]}&
$2.569\times10^{-2}$&
$5.976\times10^{-2}$\tabularnewline
\hline 
$G^{\max}$ {[}dBi{]}&
$13.7$&
$20.4$\tabularnewline
\hline
\end{tabular}\end{center}

\begin{center}~\vfill\end{center}

\begin{center}\textbf{Table I - G. Oliveri et} \textbf{\emph{al.}}\textbf{,}
\textbf{\emph{{}``}}Generalized Analysis and Unified Design of \emph{EM}
Skins ...''\end{center}
\end{document}